\documentclass[11pt]{article}
\usepackage{epsfig}
\usepackage{amsfonts}
\usepackage{amsmath}
\usepackage{amssymb}

\newcommand{\slashchi}{\slash \hspace{-0.4em}\chi}
\newcommand{\simge}{\hspace*{0.2em}\raisebox{0.5ex}{$>$}
     \hspace{-0.8em}\raisebox{-0.3em}{$\sim$}\hspace*{0.2em}}
\newcommand{\simle}{\hspace*{0.2em}\raisebox{0.5ex}{$<$}
     \hspace{-0.8em}\raisebox{-0.3em}{$\sim$}\hspace*{0.2em}}
\newcommand{\ep}{\epsilon}

\newcommand{\al}{\alpha}
\newcommand{\bt}{\beta}
\newcommand{\g}{\gamma}

\newcommand{\dt}{\delta}

\newcommand{\la}{\lambda}
\newcommand{\simu}{\sigma^{\mu\nu}}
\newcommand{\Fmu}{F_{\mu\nu}}
\newcommand{\Gmu}{G^a_{\mu\nu}}

\newcommand{\slashT}{\slash\hspace{-0.4em}T}

\newcommand{\slashPT}{\slash\hspace{-0.6em}P\slash\hspace{-0.5em}T}

\newcommand{\slashTsub}{\slash\hspace{-0.4em}T}
\newcommand{\h}{\frac{1}{2}}

\newcommand{\qb}{\bar q}
\newcommand{\Nb}{\bar N}
\newcommand{\Fp}{F_\pi}

\newcommand{\mb}{\bar m}
\newcommand{\tb}{\bar \theta}
\newcommand{\rb}{r(\bar \theta)}
\newcommand{\rbm}[1]{r^{-#1}(\bar \theta)}
\newcommand{\ms}{m_{\ast}}
\newcommand{\mpi}{m_{\pi}}
\newcommand{\MQCD}{M_{\mathrm{QCD}}}

\newcommand{\Or}{\mathcal O}

\newcommand{\dslash}[1]{#1 \llap{/\kern-0.5pt}}
\newcommand{\Dslash}[1]{#1 \llap{/\kern+1.2pt}}
\newcommand{\DDslash}[1]{#1 \llap{/\kern+2.3pt}}
\newcommand{\dslashh}[1]{#1 \llap{/\kern+1pt}}
\newcommand{\abs}[1]{|#1|}

\newcommand{\boldtau}{\mbox{\boldmath $\tau$}}
\newcommand{\boldpi}{\mbox{\boldmath $\pi$}}
\newcommand{\boldzeta}{\mbox{\boldmath $\zeta$}}

\def\slashchar#1{\setbox0=\hbox{$#1$}           
  \dimen0=\wd0                                    
  \setbox1=\hbox{/} \dimen1=\wd1                  
  \ifdim\dimen0>\dimen1                           
    \rlap{\hbox to \dimen0{\hfil/\hfil}}            
    #1                                             
  \else                                          
    \rlap{\hbox to \dimen1{\hfil$#1$\hfil}}        
    /                                           
 \fi}                                           %

\begin{document}

\begin{titlepage}

\vspace{2.0cm}

\begin{center}
{\Large\bf 
The Effective Chiral Lagrangian From \\
Dimension-Six Parity and Time-Reversal Violation}

\vspace{1.7cm}

{\large \bf   J. de Vries$^{1,2}$, E. Mereghetti$^{3}$, R. G. E. Timmermans$^1$, 
and U. van Kolck$^{4,5}$} 

\vspace{0.5cm}

{\large 
$^1$ 
{\it KVI, Theory Group, University of Groningen,\\
9747 AA Groningen, The Netherlands}}

\vspace{0.25cm}
{\large 
$^2$ 
{\it Nikhef, Science Park 105, \\ 
1098 XG Amsterdam, The Netherlands}}

\vspace{0.25cm}
{\large 
$^3$ 
{\it Ernest Orlando Lawrence Berkeley National Laboratory,
University of California, \\ 
Berkeley, CA 94720, USA}}

\vspace{0.25cm}
{\large 
$^4$ 
{\it Institut de Physique Nucl\'{e}aire, 
Universit\'e Paris Sud, CNRS/IN2P3,\\
91406 Orsay, France}}

\vspace{0.25cm}
{\large 
$^5$ 
{\it Department of Physics, University of Arizona,\\
Tucson, AZ 85721, USA}}

\end{center}

\vspace{1.5cm}

\begin{abstract}
We classify the parity- and time-reversal-violating operators involving 
quark and gluon fields that have effective dimension six:
the quark electric dipole moment, the quark and gluon chromo-electric dipole
moments, and four four-quark operators.  
We construct the effective chiral Lagrangian with hadronic and 
electromagnetic interactions that originate from them, which serves as the
basis for calculations of low-energy observables.
The form of the effective
interactions depends on the chiral properties of these operators.
We develop a power-counting scheme and calculate within this scheme,
as an example, the parity- and time-reversal-violating pion-nucleon 
form factor. 
We also discuss the electric dipole moments of the nucleon
and light nuclei.
\end{abstract}

\vfill
\end{titlepage}


\section{Introduction}
The Standard Model (SM) of particle physics contains, in its
minimal version, two sources of
time-reversal ($T$), or, equivalently, $C\!P$ violation.
In the electroweak
sector, the phase in the quark mixing matrix \cite{Kobayashi:1973fv} is
associated with the Jarlskog parameter 
$J_{\mathrm {CP}} \simeq 3\times 10^{-5}$ \cite{Jarlskog:1985ht}.
The strong sector contains the QCD vacuum angle $\tb$
\cite{'tHooft:1976up}, but the experimental
upper limit on the neutron electric dipole moment (EDM) shows that $\tb$ is
unnaturally small, $\tb \simle 10^{-10}$~\cite{CDVW79}. 
$C\!P$ violation within the SM is believed to be 
insufficient for a successful baryogenesis scenario \cite{Riotto:1999yt}, and
therefore new sources of $C\!P$ violation are expected in order to explain
the cosmological matter-antimatter asymmetry. 
This is not a surprise since the SM is likely but the dimension-four
part of an effective field theory (EFT) that contains higher-dimensional
operators, some of which will violate $C\!P$.

Powerful probes of such
$C\!P$ violation beyond the SM are EDMs of 
nucleons, nuclei, 
atoms, and molecules 
\cite{KhripLam1997,Pospelov:2005pr},
which violate both parity and time reversal ($\slashPT$).
Since the SM predictions from the quark mixing matrix \cite{McKellar:1987tf} 
are orders of magnitude away 
from current experimental limits, a finite EDM in upcoming
experiments would be an unambiguous sign of new
physics. In addition to impressive improvements \cite{expts} on
the time-honored EDM experiments with neutrons and neutral atoms,
in particular $^{199}$Hg,
which have resulted in very precise limits~\cite{dnbound,hgbound},
novel ideas exist for the measurement of EDMs of charged
particles, such as the proton, deuteron and helion, in storage rings
\cite{storageringexpts}.
An important question that comes up is whether, when future experiments 
measure nonzero EDMs, we will be 
able to pinpoint the microscopic source of $P$ and $T$ violation.

EDMs of strongly interacting particles 
arise from the higher-dimensional $\slashPT$
operators at the quark-gluon level. These non-renormalizable
operators might have their origin in a renormalizable theory at a higher-energy
scale, such as, for example, supersymmetric (SUSY) extensions of the SM.
At the SM scale, the most important higher-dimensional $\slashPT$ operators
should be those of dimension six, as we are not concerned
here with $C\!P$ violation in the dimension-five leptonic operator
\cite{Weinberg:1979sa}
that gives rise to neutrino masses and mixings.
{}From symmetry considerations it is found 
\cite{Buchmuller:1985jz,Rujula,RamseyMusolf:2006vr,Grzadkowski:2010es}
that the following flavor-diagonal $\slashPT$
operators appear at an effective dimension six: the quark electric dipole moment
(qEDM)~\cite{qEDM}, which couples quarks and photons; the quark chromo-electric
dipole moment (qCEDM)~\cite{cEDM}, which couples quarks and gluons; the
Weinberg operator \cite{Weinberg:1989dx}, which couples three gluons and gives
rise to a gluon chromo-electric dipole moment (gCEDM) \cite{Braaten},  
and four four-quark operators \cite{RamseyMusolf:2006vr,Ng:2011ui}.

Since it is not feasible to calculate hadronic and nuclear
properties directly from a
Lagrangian at the quark-gluon level, we use 
chiral EFT \cite{original}
---a generalization to more than one nucleon of 
chiral perturbation theory ($\chi$PT) \cite{weinberg79,Gasser}---
to translate
microscopic operators into operators that include nucleons, pions, and photons.
(For reviews, see Refs. \cite{Weinberg,Bernard:1995dp,Scherer,paulo}.)
After the translation, we are able to calculate hadronic properties directly 
from the effective Lagrangian. For the dimension-four $\tb$ term, this method
was first employed in Ref. \cite{CDVW79}, and later extended in the
context of  $SU(2)\times SU(2)$ 
\cite{Thomas:1994wi,BiraHockings,Hockings,Narison,BiraEmanuele, Mer11, Vri11b, 
Mae11, Vri12, Liu12,bsaisou} 
and of $SU(3)\times SU(3)$ \cite{su3,ottnad} $\chi$PT.

In this paper we extend the method further
to include dimension-six operators in the
framework of $SU(2)\times SU(2)$ $\chi$PT. 
(Generalization to $SU(3)\times SU(3)$ is straightforward.)
 The effective chiral Lagrangian
includes not only interactions that stem from spontaneous chiral-symmetry
breaking and are therefore chiral invariant, but also interactions
that break chiral symmetry in the same way as chiral-symmetry-breaking
operators at the QCD level.
Since the dimension-six operators break chiral symmetry differently 
from each other and from
the $\tb$ term, they will generate different low-energy hadronic interactions. 
Given enough observables it should be thus possible to separate
the various $\slashPT$ sources.

In addition to constructing the Lagrangian, we need to organize 
in leading order (LO), next-to-leading order (NLO), {\it etc.}
the various effective $\slashPT$ operators that appear. This is done 
according to the estimated size of their contributions to observables.
In order to get a consistent, manifest power counting we work in the 
heavy-baryon framework \cite{Jenkins:1990jv} wherein the nucleon mass
has been eliminated from the nucleon propagator. This framework has a
transparent power counting and greatly simplifies loop calculations, 
but there are some complications when one goes to subleading orders
in the Lagrangian. These problems can be solved by demanding that the 
Lagrangian obeys reparametrization invariance (RPI) 
\cite{ManoharLuke}.
This puts constraints on certain coefficients of operators,
which we construct up to NNLO. 

In Refs. \cite{Vri11a, Mer11, Vri11b, Mae11, Vri12} 
we have used some of the effective interactions 
to calculate
the EDMs of the nucleons and the lightest nuclei. 
It was found that 
in LO 
they depend on six low-energy constants (LECs),
which for each fundamental $\slashPT$ source have different relative sizes. 
The idea is that, from EDM measurements, the LECs can be inferred and 
the dominant fundamental source identified. 
For example,
if the deuteron EDM is significantly larger than the sum of the nucleon EDMs,
this points towards new physics in form of a qCEDM \cite{Vri11b, Vri12} 
or, as we will demonstrate here, a particular isospin-breaking $\slashPT$ 
four-quark operator.
If the deuteron EDM is well approximated by the nucleon EDMs, 
but the helion (triton) EDM is far away from the neutron (proton) EDM, 
it is a hint for the SM $\tb$ term.
The deuteron magnetic quadrupole moment (MQM), 
if experimentally accessible, could
play an important role as well \cite{Vri11b, Liu12}.

In this article, we construct the full Lagrangian necessary to perform 
these and other low-energy $\slashPT$ calculations.
We follow the approach outlined in Refs. \cite{Hockings,BiraEmanuele}, 
where the $\chi$PT Lagrangian originating from the $\tb$ term was derived,
and the higher-dimensional interactions were examined. 
We identify, for each source, the size of the six
$\slashPT$ interactions relevant for light nuclear EDMs. 
Apart from that, we construct also all other
operators that appear at the same order. 
These operators could play a role in the calculation
of other observables that violate $P$ and $T$ such as the EDMs of 
heavier nuclei 
or of ions and atoms. $\slashPT$ form factors 
and scattering observables could depend on these operators as well. 
An important ingredient in calculating hadronic and nuclear
$\slashPT$ observables is the $\slashPT$ pion-nucleon form factor. 
We investigate this form factor here.
Finally, we extend our earlier light-nuclear EDM calculations
to a four-quark operator that has recently been shown to arise
below the SM scale from weak-boson exchange \cite{Ng:2011ui}.

Our paper is organized as follows. In Secs.~\ref{II} and \ref{Match} 
we discuss the
QCD Lagrangian, the $\tb$ term, and the possible higher-dimensional
$\slashPT$ operators. In Sec.~\ref{III} we briefly discuss
$SU(2)\times SU(2)$ $\chi$PT, and how to incorporate the $\slashPT$
operators in this framework. 
The bulk of the article is Secs.~\ref{pionsector}, \ref{piNsector},
and \ref{Ngammasector},
where we construct the 
$\chi$PT Lagrangian up to NNLO
including the removal of pion tadpoles that proliferate when $\slashPT$ appears
together with isospin violation.
In Sec.~\ref{PNFF} 
we use the constructed Lagrangian to calculate
the $\slashPT$ pion-nucleon form factor (PNFF), and
in Sec.~\ref{FQLREDM} we give the
electric dipole form factor (EDFF) of the nucleon
in case of the specific
$\slashPT$ four-quark operator of Ref. \cite{Ng:2011ui}. 
In Sec.~\ref{NNsector} the extra elements that arise in nuclear systems
are addressed.
We discuss our results and conclude in Sec.~\ref{discussion}.

\section{The underlying quark-gluon Lagrangian}
\label{II}

Before we discuss higher-dimensional $\slashPT$ operators,
we set our notation by recalling the main SM ingredients.
The SM Lagrangian is completely determined 
by gauge symmetry
with gauge group $SU_c(3) \times SU_L(2) \times U_Y(1)$,
by the matter content,
and by the requirement of renormalizability.

In its minimal form, which we assume in this paper,
the matter content consists of three generations of 
leptons and quarks, and one scalar doublet. 
Quark and lepton fields carry a generation index $r=1,2,3$, which we 
will often leave implicit, 
running over the three generations of up-type quarks $u = (u,c,t)$, 
down-type quarks $d = (d,s,b)$, charged
leptons $e = (e,\mu,\tau)$ and 
neutrinos $\nu = (\nu_e,\nu_\mu,\nu_{\tau})$.
The left-handed fermions
are doublets of $SU_L(2)$,
\begin{equation}
q_L = \left(\begin{array}{c}
u_L \\
d_L
\end{array}
\right), \qquad 
l_L = \left(\begin{array}{c}
\nu_L \\
e_L
\end{array}\right),
\end{equation}
while the right-handed fields $u_R$, $d_R$ and $e_R$ are singlets.
The field $\varphi$ denotes an $SU_L(2)$ doublet of scalar fields 
$\varphi^J$, $J=1,2$. 
For convenience we define
$\tilde \varphi^I = \ep^{IJ} \varphi^{J*}$,
where $\ep^{IJ}$ is the antisymmetric tensor in two dimensions
($\ep^{12}=+1$).
Left- and right-handed quarks are in the fundamental representation of 
$SU_c(3)$, while leptons and scalars are singlets.
We sometimes group the right-handed 
up- and down-type quarks in a doublet $q_R$,
and define quark doublets $q = q_L + q_R$.
The hypercharge assignments under the group $U_Y(1)$ are 
$1/6$, $2/3$, $-1/3$, $-1/2$, $-1$, and $1/2$ 
for $q_L$, $u_R$, $d_R$, $l_L$, $e_R$, and $\varphi$, respectively.

We denote the gauge bosons associated with the gauge groups
$SU_c(3)$, $SU_L(2)$, and $U_Y(1)$ by, respectively,
$G^a_{\mu}$, $W^i_{\mu}$, and $B_{\mu}$,
with $a=1, \ldots, 8$ and $i=1,2,3$.
Gauge invariance is most easily imposed by employing covariant combinations
of the gauge and matter fields.
The covariant derivative of matter fields is
\begin{eqnarray}\label{Cov.1}
D_{\mu} = \partial_{\mu}  - i \frac{g_s}{2}\, G^a_{\mu} \lambda^a 
- i \frac{g}{2}\, W^i_{\mu} \tau^i - i g^{\prime} Y B_{\mu},
\end{eqnarray}
where $g_s$,
$g$, and $g^{\prime}$ are the $SU_c(3)$, $SU_L(2)$, and $U_Y(1)$
coupling constants; 
and $\lambda^a/2$ and $\tau^i/2$ are $SU(3)$ and $SU(2)$ generators, 
in the
representation of the field on which the derivative acts.   
For example, for left-handed quarks 
$\lambda^a$ and $\tau^i$ are, respectively,
the Gell-Mann color 
and Pauli isospin matrices. 
The 
field strengths are
\begin{eqnarray}
G^a_{\mu \nu} &=& \partial_{\mu} G^a_{\nu} - \partial_{\mu} G^a_{\mu}
- g_s f^{a b c} G^b_{\mu} G^c_{\nu}, \\
W^i_{\mu \nu} &=& \partial_{\mu} W^i_{\nu} - \partial_{\nu} W^i_{\mu} -
g \ep^{i j k} W^j_{\mu} W^k_{\nu}, \\
B_{\mu \nu} &=& \partial_{\mu} B_{\nu} - \partial_{\nu} B_{\mu},
\end{eqnarray}
with
$f^{abc}$ and $\ep^{ijk}$ denoting 
the $SU(3)$ and $SU(2)$ structure constants.

The SM Lagrangian is expressed in terms of all possible dimension-four 
gauge-invariant operators: 
\begin{eqnarray}
\mathcal L_{\rm SM} &=&  
-\frac{1}{4} 
\left(G^a_{\mu \nu} G^{a\, \mu \nu}
+W^{i}_{\mu\nu} W^{i\, \mu \nu}
+B_{\mu \nu} B^{\mu \nu}\right)
\nonumber \\ 
& &
+ \bar q_L i \slashchar{D}\, q_L + \bar u_R i \slashchar{D}\, u_R 
+ \bar d_R i \slashchar{D}\, d_R
+ \bar l_L i \slashchar D\, l_L + \bar e_R i \slashchar D\, e_R 
+  D_{\mu} \varphi^{\dagger} D^{\mu} \varphi 
\nonumber \\  
& & + \mu^2 \varphi^{\dagger} \varphi -
\frac{\lambda}{2} (\varphi^{\dagger} \varphi)^2 
\nonumber \\
& &  
- \bar q_L Y^u \tilde \varphi u_R -  \bar q_L Y^d \varphi d_R  
- \bar l_L Y^e \varphi e_R  + \textrm{H.c.} 
\nonumber \\ 
& & -\frac{\ep^{\mu \nu \alpha \beta}}{64 \pi^2} \left(
g_s^2 \theta \, G^a_{\mu \nu} G^{a}_{\alpha \beta}
+g^2 \theta_w \, W^i_{\mu \nu} W^{i}_{\alpha \beta}
+g^{\prime\, 2} \theta_b \, B_{\mu \nu} B_{\alpha \beta}  
\right),
\label{SM}
\end{eqnarray}
where $\ep^{\mu \nu \alpha \beta}$ is the totally antisymmetric symbol
in four dimensions ($\ep^{0123}=1$).

The first line of Eq. \eqref{SM} 
contains the kinetic terms and self-interactions of the $SU_c(3)$,
$SU_L(2)$, and $U_Y(1)$ gauge bosons.
The second line 
contains the kinetic energy and the gauge couplings of fermions and scalars. 
These couplings are completely
determined by gauge invariance.
The 
terms of the third line 
form the scalar potential.
With the parameter $\lambda>0$ for stability
and the parameter $\mu^2 >0$, 
the scalar field acquires a vacuum expectation value
$v = \sqrt{\mu^2/\lambda}$;
the Higgs boson $h(x)$ represents fluctuations around this vacuum,
\begin{equation}
\varphi  = \frac{v}{\sqrt{2}} U(x) 
\left(\begin{array}{c} 0 \\ 1 + \frac{h(x)}{v} \end{array}\right),
\end{equation}
where $U(x)$ is an $SU(2)$ matrix, which encodes the three Goldstone bosons. 
The Goldstone
bosons are not physical degrees of freedom, so that with a particular choice 
of gauge, the unitarity gauge, $U(x)$ can be set to one. 
In this gauge, the Goldstone bosons are ``eaten'' by the longitudinal
polarizations of the massive vector bosons. 
We will often use this gauge to discuss the structure of dimension-six 
operators.
After electroweak symmetry breaking, the scalar field kinetic energy provides 
a mass term for the weak gauge bosons. 
It is convenient to express the fields $W^3_{\mu}$ and $B_{\mu}$  in terms
of the physical photon and $Z$-boson fields, 
\begin{eqnarray}
W^3_{\mu}  &=& \cos\theta_W Z_{\mu} + \sin\theta_W A_{\mu}, 
\\
B_{\mu} & = & \cos\theta_W A_{\mu} -\sin\theta_W Z_{\mu}, 
\label{Zphoton}
\end{eqnarray}
where the weak mixing angle $\theta_W$ is given,
together with the proton charge $e>0$, by
the couplings $g$ and $g^{\prime}$ via
\begin{equation}\label{eq:2.2.24}
g = -\frac{e}{\sin\theta_W}, \qquad g^{\prime} = -\frac{e}{\cos\theta_W}.
\end{equation}
The charged $W$-boson fields are defined as
\begin{equation}
W^\pm_{\mu}  = \frac{1}{\sqrt{2}}\left(W^1_\mu \mp i W^2_\mu\right).
\end{equation}

The next dimension-four operators one can write are the Yukawa couplings 
of the fermions to
the scalar boson in the fourth line of Eq. \eqref{SM},
via matrices $Y^{u,d,e}$. 
After electroweak symmetry breaking, they generate
the quark and lepton masses. 
By means of unitary transformations on the quark and lepton fields, 
it is always possible to 
make the fermion mass matrices diagonal and real, up to a common phase.
For leptons, these transformations do not leave any trace, while for quarks
the price to pay for the diagonalization of the mass matrix is
that the interaction of the charged $W$ bosons  
with the quarks is no longer 
flavor diagonal, an effect that can be obtained by replacing  
$d_L^r$ in Eq. \eqref{SM} by $V_{rs} d^s_L$,
where $V_{rs}$ is the unitary Cabibbo-Kobayashi-Maskawa (CKM) matrix. 
For three generations
of quarks, the CKM matrix has one complex phase
\cite{Kobayashi:1973fv}, which is responsible for the observed $C\!P$ 
violation in the kaon and $B$-meson systems. 
However, the contribution of the CKM phase to
nuclear EDMs is orders of magnitude smaller than the current experimental 
sensitivity, and we
will neglect it in the rest of the paper.
The second $\slashPT$ parameter in the SM Lagrangian
is the global phase of the quark mass matrices. 
It can be eliminated  \cite{Baluni} by an axial rotation of all the
quark fields $q_L \rightarrow e^{i\rho} q_L$, $q_R \rightarrow e^{-i\rho} q_R$. 
Such transformation
is anomalous \cite{Fujikawa:1979ay}, 
and its net effect is to shift the coefficients of the 
$P$- and $T$-odd gluon operator in
the fifth line of Eq. \eqref{SM}
from $\theta$ to $\bar\theta =\theta +n_f\rho$, 
where $n_f$ is the number of quark flavors.
Operators in that line are total derivatives, but, for non-Abelian gauge fields,
they  contribute to the action through extended field configurations,
instantons \cite{'tHooft:1976up}.
The contribution of instantons 
is proportional to $\exp(-8\pi^2/g^2_i)$.
For QCD instantons, the coupling constant $g_s$ is large at 
low energies,
and the instanton contribution to the action is not negligible. 
On the other hand, 
the electroweak coupling $g$ is small, 
and electroweak instantons are extremely suppressed,
negligible for all practical purposes. 
We will neglect the 
electroweak theta term.

The SM terms related to $C\!P$ violation which
we will be concerned with below 
can be summarized by
\begin{equation}
\mathcal L_{4} = - \bar\theta\frac{g_s^2}{64 \pi^2}  
\ep^{\mu \nu \alpha \beta} G^a_{\mu \nu}  G^{a}_{\alpha \beta}
-  \left(1 + \frac{h}{v}\right)  
\bar q_L \left(M_0 + M_3 \tau_3\right)q_R
+ \textrm{H.c.}, 
\label{dim4}
\end{equation}
where
$M_{0,3}$ are real, diagonal quark mass matrices. 

There are good experimental and theoretical reasons to believe that the SM
is an EFT, at the electroweak scale $M_W$,
of a more fundamental theory, and it is renormalizable
in the more general sense of including 
all higher-dimensional operators allowed by the symmetries. 
In the form of Eq. \eqref{SM}
the SM does not include neutrino masses and mixings. 
These can be accounted for by introducing
the only gauge-invariant dimension-five operator that can be written
in terms of SM fields \cite{Weinberg:1979sa}. 
The lightness of neutrinos
compared to other fermions
is explained if the new-physics scale that suppresses this operator is 
of the order of $10^{15}$ GeV.  
Any new physics between the electroweak and the GUT or Planck scales 
will manifest itself in higher-dimensional operators. 
A promising strategy to probe such operators is to look for
processes where the SM contribution is extremely small. 
Examples are 
rare processes like the lepton flavor-changing
process $\mu \rightarrow e \gamma$
or neutrinoless double beta decay.  

Our work focuses on
EDMs, which signal $C\!P$ violation in the flavor-diagonal sector and are
insensitive to the phase of the CKM matrix. 
Since $\bar\theta$ is very small,
it is 
possible that higher-dimensional $\slashPT$ operators
are competitive with
the $\tb$ term.  
These higher-dimensional operators might eventually 
be linked to an underlying, ultraviolet
complete $\slashPT$ theory. 
We denote the scale characteristic of  this theory by $M_{\slashT}$.
Well below the scale $M_{\slashT}$ we expect $\slashPT$ effects 
to be captured by the
lowest-dimensional interactions among SM fields that respect the 
theory's gauge symmetry.
In general, operators of dimension $(4+n)$ at the SM scale 
are suppressed by powers of $M^{-n}_{\slashT}$. 
The next-to-lowest-dimension $\slashPT$ operators involving quark and
gluon fields that can be added to the 
SM Lagrangian have effective dimension six. Their
complete set was constructed in 
Refs. \cite{Buchmuller:1985jz, Rujula, Weinberg:1989dx,RamseyMusolf:2006vr} 
and recently reviewed in Ref. \cite{Grzadkowski:2010es}.

We focus here on flavor-diagonal $\slashPT$ operators involving 
quarks and gauge bosons only.
Following Ref. \cite{Grzadkowski:2010es}, 
we organize the dimension-six 
operators we need at the electroweak scale 
according to their field content: 
three vector bosons ($X^3$), 
two gauge bosons and two scalars ($X^2\varphi^2$), 
two quarks, a scalar, and a vector boson ($q^2 \varphi X$),
two quarks, two scalars, and a derivative ($q^2\varphi^2D$), 
two quarks and three scalars ($q^2\varphi^3$), 
and four quarks ($q^4$). 
The coefficients of these operators 
are all proportional to $M_{\slashT}^{-2}$.

At dimension six, the interactions of a quark with gauge bosons
gain in complexity. With the aid of the scalar boson we can write
\begin{eqnarray}
\mathcal L_{q^2 \varphi X}  &=&
- \frac{1}{\sqrt{2}}\bar q_L \sigma^{\mu \nu} 
\left(\tilde\Gamma^{u} \lambda^a G^a_{\mu \nu} + \Gamma^u_{B} B_{\mu \nu} 
+ \Gamma^u_{W} \tau^i W^i_{\mu \nu}  \right) 
\frac{\tilde\varphi}{v} \, u_R  
\nonumber \\ 
&& 
- \frac{1}{\sqrt{2}} \bar q_L \sigma^{\mu \nu} 
\left(\tilde\Gamma^{d} \lambda^a G^a_{\mu \nu} + \Gamma^{d}_B  B_{\mu \nu} 
+ \Gamma^{d}_W \tau^i W^i_{\mu \nu} 
\right) \frac{\varphi}{v}  d_R 
+ \mathrm{H. c.}\,, 
\label{qqHX}
\end{eqnarray}
where, in the most general case, 
the couplings 
$\tilde\Gamma^{u,d}$ and $\Gamma^{u,d}_{B,W}$, 
are 3 $\times$ 3 complex-valued matrices. 
After electroweak symmetry breaking, the operators in Eq. \eqref{qqHX} 
can be expressed in terms of dipole moment operators,
\begin{eqnarray}
\mathcal L_{q^2\varphi X} &=& 
- \frac{1}{2} \bar q_L \sigma^{\mu \nu} 
\left( \tilde\Gamma_0 + \tilde\Gamma_3 \tau_3 \right) 
\lambda^a q_R \, G^a_{\mu \nu} 
- \frac{1}{2} \bar q_L \sigma^{\mu \nu} 
\left( \Gamma_0 + \Gamma_3 \tau_3 \right)  q_R \, F_{\mu \nu} 
\nonumber\\ 
& &
- \frac{1}{2} \bar q_L \sigma^{\mu \nu} 
\left( \Gamma_{Z\, 0} + \Gamma_{Z\, 3} \tau_3 \right)  q_R \, Z_{\mu \nu}
\nonumber \\ 
& & 
- \frac{1}{\sqrt{2}} \bar d_L \sigma^{\mu \nu}  \Gamma^{u}_W  u_R\, W^{-}_{\mu \nu}
- \frac{1}{\sqrt{2}} \bar u_L \sigma^{\mu \nu}  \Gamma^{d}_W  d_R\, W^{+}_{\mu \nu}
+ \textrm{H.c.}
+\ldots,
\label{qqHX2}
\end{eqnarray}
where, at a renormalization scale $\mu \simeq M_W$,
\begin{eqnarray}
\tilde\Gamma_{0,3} &=& 
\frac{1}{2} \left( \tilde\Gamma^u \pm \tilde\Gamma^d \right),
\label{coeffqCEDM}
\\
\Gamma_{0,3}  &=& 
\frac{1}{2} \left[ \left(\Gamma^u_B \pm \Gamma^d_B\right) \cos\theta_W 
+ \left(\Gamma^u_W \mp \Gamma_W^d\right) \sin\theta_W \right],
\label{coeffqEDM} 
\\
\Gamma_{Z\, 0,3}  &=& 
\frac{1}{2} \left[ \left(\Gamma^u_W \mp \Gamma^d_W\right) \cos\theta_W 
- \left(\Gamma^u_B \pm \Gamma_B^d\right) \sin\theta_W \right]. 
\label{coeffqWEDM}
\end{eqnarray}
The first two operators in Eq. \eqref{qqHX2} are the most interesting for 
low-energy applications.
The imaginary parts of the diagonal entries of $\Gamma_{0,3}$ and 
$\tilde \Gamma_{0,3}$ generate
the quark electric and chromo-electric dipole moments (qEDM and qCEDM).
Since these operators flip the chirality of the quark field,  
we assume these matrices to be proportional to the Yukawa couplings 
in the SM Lagrangian, and thus to the masses:
\begin{equation}
\tilde\Gamma_{0,3} =
\mathcal O\left( 4\pi \tilde\delta_{0,3} 
\frac{M_{0}}{M^2_{\slashT}}\right), 
\qquad
\Gamma_{0,3} = 
\mathcal O\left( e\delta_{0,3} \frac{M_{0}}{M^2_{\slashT}}\right),
\label{Gammas}
\end{equation}
where $\tilde \delta_{0,3}$ and $\delta_{0,3}$ are dimensionless constants 
that parameterize any deviation from this assumption, 
and contain all the information on physics beyond the SM. 
The non-diagonal entries are also of considerable interest, 
since they produce flavor-changing
neutral currents. In the $C\!P$-odd sector, for example, 
the $uc$ entries were found \cite{Isidori:2011qw} to be the less
constrained dimension-six operators that contribute to 
the recently observed $C\!P$ violation
in charm decays \cite{Aaij:2011in}. 
Since the flavor-changing operators  contribute
to nuclear EDMs only via additional loops involving weak-boson exchange, 
for our current purposes we can neglect them. 
The next three operators in Eq. \eqref{qqHX2} are weak dipole moments.
Their contribution to the qEDM and qCEDM is suppressed by $(g/4\pi)^2$.
In the ``$\ldots$'' in Eq. \eqref{qqHX2} we find couplings of
the Higgs boson, which we can also neglect. 

The self-interaction of gauge bosons also exhibits new structures
at dimension six. $C\!P$ violation can be found in
\begin{equation}
\mathcal L_{X^3}= 
\frac{d_{W}}{6} f^{a b c} \ep^{\mu \nu \alpha \beta} 
G^a_{\alpha \beta} G_{\mu \rho}^{b} G^{c\, \rho}_{\nu} 
+ \frac{d_{w}}{6} \ep^{ijk} \ep^{\mu \nu \alpha \beta} 
W^i_{\alpha \beta} W_{\mu \rho}^{j} W^{k\, \rho}_{\nu}.
\label{XXX}
\end{equation}
The first operator 
is the Weinberg three-gluon operator \cite{Weinberg:1989dx}
which can be interpreted as the gluon chromo-electric dipole moment (gCEDM) 
\cite{Braaten}.
Similarly, the second operator is the $W$-boson weak electric dipole moment. 
After electroweak symmetry breaking, this
operator generates interactions containing at least two 
heavy gauge bosons \cite{Rujula}, which
can contribute to q(C)EDMs through loop corrections. 
Again such contributions are suppressed
by $ (g/4\pi)^2$ \cite{Rujula}. 
For our purposes we can neglect the weak dipole moment
and focus on the gCEDM,
\begin{equation}\label{dW}
d_W = \mathcal O\left(4\pi\frac{w}{M^2_{\slashT}}\right),
\end{equation}
with $w$ a dimensionless constant.

Just as there is $C\!P$ violation in multi-gluon operators,
so there is in multi-quark operators.
In agreement with Ref. \cite{RamseyMusolf:2006vr}, 
we find just two four-quark interactions,
\begin{equation}
\mathcal L_{q^4} =
\Sigma_1 \left(\bar q^J_L  u_R\right) \, \ep_{JK} \,
\left(\bar q_L^K d_R\right) +
\Sigma_8 \left(\bar q^J_L \lambda^a u_R\right) \, \ep_{JK} \, 
\left(\bar q_L^K \lambda^a d_R\right)+ \textrm{H.c.},
\label{qqqq}
\end{equation}
where the couplings $\Sigma_{1,8}$ 
are four-index tensors in flavor space.
They scale as
\begin{equation}\label{sigma}
\Sigma_{1,8} = \mathcal O\left((4\pi)^2\frac{\sigma_{1,8}}{M^2_{\slashT}}\right), 
\end{equation}
where $\sigma_{1,8}$ 
are dimensionless constants. 
The operators in Eq. \eqref{qqqq} are not
affected by electroweak symmetry breaking. 
For later convenience, we rewrite Eq. \eqref{qqqq} in terms of quark doublets
$q = q_L + q_R$, and focus on the flavor-diagonal $\slashPT$ terms only,
\begin{eqnarray}
\mathcal L_{q^4}  &=& 
\frac{1}{4} \left( \textrm{Im}{\Sigma_1} \right)_{1 1 1 1}
\left( \bar q q\, \bar q i \gamma_5 q 
- \bar q \boldtau q\, \cdot \bar q \boldtau i \gamma_5 q \right) \nonumber
\\ & &+  \frac{1}{4} \left( \textrm{Im}{\Sigma_8} \right)_{1 1 1 1} 
\left( \bar q \lambda^a q\, \bar q i \gamma_5 \lambda^a q 
- \bar q \boldtau \lambda^a q\, \cdot \bar q \boldtau i \gamma_5 \lambda^a q 
\right)
+\ldots
\label{qqqq2}
\end{eqnarray}
These four-quark operators, originating directly at the
electroweak scale and constrained by exact $SU_L(2)$ gauge invariance,
are chiral invariant.
The remaining parts of Eq. \eqref{qqqq} and other dimension-six $\slashPT$ 
four-quark operators listed in Ref. \cite{Grzadkowski:2010es} 
necessarily involve flavor-changing effects, and contribute 
to flavor-diagonal $C\!P$ violation only at the loop level. 
We will neglect them in what follows.

The remaining $\slashPT$ sources involve two or more scalar fields.
The most important of them is
\begin{equation}
\mathcal L_{q^2\varphi^2D} =  \bar u_R \Xi_1 \gamma^\mu d_R  \;
\tilde \varphi^\dagger iD_\mu \varphi
+ \mathrm{H.c.},
\label{qqHH}
\end{equation}
where in general $\Xi_1$ is a 
complex $3\times 3$ matrix in flavor space. 
Here we focus, again, on the flavor-diagonal parts only, given
after electroweak symmetry breaking by
\begin{equation}
\mathcal L_{q^2\varphi^2D} = 
\frac{gv^2}{2 \sqrt{2}} 
\left(W_\mu^{+} \bar u_R \Xi_1 \g^\mu d_R
+ W_\mu^{-} \bar d_R \Xi_1 \g^\mu u_R\right)
+\ldots
\label{qqW}
\end{equation}
Contrary to the operators in Eq. \eqref{qqHX}, 
the operators in Eq. \eqref{qqW} do not change
chirality and we do not expect them to be proportional to the quark mass, 
so we parameterize
\begin{eqnarray}
\Xi_1 = \mathcal O\left((4\pi)^2 \frac{\xi}{M^2_{\slashT}}\right).
\end{eqnarray} 
At low energy, after we integrate out the $W$ boson, 
the imaginary part of $\Xi_1$ contributes to
$\slashPT$ four-quark operators, 
which are particularly interesting because they are not suppressed
by the light quark masses \cite{Ng:2011ui}. 
As detailed in Sec. \ref{Match}, $\textrm{Im}\,\Xi_1$ generates
four-quark operators of the same importance as the chiral-invariant 
four-quark operators that are
generated directly at the electroweak scale, 
which we introduced in Eq. \eqref{qqqq}.

Finally, there are other terms that are closely related to interactions in
Eq. \eqref{SM},
\begin{eqnarray}
\mathcal L_{X^2\varphi^2,q^2\varphi^3}&=& 
-\frac{\ep^{\mu\nu\al\bt}}{32\pi^2} \left( 
g_s^2 \theta' \, G^a_{\mu\nu}G^a_{\al\bt} 
+g^2 \theta^{\prime}_w \, W^i_{\mu\nu}W^i_{\al\bt} 
+g^{\prime\, 2} \theta^{\prime}_b\, B_{\mu\nu}B_{\al\bt}
\right)\frac{\varphi^\dagger \varphi}{v^2}
\nonumber\\
&& 
+\frac{\ep^{\mu\nu\al\bt} }{32\pi^2}g g^{\prime}\theta^{\prime}_{w b} \, 
W^i_{\mu\nu}B_{\al\bt} \frac{\varphi^\dagger \tau^i \varphi}{v^2}
-2 \frac{\varphi^{\dagger} \varphi}{v^2}  
\left( \bar q_L  Y^{\prime\, u} \tilde \varphi u_R 
+ \bar q_L Y^{\prime\, d} \tilde \varphi d_R \right),
\label{XXHH}
\end{eqnarray}
where the angles $\theta^{\prime}$, $\theta_{w,\, b,\, w b}^{\prime}$
and the Yukawa couplings $Y^{\prime \, u,d}$, which are symmetric, complex
matrices in flavor space, 
scale as
\begin{equation}
\theta^{\prime}, \theta^{\prime}_{w,\, b,\, w b}  =
\mathcal O \left( \frac{v^2}{M^2_{\slashT}} \right) , 
\qquad 
Y^{\prime\, u,d} = \mathcal O\left(\frac{v^2}{M^2_{\slashT}}\right).
\end{equation}
Equation \eqref{XXHH} is reminiscent of Eq. \eqref{SM}.  
Indeed, if one rewrites
$2 \varphi^{\dagger} \varphi/v^2 = 1 + (2 \varphi^{\dagger} \varphi/v^2 -1 )$,
the pieces 
of the first three and last two operators 
that do not contain the Higgs boson can be absorbed in a 
redefinition of the
couplings $\theta$, $\theta_{w,\, b}$, and $Y^{u,d}$ 
in Eq. \eqref{SM}. 
In some sense this aggravates the strong $C\!P$ problem 
since even if the bare QCD vacuum angle
is tuned to zero, we need to explain why the 
contributions induced by these higher-dimensional terms
are small as well. 
Of the first four terms in Eq. \eqref{XXHH}, only the operator 
$\theta^{\prime}_{w b}$ obtains after 
electroweak breaking a nontopological piece 
without a Higgs boson \cite{Rujula}. 
This piece connects three or more electroweak gauge bosons and 
contributes through loop diagrams to the quark electric 
and weak dipole moments, but such 
contributions are suppressed by $(g/4\pi)^2$.
The remainder of Eq. \eqref{XXHH} then includes $C\!P$-odd 
operators with at least one Higgs boson:
in the unitarity gauge,
\begin{eqnarray}
\mathcal L_{X^2\varphi^2,q^2\varphi^3}&=& 
-\left\{ \frac{\ep^{\mu\nu\al\bt} }{32\pi^2} 
\left( g_s^2 \theta' \, G^a_{\mu\nu}G^a_{\al\bt} 
+ g^2 \theta^{\prime}_w  \, W^i_{\mu\nu}W^i_{\al\bt} 
+ g^{\prime\, 2}\theta_b^{\prime}  \, B_{\mu\nu}B_{\al\bt} 
+  g g^{\prime}\theta^{\prime}_{w b} \, W^3_{\mu\nu}B_{\al\bt}\right)  
\right.
\nonumber\\
&& \left.   
+ \sqrt{2} v \left(1 + \frac{h}{v} \right)  
\left( \bar u_L   Y^{\prime\, u}  u_R + \bar d_L  Y^{\prime\, d} d_R \right) 
\right\}
\frac{h}{v} \left(1 + \frac{h}{2 v}\right).
\label{XXHH2}
\end{eqnarray}
In the first line of Eq. \eqref{XXHH2} are 
$C\!P$-odd 
interactions of the Higgs to two gluons, two photons, or two weak bosons.
The imaginary parts of the Yukawa couplings generate 
flavor-diagonal and flavor-changing  $\slashPT$ 
Higgs-quark interactions.
At low energy, where the Higgs boson is integrated out, these
interactions manifest themselves in loop corrections to the qEDM, qCEDM 
and gCEDM.
At tree level, $\slashPT$ Higgs-quark interactions generate  
$\slashPT$ four-quark operators, which, as discussed below, are relatively
small.

Of course, we could continue the procedure to higher-dimensional
$\slashPT$ sources. Those would include, for example, further four-quark
operators \cite{ji} at dimension eight.
We limit ourselves here to a systematic study of the effects of 
dimension-six sources to low-energy observables.

\section{Tree-level matching onto the QCD scale}
\label{Match}

The $\slashPT$ Lagrangian of dimension up to six 
that is relevant for the calculation 
of hadronic and nuclear EDMs is summarized
in Eqs. \eqref{dim4}, \eqref{qqHX2}, \eqref{XXX},
\eqref{qqqq2}, \eqref{qqW}, and \eqref{XXHH2}. 
For low-energy applications, it is important to
evolve the $\slashPT$ Lagrangian from the electroweak scale down to the 
typical hadronic scale $\mu \sim M_{\textrm{QCD}} \simeq 1$ GeV. 
In the process, one has to integrate out the effects of heavy SM particles
\cite{Rujula, CorderoCid:2007uc};
at the same time, one has to
evaluate the running of the coupling constants and 
account for the possible mixing of the
dimension-six operators \cite{Braaten, gCEDMrunning, ji, Dekens:2013zca}.
A detailed account of the matching and 
evolution of the complete dimension-six $\slashPT$ Lagrangian
is beyond the scope of this work.  
Here we limit ourselves to tree-level matching, and we turn off (most of) the
running of the couplings.

At the QCD scale, the theory involves only light quarks, gluons and photons. 
Focusing on the first generation,
\begin{equation}
\mathcal L_{\rm cl} =  
-\frac{1}{4} 
\left(G^a_{\mu \nu} G^{a\, \mu \nu}+F_{\mu \nu} F^{\mu \nu} 
\right) 
+ \bar q \left(i\slashchar{\partial} 
+\frac{g_s}{2} \slashchar{G}^a\lambda^a \right)q
\label{cl}
\end{equation}
has a global, chiral $SU_L(2)\times SU_R(2)\sim SO(4)$
symmetry under independent $SU(2)$ rotations of left- and
right-handed quarks.
Remaining dimension-four (and most of the higher-dimensional) interactions
break chiral symmetry explicitly.
Because all the breaking is due to relatively small quantities,
it is useful to classify interactions according to
their transformation pattern under $SO(4)$.

For the construction of $\slashPT$ electromagnetic operators in the 
chiral Lagrangian, we will 
need to
consider the chiral properties of the $PT$ electromagnetic couplings of 
the quarks in
\begin{equation}
\mathcal{L}_{e}= 
-e A_\mu  \qb \g^\mu Q q,
\label{Le}
\end{equation}
where 
\begin{equation}
Q = \frac{1}{6} + \frac{\tau_3}{2}
\label{qQ}
\end{equation}
is the quark charge matrix.
Introducing a chiral-invariant 
$I^{\mu}$ and an antisymmetric tensor $T^{\mu}$ through
\begin{equation}
I^{\mu}=\qb \g^\mu q,
\qquad 
T^{\mu}=\left(\begin{array}{cc}
\ep^{ijk}\qb \g^{\mu}\g_5 \tau^k q & \qb \g^\mu \tau^j q\\
- \qb \g^\mu \tau^i q              &0
\end{array} \right),
\label{eq:tensors}
\end{equation}
we rewrite Eq. \eqref{Le} as
\begin{equation}
\mathcal{L}_{e}=
-\frac{e}{2} A_\mu \left(\frac{I^{\mu}}{3}+T^{\mu}_{34}\right). 
\label{Leprime}
\end{equation}

At low energy, we can neglect the effects of the Higgs 
and of the heavy quarks, and Eq. \eqref{dim4} is rewritten as
\begin{equation}
\mathcal{L}_{4}=
- \qb_L M q_R - \qb_R M q_L
-\bar\theta\frac{g^2_s}{64\pi^2} \ep^{\mu\nu\al\bt} \, G^a_{\mu\nu}G^a_{\al\bt} \ ,
\end{equation}
where
\begin{equation}
M = \left(\begin{array}{cc} 
                       m_u & 0 \\
                       0 & m_d
                       \end{array}\right)
= \bar{m} \left(1- \varepsilon \tau_3\right),
\end{equation}
with real parameters 
$m_{u,d}$, or alternatively 
\begin{equation}
\bar{m}=\frac{m_u+m_d}{2},\qquad 
\varepsilon=\frac{m_d-m_u}{m_u+m_d}.
\label{qmass}
\end{equation}
For a $\chi$PT treatment, it is more convenient to eliminate the $\tb$ term 
with an axial $U(1)$ rotation
on the quark fields, and move all $C\!P$ violation to the quark mass term. 
After vacuum alignment \cite{Baluni}, the QCD $\tb$ term becomes,
in the notation of Ref. \cite{BiraEmanuele},
\begin{equation}
\mathcal{L}_{4}=
\mb \rb S_4-\varepsilon \mb\,\rbm 1 P_3-\ms\sin\tb\,\rbm 1 P_4 \ ,
\label{Ltprime}
\end{equation}
where we introduced two $SO(4)$ vectors
\begin{equation}
\begin{array}{lcr}
S=\left(\begin{array}{c}- i\qb \g^5 \boldtau q\\
\qb q\end{array}\right),
\qquad 
P=\left(\begin{array}{c}\qb \boldtau q\\
i\qb\g^5 q\end{array}\right),
\end{array}
\label{eq:vectors}
\end{equation}
a function $\rb$ that goes to 1 in the limit of small $\tb$,
\begin{equation}
\rb = \left(\frac{1+\varepsilon^2 \tan^2\h\tb}{1+\tan^2\h\tb}\right)^{1/2} 
\simeq 1 + \mathcal O(\tb^2),
\end{equation}
and the parameter
\begin{equation}
\ms =\frac{m_u m_d}{m_u+m_d}=\frac{1-\varepsilon^2}{2}\bar{m}.
\label{mstar}
\end{equation}
From these ingredients one can derive the form of low-energy 
interactions originating in the QCD vacuum angle,
as presented, for example, in Ref. \cite{BiraEmanuele}.

We will now consider the dimension-six Lagrangian,
which breaks chiral symmetry in new ways.
The light-flavor components of the dipole operators in Eq. \eqref{qqHX2} 
match onto the light-quark EDM and CEDM,
\begin{equation}
\mathcal L_{\mathrm{q(C)EDM}} = 
-\frac{1}{2} \, \bar q \,  i \sigma^{\mu \nu}  \gamma^5 
\left(  d_0 + d_3 \tau_3 \right) q \, F_{\mu \nu} 
-\frac{1}{2} \, \bar q \,  i \sigma^{\mu \nu}  \gamma^5 
\left( \tilde{d}_{0} + \tilde{d}_{3} \tau_3  \right)\lambda^a q \, G^a_{\mu \nu}, 
\label{qqHX3}
\end{equation}
where, at tree level,
\begin{equation}\label{d0d3}
d_{0,3} = \textrm{Im} (\Gamma_{0,3})_{11}, \qquad \, 
\tilde d_{0,3} = \textrm{Im} (\tilde\Gamma_{0,3})_{11}.
\end{equation}
The non-diagonal components of the qEDM and qCEDM operators in 
Eq. \eqref{qqHX2} correct Eq. \eqref{d0d3} at the loop level. 
The weak EDMs also generate loop corrections to the qEDM and qCEDM. 
At tree level, they
generate four-quark operators, which, however, involve one power of the 
quark momentum, and are thus dimension-seven.
The couplings of such operators are suppressed by an additional 
factor $m_{u,d}/M_W^2$ compared to the couplings
of the dimension-six operators that we keep.

The gCEDM in Eq. \eqref{XXX} and the four-quark
operators in Eq. \eqref{qqqq2} match onto themselves.
As pointed out in Ref. \cite{Ng:2011ui},
at tree level, the only non-trivial result comes from the operator 
in Eq. \eqref{qqW}.
The largest effect comes from a 
$W$ exchange between light quarks, the suppression by
$M_W^2$ from the $W$ propagator being compensated by 
the factor $g^2 v^2$ from the vertices. 
The resulting $\slashPT$ four-quark operator, 
when evolved from $M_W$ to $\MQCD$, 
induces another four-quark operator of similar form 
but with additional color structure\footnote{This was pointed out to us by W. Dekens. 
We thank him for useful discussions on this subject.}.
The latter does not appear directly at the electroweak scale due 
to its gauge-symmetry-breaking properties.
We find the following $\slashPT$ operators, which 
because of their left-right mixing 
we abbreviate 
as FQLR:
\begin{eqnarray}
\mathcal L_{\mathrm{LR}} &=&
i \, \textrm{Im}(\Xi_1)_{11} \, V_{ud} 
\left(\bar u_R \g^\mu d_R \, \bar d_L \g_\mu u_L
- \bar d_R \g^\mu u_R \, \bar u_L \g_\mu d_L\right)\nonumber\\
&&+ i \, \textrm{Im}(\Xi_{8})_{11} \, V_{ud} 
\left(\bar u_R \g^\mu \lambda^a d_R \, \bar d_L \g_\mu \lambda^a u_L
- \bar d_R \g^\mu \lambda^a u_R \, \bar u_L \g_\mu  \lambda^a d_L\right), 
\label{LR}
\end{eqnarray} 
where $V_{ud}\simeq 1$ is a CKM element. 
Here, $(\Xi_8)_{11}$ is not an independent coupling,
but it depends on $(\Xi_1)_{11}$ and on QCD renormalization-group factors,
which are calculable  \cite{Dekens:2013zca}. 
Thus $(\Xi_1)_{11}$ and $(\Xi_8)_{11}$ 
both depend on the same dimensionless parameter $\xi$. 
In terms of light-quark doublets, we can rewrite  Eq. \eqref{LR} as
\begin{equation}
\mathcal L_{\mathrm{LR}} = 
\frac{1}{4} \,\textrm{Im}(\Xi_1)_{11} \, \ep^{3ij}\, 
\bar q  \tau^{i}\gamma^{\mu}  q\, \bar q \tau^j \gamma_{\mu} \gamma_5 q 
+ \frac{1}{4} \,\textrm{Im}(\Xi_8)_{11} \, \ep^{3ij}\, 
\bar q  \tau^{i}\gamma^{\mu} \lambda^a q\, 
\bar q \tau^j \gamma_{\mu} \gamma_5 \lambda^a q. 
\end{equation}
By a Fierz rearrangement we can further rewrite the first term 
of this equation as 
\begin{eqnarray}\label{Fierz}
\frac{1}{12} \,\textrm{Im}(\Xi_1)_{11} \left[
2\left(\bar q q \, \bar q i \gamma_5 \tau_3 q 
-\bar q \tau_3 q\, \bar q i \gamma_5 q\right) 
+ 3 \left(\bar q \lambda^a q \, \bar q i \gamma_5 \tau_3 \lambda^a q
- \bar q \tau_3 \lambda^a  q\, \bar q i \gamma_5 \lambda^a q \right) \right].
\end{eqnarray}
This operator was studied in Ref. \cite{Xu:2009nt} 
in the framework of left-right models. Although a large list
of four-quark operators is presented there, 
only one combination ($O_{11}-O_{12}+6 O_{21}-6 O_{22}$, in 
their notation) is generated at the electroweak scale. 
This combination is identical to Eq. \eqref{LR}.
It is interesting to point out that we could have coupled the operator 
in Eq. \eqref{qqW} to the left-handed lepton current. 
The operator created this way causes $\slashT$ in $\bt$ decay 
through contributions to the triple correlation 
$\sim D\,\vec J\cdot(\vec p_e\times \vec p_\nu)$. 
In Ref. \cite{Ng:2011ui} it is argued that, 
with current experimental accuracy, 
the best limit on the FQLR operator comes from EDM experiments. 

All other $\slashPT$ effects have extra suppression at low energies.
The operators in Eq. \eqref{XXHH2} contain at least one Higgs field,
and generate $\slashPT$ effects at either tree or loop level.
For example, at tree level, the operators in the second line contribute to 
four-quark 
operators.
But, since the Higgs couples to the light quark mass, the
resulting $\slashPT$ operators have a suppression of at least 
the small ratio of the light-quark mass over the electroweak scale.

In summary, the (effectively) dimension-six $\slashPT$ Lagrangian is
\begin{eqnarray}\label{QCDscale}
\mathcal L_{6} &=&
-\frac{1}{2}\qb \left(d_0+d_3 \tau_3\right)i \simu \g_5 q \; \Fmu 
-\frac{1}{2}\qb \left(\tilde{d}_0+\tilde{d}_3 \tau_3\right)
i \simu\g_5\la^a q \; \Gmu\nonumber\\
&&
+ \frac{d_{W}}{6} f^{a b c} \ep^{\mu \nu \alpha \beta} 
G^a_{\alpha \beta} G_{\mu \rho}^{b} G^{c\, \rho}_{\nu} \nonumber\\
&& 
+ \frac{\textrm{Im}\,\Xi_1}{4} \ep^{3ij} \bar q \tau^i \gamma^{\mu}q \, 
\bar q \tau^j \gamma_{\mu} \gamma_5 q  
+ \frac{\textrm{Im}\,\Xi_8}{4} \ep^{3ij} \bar q \tau^i  \gamma^{\mu}\lambda^a q \, 
\bar q \tau^j  \gamma_{\mu} \lambda^a \gamma_5 q
\nonumber\\
&&
+ \frac{\textrm{Im}{\Sigma_1}}{4}  \left( \bar q q\, \bar q i \gamma_5 q 
- \bar q \boldtau q\, \cdot \bar q \boldtau i \gamma_5 q \right)
+  \frac{\textrm{Im}{\Sigma_8}}{4}  
\left( \bar q \lambda^a q\, \bar q i \gamma_5 \lambda^a q 
- \bar q \boldtau \lambda^a q\, \cdot \bar q \boldtau i \gamma_5 \lambda^a q 
\right),
\end{eqnarray}
where all coupling constants have been redefined in order to absorb effects 
from operator mixing and renormalization-group running,
and to drop the generation indices in the case of four-quark operators. 
As discussed above,
one can imagine additional $\slashPT$ four-quark operators, 
but such operators 
are suppressed by 
weak gauge couplings in the typical combination $(g/4\pi)^2$, 
small off-diagonal CKM elements, 
and/or 
powers of $m_{u,d}/M_W$. We therefore expect the operators in 
Eqs. (\ref{Ltprime}) and (\ref{QCDscale}) to give rise to the dominant 
$P$ and $T$ violation in hadronic and nuclear
systems at low energy.

The redefined constants in Eq. (\ref{QCDscale}) scale as 
\begin{eqnarray}
&&d_{0,3} =\Or\!\left(\frac{e\delta_{0,3} \bar m}{M^2_{\slashT}}\right),
\qquad
\tilde d_{0,3} =
\Or\!\left(4\pi \frac{\tilde\delta_{0,3}\bar m}{M^2_{\slashT}}\right),
\qquad 
d_W = \Or\!\left(4\pi\frac{w}{M^2_{\slashT}}\right),
\nonumber\\
&&\Xi_{1,8} = \Or\!\left(\frac{(4\pi)^2\xi}{M^2_{\slashT}}\right),
\qquad
\Sigma_{1,8} = \Or\!\left(\frac{(4\pi)^2\sigma_{1,8}}{M^2_{\slashT}}\right),
\label{scalingofdim6}
\end{eqnarray}
in terms of dimensionless numbers 
$\delta_{0,3}$, $\tilde \delta_{0,3}$, $w$, $\xi$, and $\sigma_{1,8}$.
The sizes of 
these dimensionless parameters
depend on the exact mechanisms of electroweak and $P$ and $T$ breaking,
and on the running to the low energies where non-perturbative QCD effects
take over.
The minimal assumption is that
$\delta_{0,3}$, $\tilde{\delta}_{0,3}$, $w$, $\xi$, and $\sigma_{1,8}$
are 
$\Or(1)$, $\Or(g_s/4\pi)$, $\Or((g_s/4\pi)^3)$, $\mathcal O(1)$, and 
$\mathcal O(1)$, respectively.
However, they can be 
much smaller or larger,
depending on the parameters encoding $\slashPT$ beyond the Standard Model.
In the SM itself,
where $M_{\slashT}=M_{W}$, $\delta_{0,3}$, $\tilde \delta_{0,3}$ 
and $w$ are suppressed 
not only by the Jarlskog parameter \cite{Jarlskog:1985ht}
$J_{\mathrm {CP}} \simeq 3 \times 10^{-5}$, but
also by additional powers of
small gauge coupling constants
and ratios of quark-to-$W$ masses
\cite{Pospelov:1994uf,
Pospelov:2005pr}.
In certain supersymmetric models with various simplifying
universality assumptions 
of a soft-breaking sector with a common scale $M_{\mathrm{SUSY}}$, 
one has $M_{\slashT}=M_{\mathrm{SUSY}}$ and 
the dimensionless parameters are of the size of the minimal 
assumption times
a factor which is \cite{Pospelov:2005pr,Ibrahim:2007fb}, 
roughly,
$A_{C\!P}=(g_s/4\pi)^2 \sin \phi$ (neglecting electroweak parameters),
with $\phi$ a phase encoding $T$ violation.
If in the soft-breaking sfermion mass matrices non-diagonal terms are allowed,
enhancements of the type $m_b/m_d\sim 10^3$ or even $m_t/m_u\sim 10^5$
become possible, although
they are usually associated with
other, smaller phases \cite{Pospelov:2005pr}.
In other models, the relative sizes of the dimensionless parameters could
be different still.

Below the hadronic scale $\MQCD$,
the dimension-six $\slashPT$ sources generate further effective 
interactions, which break chiral symmetry in their own ways.
Introducing the $SO(4)$ singlets
\begin{eqnarray}
I_W &=&\frac{1}{6} f^{a b c} \ep^{\mu \nu \alpha \beta} 
G^a_{\alpha \beta} G_{\mu \rho}^{b} G^{c\, \rho}_{\nu} , 
\nonumber \\
I^{(1)}_{qq} &=& \frac{1}{4} \left( \bar q q \, \bar q i \gamma_5 q 
- \bar q \boldtau q \cdot \bar q \boldtau i \gamma_5 q \right)
=\frac{1}{4} S \cdot P,
\nonumber \\
I^{(8)}_{qq} &=& \frac{1}{4} 
\left( \bar q \lambda^a q\, \bar q i \gamma_5 \lambda^a q 
- \bar q \boldtau \lambda^a q \cdot \bar q \boldtau i \gamma_5 \lambda^a q 
\right),
\label{Is}
\end{eqnarray}
the $SO(4)$ vectors
\begin{eqnarray}
\begin{array}{lcr}
W=\h\left(\begin{array}{c}-i\qb \simu \g_5 \boldtau q\\
\qb\simu q\end{array}\right)\Fmu,
& &
V=\h\left(\begin{array}{c}\qb \simu \boldtau q\\
i\qb\simu \g_5 q\end{array}\right)\Fmu,
\label{VW}
\end{array}
\end{eqnarray}
and
\begin{eqnarray}
\begin{array}{lcr}
\tilde{W}=\h\left(\begin{array}{c}-i\qb \simu \g_5\boldtau \la^a q\\
\qb\simu \la^a q\end{array}\right)\Gmu,
& &
\tilde{V}=\h\left(\begin{array}{c}\qb \simu \boldtau \la^a q\\
i\qb\simu \g_5 \la^a q\end{array}\right)\Gmu,
\label{tildeVW}
\end{array}
\end{eqnarray}
and the symmetric $SO(4)$ tensors
\begin{equation}
X^{(1)}=\frac{1}{4}\left(\begin{array}{cc}
\qb \tau^i \g^\mu q \, \qb \tau^j \g_\mu q 
- \qb \tau^i \g^\mu \g_5 q\, \qb \tau^j \g_\mu \g_5 q  
&-\ep^{jkl}\qb \tau^k \g^\mu q\, \qb \tau^l \g_\mu \g_5 q
\\
-\ep^{ikl}\qb \tau^k \g^\mu q\, \qb \tau^l \g_\mu \g_5 q       
& \qb \boldtau \g^\mu q \cdot \qb \boldtau \g_\mu q 
- \qb \boldtau \g^\mu \g_5 q \cdot \qb \boldtau \g_\mu \g_5 q 
\end{array} \right),
\label{tensorXi}
\end{equation}
and
\begin{equation}
X^{(8)}=
\frac{1}{4}\left(\begin{array}{cc}
\qb \tau^i \g^\mu\lambda^a q \, \qb \tau^j \g_\mu\lambda^a q 
- \qb \tau^i \g^\mu \g_5\lambda^a q\, \qb \tau^j \g_\mu \g_5\lambda^a q  
&-\ep^{jkl}\qb \tau^k \g^\mu\lambda^a q\, \qb \tau^l \g_\mu \g_5\lambda^a q
\\
-\ep^{ikl}\qb \tau^k \g^\mu\lambda^a q\, \qb \tau^l \g_\mu \g_5 \lambda^aq       
& \qb \boldtau \g^\mu\lambda^a q \cdot \qb \boldtau \g_\mu\lambda^a q 
- \qb \boldtau \g^\mu \g_5\lambda^a q \cdot \qb \boldtau \g_\mu \g_5\lambda^a q 
\end{array} \right),
\label{tensorXic}
\end{equation}
we summarize the $\slashPT$ Lagrangian as
\begin{equation}
\mathcal{L}_{6} =
-d_0 V_4 + d_3 W_3  -\tilde d_0 \tilde{V}_4  + \tilde d_3 \tilde{W}_3
+ d_W I_W - \textrm{Im}\,\Xi_1 \,  X^{(1)}_{34} 
- \textrm{Im}\,\Xi_8 \,  X^{(8)}_{34} 
+ \textrm{Im} \Sigma_1 \, I^{(1)}_{qq} + \textrm{Im} \Sigma_8\, I^{(8)}_{qq}. 
\label{dim6}
\end{equation}

We already emphasized that the dimensionless coefficients 
$\delta_{0,3}$, $\tilde \delta_{0,3}$, $w$, $\xi$, and $\sigma_{1,8}$ 
that we have introduced, and their relative sizes, strongly depend on the 
particular high-energy model, making it difficult to compare the relative 
contributions of different $\slashPT$ sources to the same observable 
in a way that is independent of the details of the physics at the 
high-energy scale $M_{\slashT}$. 
We therefore adopt the approach that we construct the low-energy $\slashPT$ 
Lagrangian for each dimension-six source separately.
These separate Lagrangians can be used to calculate 
for each source their contribution to hadronic $\slashPT$ observables. 
Each $\slashPT$ source generates a characteristic pattern of relations 
between different observables, rooted in its field content and its 
transformation properties under chiral symmetry. The observation of such 
pattern in the current generation of, for example, nucleon and nuclear 
EDM experiments would then effectively pinpoint the dominant $\slashPT$ 
mechanism at the QCD scale \cite{Vri11a, Mer11, Vri11b, Mae11, Vri12, Liu12}, 
if one exists. Once this is known, the next step would be 
to trace the dominant effects
up to the electroweak scale.

On the other hand, the formalism we develop can be easily adjusted to 
specific extensions of the SM. 
Any low-energy observable can be obtained by combining
hadronic contributions from each separate source.
Once the values
of $\delta_{0,3}$, $\tilde\delta_{0,3}$, $w$, $\xi$, and $\sigma_{1,8}$
in a given model, and their running from $M_{\slashT}$ to $M_{\textrm{QCD}}$, 
are known, 
then the relative importance of the interactions constructed in the 
next sections can be reassessed
to accommodate, for instance, a large hierarchy between these parameters.

\section{SO(4) chiral framework}
\label{III}

Below the QCD scale quarks and gluons are no longer convenient 
degrees of freedom to describe strong interactions,
which can be rewritten in terms of hadronic EFTs.
The implications of the Lagrangian in Eqs. 
\eqref{cl},  
\eqref{Leprime},
\eqref{Ltprime}, and \eqref{dim6} 
for the interactions among pions
and nucleons (and 
Delta isobars, since they are not much more massive than nucleons)
at low momentum $Q\sim m_\pi\ll \MQCD$,
where $m_\pi\simeq 140$ MeV is the pion mass,
are described by 
chiral EFT, an extension to arbitrary number of nucleons of
$\chi$PT. At such momenta, pions must explicitly be accounted for in the
theory, while other mesons can be integrated out. 
At lower momenta, $Q\ll m_\pi$, even pions (and Delta isobars)
can be integrated out, the corresponding EFT being called 
pionless EFT. This EFT finds applications in loosely bound nuclei,
where it is much simpler to deploy than chiral EFT for
it involves only short-range inter-nucleon interactions. However,
the constraints from chiral symmetry, and consequently
some predictive power, are lost.
The {\it form} of the relevant pionless EFT interactions can be read from
the following by discarding pion fields.

The special role of the
pion is a consequence of the approximate invariance of the QCD Lagrangian
under the chiral symmetry $SU_L(2)\times SU_R(2)\sim SO(4)$. 
Because it is not manifest in the spectrum, which
only exhibits approximate isospin symmetry, chiral symmetry must be
spontaneously broken down to its diagonal, isospin subgroup,
$SU_{V}(2)\sim SO(3)$. From Goldstone's theorem, 
one expects to find in the spectrum three massless
Goldstone bosons that live on the ``chiral circle'' $S^3\sim SO(4)/SO(3)$.
The explicit breaking of chiral symmetry gives pions a mass,
but the way interactions at the quark/gluon level
transform constrains pion interactions.
In this section we show how these constraints arise in the $C\! P$-even
sector for one particular choice of pion fields,
and in the next section apply this technique
to $\slashPT$ interactions. (Because observables are independent
of the choice of fields, the method we use can be straightforwardly
reproduced for other choices without affecting the physics.)
Generalization to $SU_L(3)\times SU_R(3)$ is possible,
but the $SU_L(2)\times SU_R(2)$ case is best adapted to 
nuclear physics, on which scale the strange quark 
is not particularly light. 
The method used in this section stems from Refs. \cite{Weinberg,vanKolck}.

We parametrize the chiral circle
with stereographic coordinates \cite{Weinberg}, whose 
dimensionless fields we denote by an isovector field $\boldzeta$.
(The relation of this parametrization to other commonly used choices of 
the pion field is detailed, for example, in App. D of 
Ref. \cite{Bernard:1995dp}.)
We can identify these degrees of freedom with canonically normalized pion
fields $\boldpi= F_\pi \boldzeta$, where $F_\pi\simeq 186$ MeV, the pion
decay constant, is the diameter of the chiral circle. 
Such fields transform in a complicated way under chiral symmetry.
However, a pion covariant derivative can be defined by 
\begin{equation}
D_\mu \boldpi =D^{-1} \partial_\mu \boldpi,
\label{picovder}
\end{equation}
with 
\begin{equation}
D=1+ \frac{\boldpi^2}{F_\pi^2},
\label{D}
\end{equation}
which transforms under chiral transformations
as under an isospin transformation, but with a field-dependent parameter.
Similarly, we can use an isospin-1/2 nucleon field $N=(p \; n)^T$ 
that transforms
in an analogous way, and a nucleon covariant derivative,
\begin{equation}
{\mathcal D}_\mu N
=\left(\partial_\mu 
 +\frac{i}{F_\pi^2} \boldtau\cdot \boldpi \times D_\mu\boldpi\right)N. 
\label{nucovder}
\end{equation}
We define ${\mathcal D}^\dagger$ through
$\bar N \mathcal D^\dagger\equiv \overline{\mathcal D N}$ and 
use the short-hand notation:
\begin{eqnarray}
\mathcal D_\pm^\mu \equiv \mathcal D^\mu \pm \mathcal D^{\dagger\mu} , 
&&\qquad 
\mathcal D_\pm^\mu \mathcal D_\pm^\nu \equiv
\mathcal D^\mu \mathcal D^\nu
+\mathcal D^{\dagger \nu} \mathcal D^{\dagger \mu}
\pm \mathcal D^{\dagger\mu} \mathcal D^\nu
\pm \mathcal D^{\dagger\nu} \mathcal D^\mu , 
\nonumber \\
\tau_i \mathcal D_\pm^\mu \equiv 
\tau_i\mathcal D^\mu \pm\mathcal D^{\dagger \mu} \tau_i,
&&\qquad
\tau_i \mathcal D_\pm^\mu \mathcal D_\pm^\nu\equiv 
\tau_i \mathcal D^\mu \mathcal D^\nu
+\mathcal D^{\dagger \nu} \mathcal D^{\dagger \mu}\tau_i 
\pm \mathcal D^{\dagger\mu} \tau_i \mathcal D^\nu
\pm \mathcal D^{\dagger\nu} \tau_i \mathcal D^\mu .
\end{eqnarray}
Covariant derivatives of covariant derivatives
can be constructed similarly,
for example 
\begin{equation}
(\mathcal D_{\mu} D_{\nu} \pi)_i = \left(\partial_{\mu} \delta_{i j} 
- \frac{2}{F^2_{\pi}} \ep_{i k j} (\boldpi \times D_{\mu} \boldpi)_k 
\right) D_{\nu} \pi_j.
\end{equation}
For simplicity we omit the Delta isobar here,
but one can introduce an isospin-3/2 field for it
along completely analogous lines \cite{vanKolck}.

Since nucleons are essentially nonrelativistic for 
$Q\ll m_N$, the nucleon mass, we work in the 
heavy-baryon framework \cite{Jenkins:1990jv}
where, instead of gamma matrices, it is the nucleon 
velocity $v^\mu$ and spin $S_\mu$
($S=(\vec{\sigma}/2, 0)$ in the rest frame $v=(\vec{0}, 1)$)
that appear in interactions.
Below we use a subscript $\perp$ to denote the component of a four-vector 
perpendicular to the velocity, for example
\begin{equation}
\mathcal D^{\mu}_{\perp} =\mathcal D^{\mu} - v^{\mu} v \cdot \mathcal D.
\end{equation}
The interactions we construct are 
manifestly invariant under 
rotations and translations, but not under Lorentz boosts. 
Nevertheless, Lorentz invariance imposes non-trivial constraints 
on the interactions in the effective Lagrangian and on their coefficients
\cite{ManoharLuke}.

The first step in describing QCD at low energies is
to construct the most general Lagrangian that transforms under the
symmetries of QCD in the same way as the QCD Lagrangian itself.  
Along with this, one needs a power-counting scheme so that interactions 
can be ordered according to the expected size of their contributions.  
The Lagrangian contains an infinite number of terms
that we group using an integer ``chiral index'' $\Delta$ \cite{original}
and the (even) number of fermion fields $f$,
\begin{equation}
 {\cal L}=\sum_{\Delta=0}^{\infty}\sum_{f/2}{\cal L}^{(\Delta)}_{f}.
\label{ChiralL}
\end{equation}
We will restrict ourselves here mostly to $f\le 2$,
leaving a discussion of
the case $f\ge 4$ 
for Sec. \ref{NNsector}.

The technology for constructing the Lagrangian is well known, see, 
for example, Ref. \cite{Weinberg,Bernard:1995dp,Scherer}. When we neglect 
${\mathcal L}_{e}$ \eqref{Leprime}, 
${\mathcal L}_{4}$ \eqref{Ltprime}, and 
$\mathcal L_6$ \eqref{dim6},
the EFT Lagrangian includes all chiral-invariant interactions made out of 
$D_\mu\boldpi$, $N$, and their covariant derivatives.
In this case, $f\le 2$ interactions have chiral index \cite{original}
\begin{equation}
\Delta = d+f/2-2 \ge 0,
\label{Delta}
\end{equation}
in terms of the number  $d$ of derivatives (and powers of 
the Delta-nucleon mass difference).
The coefficients of the effective operators, the so-called low-energy constants
(LECs), cannot yet be calculated directly from QCD, but they 
can be estimated using naive dimensional analysis (NDA) 
\cite{NDA,Weinberg:1989dx},
in which case the index $\Delta$ tracks the number of inverse powers
of $\MQCD\sim 2\pi F_\pi\simeq 1.2$ GeV associated with an interaction.
(Note that 
since NDA associates the LECs of chiral-invariant operators to $g_s/4\pi$, 
for consistency one should take $g_s\sim 4\pi$.)
For the purposes of the present work, we need explicitly 
only the leading 
$C\! P$-even interactions,
\begin{equation}
{\mathcal L}_{\chi, f\le 2}^{(0)} = 
\frac{1}{2} D_{\mu} \boldpi \cdot D^{\mu} \boldpi 
+ \bar{N}\left( iv\cdot {\mathcal D}
         -\frac{2 g_A}{F_\pi}S^{\mu} \boldtau\cdot D_{\mu} \boldpi\right) N,
\label{LagrCons}
\end{equation}
where $g_A\simeq 1.267$ is the pion-nucleon axial-vector coupling. 
At this order the nucleon is static; kinetic corrections have relative
size ${\cal O}(Q/\MQCD)$ and appear in ${\mathcal L}^{(1)}$.

The formalism to include chiral-symmetry-breaking operators in the 
$SU(2)\times SU(2)$ $\chi$PT 
Lagrangian has been developed in Refs. \cite{Weinberg,vanKolck}.
Operators that break the symmetry as components of chiral tensors
can be obtained by rotating operators constructed with 
covariant fields $\Psi$
such as nucleon fields, and nucleon and pion covariant derivatives,
\begin{eqnarray}
\mathcal O_{\al\bt \cdots \omega}(\boldpi, \Psi)
= R_{\al \al'}(\boldpi) R_{\bt \bt'}(\boldpi) \cdots R_{\omega \omega'} (\boldpi)
\mathcal O_{\al'\bt'\cdots \omega'}(0, \Psi),
\label{rotationcb}
\end{eqnarray}
where
the chiral rotation $R$ is given in stereographic coordinates by
\begin{eqnarray}
R_{\al\bt}(\boldpi)=\left(\begin{array}{cc}\delta_{ij}
-\frac{2}{D}\frac{\pi_i\pi_j}{\Fp^2}&\frac{2}{D}\frac{\pi_j}{\Fp}\\
-\frac{2}{D}\frac{\pi_i}{\Fp}&\frac{1}{D}
\left(1-\frac{\boldpi^2}{\Fp^2}\right)\\
\end{array}\right).
\end{eqnarray}
In the EFT these operators generate interactions, now involving 
$\boldpi$ directly,
that transform as tensors and their tensor products.
The strengths of these interactions are proportional to powers 
of the symmetry-breaking parameters,
other dimensional factors being estimated by NDA.
(The equivalent formalism, based on the introduction of spurion fields for 
chiral-symmetry-breaking interactions, is detailed  
in Refs. \cite{Bernard:1995dp,Scherer}.)

The most important chiral-breaking term is the $\bar m$ term 
in Eq. \eqref{Ltprime}. It generates in the effective Lagrangian
all possible interactions with the structures
$S_4$ , 
$S_4\otimes S_4$, {\it etc.}
with strengths proportional to, respectively,  
$\bar m r(\tb)$, $(\bar m r(\tb))^2$, {\it etc.} 
In a similar way, one can incorporate in the EFT 
\cite{vanKolck, vanKolck:1996rm, Friar:2004ca} 
the $\varepsilon \bar m$ term in Eq.~\eqref{Ltprime},
which leads to isospin violation as $P_3$ and its tensor products.
The most important terms are,
omitting a constant irrelevant for our purposes,
\begin{equation}
{\mathcal L}^{(0,2)}_{\slashchi, f=0}= 
-\frac{1}{2D} \left(m^2_{\pi}+\frac{\Delta m_\pi^2}{D}\right) \boldpi^2 
+\frac{\delta m_\pi^2}{2D^2}  \pi_3^2 
\label{eq:pimass}
\end{equation}
and
\begin{equation}
{\mathcal L}^{(1)}_{\slashchi, f= 2}= 
\Delta m_N \left(1-\frac{2 \boldpi^2}{\Fp^2 D}\right)\Nb\!N
+\frac{\delta m_N}{2} 
\Nb \left( \tau_3-\frac{ 2 \pi_3 }{\Fp^2 D}\boldtau\cdot\boldpi \right) N.
\label{eq:Nmass}
\end{equation}
Here the dominant contribution to the pion mass is
$m^2_{\pi} = {\mathcal O}(r(\bar \theta)\bar{m}\MQCD)$,
its correction 
$\Delta m^2_{\pi} = {\mathcal O}(m^4_{\pi}/\MQCD^2)$,
and the nucleon sigma term  $\Delta m_N = \Or(m^2_{\pi}/\MQCD)$,
as the respective interactions
have the structures of $S_4$, $S_4\otimes S_4$, and $S_4$.
In addition, 
the quark-mass contribution to the nucleon mass splitting is 
$\delta m_N = \Or(r(\bar \theta)^{-1}\varepsilon \bar{m})
=\Or(r(\bar \theta)^{-2}\varepsilon m^2_{\pi}/\MQCD)$
and to the squared pion mass splitting
$\delta m^2_{\pi}  = {\mathcal O}(\delta m_N^2)$,
as the respective interactions
have the structures of $P_3$ and $P_3\otimes P_3$.
(The couplings $\Delta m^2_{\pi}$ and $\delta m^2_{\pi}$ in the mesonic Lagrangian 
above are related, respectively, to the interactions with LECs $l_3$ and $l_7$ 
in Ref. \cite{Gasser}. The nucleon sigma term $\Delta m_N$ and nucleon mass 
difference $\delta m_N$ can be expressed, respectively, in terms of the LECs 
$c_1$ and $c_5$ of Ref. \cite{Bernard:1995dp}.)

When we are interested in processes with typical momenta
$Q\sim m_\pi$, it is convenient to trade $\bar m$ by $m_\pi^2/\MQCD$
in all chiral-variant terms, which then have strengths proportional to 
powers of $m^2_{\pi}$ times appropriate powers of $\MQCD$. 
Since $\varepsilon \sim 1/3$ \cite{moreweinberg},
we choose, for simplicity, to count it as ${\cal O}(1)$. 
Since by all evidence $\bar\theta$ is small,
we also take $r(\bar \theta)$ as of ${\mathcal O}(1)$.
The power counting of $f\le 2$ interactions,
Eq. \eqref{Delta}, is straightforwardly generalized 
by defining $d$ to count powers of $m_\pi$ as well \cite{original}. 

Other isospin-violating hadronic operators and interactions
with the photon field $A_{\mu}$ come from 
${\mathcal L}_{e}$, Eq.~\eqref{Leprime}.
For applications to EDMs, it is important to construct 
electromagnetic operators in which hadrons interact with soft photons
(with momenta below $\MQCD$) in a gauge-invariant way.
We can minimally couple charged pions and nucleons 
to the photon by modifying their covariant derivatives,
\begin{eqnarray}\label{eq:minimal.1}
  (D_{\mu} \pi)_i & \rightarrow & (D_{\mu,\mathrm{em}} \pi)_i = 
\frac{1}{D} \left(\partial_{\mu}\delta_{ij}+ eA_{\mu}\ep_{3ij}\right) 
\pi_j , 
\nonumber \\
 \mathcal D_{\mu} N   &\rightarrow & \mathcal D_{\mu,\mathrm{em}} N = 
\left[\partial_{\mu}+\frac{i}{F_\pi^2}\boldtau \cdot
\left(\boldpi\times D_{\mu,\mathrm{em}}\boldpi \right) 
+ \frac{ie}{2} A_{\mu} \left(1 + \tau_3 \right)\right]N. 
\end{eqnarray}
For brevity, in the following we omit the label ``em''.
In addition, we can couple the photon through the field strength $F_{\mu\nu}$,
in operators that transform as $I^\mu/6 + T_{34}^\mu/2$ or its tensor
products.
The index $\Delta$ defined in Eq. \eqref{Delta} can be generalized to label 
electromagnetic operators, by enlarging the definition of $d$ to count also 
the number of photon fields, which, having dimension one, require compensating 
powers of $M_{\rm{QCD}}$ in their coefficients. 
Finally, integrated-out hard photons give rise to purely hadronic operators   
that also transform as tensor products 
of $I^\mu/6 + T_{34}^\mu/2$ \cite{vanKolck,vanKolck:1996rm,Meissner:1997fa}, 
and are proportional to 
powers of $(e/4\pi)^2=\alpha_{\rm{em}}/4\pi$.
The most important of these operators
is the electromagnetic contribution  $\breve\delta m^2_{\pi}$
to the squared pion mass splitting \cite{vanKolck,vanKolck:1996rm},
\begin{equation}
\mathcal L_{\textrm{em}, f=0}^{(1)} = 
-\frac{\breve\delta m^2_{\pi}}{2 D^2} \left(\boldpi^2 - \pi_3^2\right),
\label{hardEMLag}
\end{equation} 
with $\breve\delta m^2_{\pi}={\cal O}(\alpha_{\rm{em}} \MQCD^2/4\pi)$
(which is proportional to the LEC $C$ in Ref. \cite{Meissner:1997fa}).
Since numerically $\alpha_{\rm{em}}/4\pi\simeq (m_\pi/\MQCD)^3$,
we assign $d=3$ for each power of $\alpha_{\rm{em}}/4\pi$ \cite{vanKolck}.
This counting reflects both the smallness and the presumed
electromagnetic origin of the pion mass splitting.
 
The remaining dimension-four term, the $m_* \sin \tb$ term 
in Eq. \eqref{Ltprime}, which is $\slashPT$, can be treated in the same way.
This was done in some detail in Ref. \cite{BiraEmanuele},
where the implications of its $P_4$ structure were discussed.
Perhaps the most important point is that the hadronic $\slashPT$
interactions
are intrinsically linked to isospin-breaking operators that transform
like $P_3$, so that the $\slashPT$ LECs can be inferred from 
isospin-breaking couplings, when the latter are known.

In the rest of this paper we discuss the construction
of the $\slashPT$ interactions stemming from $\mathcal L_{6}$,  
Eq. \eqref{dim6}, and their power counting. 
The $\slashPT$ chiral Lagrangian can be constructed by
writing down all terms that transform in the same way under Lorentz, 
$P$, $T$, and chiral symmetry as the terms in Eq. \eqref{dim6}.
Because such
interactions are very small compared
to $C\! P$-even interactions, we restrict ourselves
to low-energy operators involving at most one power of
the dimensionless coefficients 
$\delta_{0,3}$, $\tilde \delta_{0,3}$, $w$, $\xi$, and $\sigma_{1,8}$,
even though it is straightforward to consider mixed effects.
As discussed above, these
coefficients 
are model dependent, and we construct the low-energy $\slashPT$ Lagrangian
for each source separately.
The form of the low-energy interactions is determined
by the way a source breaks chiral symmetry.

The gCEDM and the two four-quark operators in Eq. \eqref{dim6} with 
coefficients $\mathrm{Im} \Sigma_{1,8}$ do not break chiral symmetry and 
are therefore $SO(4)$ scalars,
which implies that they induce the same chiral Lagrangian
and cannot be separated on the basis of low-energy experiments alone. 
From now on we refer to these as chiral-invariant sources
($\chi$ISs) and 
use the symbol $w$ ($\bar I$) to denote collectively the 
dimensionless constants $w$ and $\sigma_{1,8}$
(the invariants $I_W$, $I^{(1)}_{qq}$, and $I^{(8)}_{qq}$)
in Eq. \eqref{scalingofdim6} (Eq. \eqref{Is}):
\begin{eqnarray}
\left\{w, \sigma_1, \sigma_8\right\} &\rightarrow& w ,\\
\left\{I_W, I^{(1)}_{qq}, I^{(8)}_{qq}\right\} &\rightarrow& \bar I.
\end{eqnarray}
Pion interactions stemming from $\chi$ISs involve the
pion covariant derivative, unless $\bar I$ is in a tensor product
with a chiral-variant tensor
({\it i.e.}, a product of $S$s and $P$s).

In contrast to the $\chi$ISs, the other sources (qEDM, qCEDM, FQLR) break
chiral symmetry explicitly and can bring in pion non-derivative
interactions by themselves, as is the case for $\bar\theta$. 
The specific dependence
on $\boldpi$ depends on the source.
The qCEDM consists of two independent components, 
the isoscalar and isovector qCEDMs.  They
transform, respectively, as the fourth and third components of 
the $SO(4)$ vectors $\tilde V$ and
$\tilde W$ defined in Eq. \eqref{tildeVW}. 
Since the QCD $\tb$ term transforms as the fourth component
of an $SO(4)$ vector as well, the $\tb$ term and 
isoscalar qCEDM generate identical chiral operators
(but with different strengths).
 
By themselves, the FQLR operators 
have the most complicated chiral structure,
transforming as the $34$-component of the symmetric tensors
$X^{(1)}$ and $X^{(8)}$
in Eqs. \eqref{tensorXi} and \eqref{tensorXic}.
As for $\chi$ISs, both operators induce the same chiral Lagrangian,
but their entanglement is not a real issue since they 
depend on the same dimensionless parameter $\xi$. 
From now on we use the symbol $X$ to denote $X^{(1)}$ and $X^{(8)}$:
\begin{eqnarray}
\left\{X^{(1)}, X^{(8)}\right\} &\rightarrow& X.
\end{eqnarray}

The structure of the qEDM, transforming as the fourth and third components of 
the $SO(4)$ vectors $V$ and
$W$ in Eq. \eqref{VW}, resembles that of the qCEDM,
but the similarity is deceiving due to the photon.
When the photon is integrated out to produce purely
hadronic operators from qEDM, one needs to take the tensor product
with the $I^\mu/6 + T_{34}^\mu/2$ from Eq. \eqref{Leprime}.
Conversely, to produce operators with a soft-photon
the qCEDM requires (as do FQLR and $\chi$ISs) a tensor product
with $I^\mu/6 + T_{34}^\mu/2$.
In both cases,
the extra $T^\mu_{34}$ produces in general interactions of more complicated
structure. 

Note that for each chiral-variant $\slashPT$ source, there are associated
quark-gluon operators that do not violate $P$ and $T$.
For example, a qEDM is associated with a quark magnetic dipole moment,
since they make the same $SO(4)$ vector, Eq. \eqref{VW}. 
Analogous statements hold for qCEDM in Eq. \eqref{tildeVW}
and for FQLR in Eqs. \eqref{tensorXi} and \eqref{tensorXic}.
The situation is similar to the $\bar\theta$ term, which is related
to the quark mass splitting \cite{BiraEmanuele}. 
However, for dimension-six sources
these relations are less useful, since the corresponding
$C\!P$-even interactions are buried among much larger, dimension-four interactions, whose coefficients, moreover, are not related in a model-independent way. Still, the strong-interaction matrix element of the $C\!P$-even partner might be easier to evaluate in lattice QCD.

The interactions stemming from the dimension-six $\slashPT$
sources can be organized according to a chiral index analogous 
to Eq. \eqref{Delta},
with the only difference that the coefficients of 
low-energy interactions must contain two
inverse powers of the high-energy scale $M_{\slashT}$, 
which replace two inverse powers of $\MQCD$. 
The powers of $\MQCD$ in a coefficient are therefore counted by
\begin{equation}
\Delta_6 = d + f/2 - 4,
\label{Deltadim6}
\end{equation}
where $d$ is the number of derivatives, and powers of the quark mass and $e$
as described above.
We will find that for the qEDM, qCEDM and $\chi$ISs 
$\Delta_6 \ge -2$, while for the
FQLR operator $\Delta_6 \ge -4$. 
The suppression by at least one power of $\alpha_{\rm{em}}/4\pi$
renders the purely hadronic operators 
from hard photons
essentially irrelevant.

The question now is to what order, for each separate source, 
do we need to construct the $\slashPT$ chiral Lagrangian. 
This obviously depends on what observable one wants to calculate. 
The most likely candidates are $\slashPT$ electromagnetic moments 
and form factors of light nuclei, which
require the use of the most important $\slashPT$ pion, pion-nucleon, 
photon-nucleon, and inter-nucleon interactions. 
In order to compare our EFT approach to more traditional approaches, 
where $\slashPT$ is implemented through three non-derivative $\slashPT$ 
pion-nucleon interactions, we construct the
pion-nucleon Lagrangian for each source up to the order where all 
three pion-nucleon interactions appear. 
As found in Refs. \cite{Mae11, Vri12}, for most sources the 
$\slashPT$ nucleon-nucleon(-photon)
interactions are subleading with respect to $\slashPT$ 
one-pion exchange, so in the nucleon-nucleon sector we only construct
LO operators.
We construct the electromagnetic sector until, for each source, 
we find the momentum dependence
of the nucleon EDFF,
as the linear term in the square
of the momentum transferred is a contribution the Schiff moment, 
which is important for the evaluation of atomic EDMs \cite{Thomas:1994wi}. 
Since operators with two explicit photons give small contributions even to 
atomic EDMs \cite{texas}, we do not construct explicitly
here operators with more than a single soft photon.

We start  in Sec. \ref{pionsector}
with the construction of operators in the purely mesonic sector
(that is, with $f=0$).
We then tackle the pion- and photon-nucleon sectors ($f=2$) 
in Secs. \ref{piNsector} and  \ref{Ngammasector},
respectively. 
We extend our analysis to include the nuclear sector in Sec. \ref{NNsector}. 
The phenomenologically most important interactions are discussed in 
Sec. \ref{discussion}, and summarized in Table \ref{table}.

\section{Pion sector}
\label{pionsector}

In contrast to $C\! P$-even dynamics, 
$\slashPT$ can generate interactions with an odd number of pions,
starting with pion tadpoles that represent the disappearance
of a neutral pion into the vacuum. 
Tadpoles cause the vacuum to become unstable because it
can create neutral pions to lower its energy. 
In the $\bar\theta$ case, imposing the condition of vacuum alignment
at the quark level  in first order
in the quark masses \cite{Baluni} ensures that the $\slashPT$ breaking
is through the $SO(4)$ vector $P$ in Eq. \eqref{eq:vectors}
and that tadpoles are absent up to that order. 
For the dimension-six sources, the more complicated
chiral structure also leads to tadpoles.
Since the tadpoles are small, they can, in principle, be
treated in perturbation theory, but a more convenient way of handling 
them is to rotate them
away by performing a field redefinition 
\cite{BiraEmanuele}.
After we consider the two leading orders for each source, we 
discuss the effects of tadpole extermination,
which has important consequences for other sectors of the theory.

\subsection{qCEDM}

As it was found in Ref. \cite{BiraEmanuele}, without any nucleon
fields it is not possible to construct an operator that transforms as the 
fourth component of an $SO(4)$ vector,
so the isoscalar qCEDM has no effect without another source
of chiral breaking. 
The isovector qCEDM, however, breaks isospin symmetry as $\tilde d_3 \tilde W_3$
and generates a pion tadpole with chiral index $\Delta_6=-2$,
\begin{equation}
\mathcal L ^{(-2)}_{\tilde q, f=0}= \bar \Delta_{\tilde q}^{(-2)} 
\frac{F_{\pi} \pi_3}{2 D},
\label{tadchromo}
\end{equation}
where, by NDA, the LEC scales as
\begin{equation}
\bar \Delta_{\tilde q}^{(-2)}=  
\Or\left(\tilde \delta_3 \frac{\mpi^2 \MQCD^2}{M_{\slashT}^2}\right).
\label{tadscalc1}
\end{equation}

The next operators appear two orders higher in the chiral expansion.
First, there is a $\tilde d_3 \tilde W_3$-type operator built from
two pion covariant derivatives.
Second,
there are contributions from the combined effect of the quark mass and 
the qCEDM, producing
pionic operators with the same chiral properties as the tensor products
$\tilde d_3 \tilde W_3 \otimes \bar m S_4$ and  
$\tilde d_0 \tilde V_4 \otimes \bar m \varepsilon P_3$.
Together, these are
\begin{equation} 
\mathcal L ^{(0)}_{\tilde q, f=0}= 
\vartheta_1\frac{\pi_3}{F_{\pi}D} \left(D_\mu \boldpi\cdot D^\mu \boldpi\right)
+\bar \Delta_{\tilde q}^{(0)} \frac{\Fp \pi_3}{2 D}
\left(1-\frac{2\boldpi^2}{\Fp^2D}\right),
\label{eq:tadpole}
\end{equation}
where 
\begin{equation}
\vartheta_1 = \Or\left(\tilde \delta_3 \frac{\mpi^2}{M^2_{\slashT}}\right),
\qquad
\bar \Delta_{\tilde q}^{(0)} = \Or\left((\varepsilon\tilde \delta_0+\tilde \delta_3)
\frac{\mpi^4}{M^2_{\slashT}}\right) .
\label{tadscalc2}
\end{equation}
The $+$ in the scaling of $\bar \Delta_{\tilde q}^{(0)}$ should not be taken 
literally, 
but as an indication that the LEC gets contributions from two sources. 

\subsection{FQLR}

In the pionic sector, $X_{34}$ also
leads to a tadpole operator, but of different form than qCEDM's:
\begin{equation}
\mathcal L ^{(-4)}_{\mathrm{LR}, f=0}= 
\bar\Delta_{\mathrm{LR}}^{(-4)} \frac{F_{\pi} \pi_3}{2 D}
\left(1-\frac{2\boldpi^2}{\Fp^2D}\right),
\label{tadFQLR}
\end{equation}
with coefficient
\begin{equation}
\bar\Delta_{\mathrm{LR}}^{(-4)} =  \Or\left(\xi \frac{\MQCD^4}{M_{\slashT}^2}\right).
\label{tadscalFQLR}
\end{equation}
Equation \eqref{tadFQLR} generates different three-pion vertices 
than Eq. \eqref{tadchromo}. As we will see, this difference
is important, because the elimination of the pion tadpole 
does not completely cancel the operator in
Eq. \eqref{tadFQLR}; instead, 
it leaves some three- and more-pion couplings behind.

Two orders down we find terms with two derivatives and terms transforming as 
${\rm Im}\,\Xi \,X_{34}\otimes \bar m S_4$,
\begin{eqnarray}
\mathcal L ^{(-2)}_{\mathrm{LR}, f=0}&=& 
\vartheta_2 \frac{ \pi_3}{\Fp D^2}\left(D_\mu \boldpi\cdot D^\mu \boldpi\right)
\left(1-\frac{\boldpi^2}{\Fp^2}\right) 
+\vartheta_3 \frac{\boldpi\cdot D_\mu \boldpi}{\Fp D}
\left(D^\mu \pi_3 -\frac{2}{\Fp^2 D}\pi_3\boldpi\cdot D^\mu \boldpi\right)
\nonumber\\
&&+ \frac{F_{\pi} \pi_3}{2 D} \left[\bar\Delta_{\mathrm{LR}1}^{(-2)}  
+ \bar\Delta_{\mathrm{LR}2}^{(-2)}\left(1-\frac{2\boldpi^2}{\Fp^2 D}\right)^2\right],
\label{tadFQLR2}
\end{eqnarray}
with the scalings 
\begin{equation}
\vartheta_{2,3} = \Or\left(\xi \frac{\MQCD^2}{M_{\slashT}^2}\right),
\qquad
\bar\Delta_{\mathrm{LR}1,2}^{(-2)} =  
\Or\left(\xi \frac{\mpi^2 \MQCD^2}{M_{\slashT}^2}\right). 
\label{tadscalFQLR2}
\end{equation}

\subsection{qEDM}

The leading pion operators induced by the qEDM transform as 
$d_3 W_3\otimes e I^\mu/6$ or $d_0 V_4\otimes e T^\mu_{34}/2$, 
and are simply given by
\begin{equation}
\mathcal L ^{(1)}_{ q, f=0}=  \bar\Delta_{q}^{(1)} \frac{F_{\pi} \pi_3}{2 D},
\label{tadedm}
\end{equation}
where the LEC scales as
\begin{equation}
\label{tadscalqEDM}
\bar\Delta_{q}^{(1)} =\Or\left(\frac{\alpha_{\mathrm{em}}}{4\pi}(\delta_0 + \delta_3)
 \frac{\mpi^2 \MQCD^2}{M_{\slashT}^2}\right).
\end{equation}
Again the $+$ in the scaling of the LEC should not be taken literally.
Because of the $\alpha_{\mathrm{em}}/4\pi$ suppression, we do not
bother to go to higher order.

\subsection{$\chi$ISs}

In the mesonic sector it is not possible to write down $\slashPT$ 
chiral-invariant operators.
Operators in this sector can be constructed by combining the effects 
of $\chi$ISs
with chiral-symmetry breaking from the quark mass difference,
which renders them identical in form to those of the qCEDM.

Indeed, at lowest order $d_W{\bar I}\otimes \varepsilon {\bar m}P_3$
gives
\begin{equation}
\mathcal L ^{(-2)}_{w, f=0}= \bar \Delta^{(-2)}_w \frac{F_{\pi} \pi_3}{2 D},
\label{tadgCEDM}
\end{equation}
with 
\begin{equation}
\label{tadscalgc1}
 \bar\Delta^{(-2)}_w =  \Or\left(w \varepsilon 
\frac{\mpi^2 \MQCD^2}{M_{\slashT}^2}\right).
\end{equation}
At $\Delta_6=0$, we get operators with two covariant derivatives and 
operators transforming as 
$d_W \bar I \otimes \bar m S_4 \otimes \bar m \varepsilon P_3$,
\begin{eqnarray} 
\mathcal L ^{(0)}_{w, f=0}= 
\vartheta_1\frac{\pi_3}{F_{\pi}D} \left(D_\mu \boldpi\cdot D^\mu \boldpi\right)
+\bar\Delta^{(0)}_w \frac{\pi_3 F_{\pi}}{2 D}
\left(1-\frac{2\boldpi^2}{\Fp^2 D}\right) ,
\label{eq:wtadpole}
\end{eqnarray}
where the LECs scale as
\begin{equation}
\vartheta_1 = \Or\left(w \varepsilon \frac{\mpi^2}{M^2_{\slashT}}\right),
\qquad 
\bar\Delta^{(0)}_w = \Or\left(w \varepsilon \frac{\mpi^4}{M^2_{\slashT}}\right).
\label{tadscalc2prime}
\end{equation}

\subsection{Tadpole extermination}
\label{tad}

We have just seen 
that the transformation properties of the $\slashPT$ dimension-six sources 
cause pion tadpoles to appear in the $\slashPT$ mesonic Lagrangian.
Since the coupling constant of the neutral pion to the vacuum is small 
compared to the pion mass, these tadpoles can be dealt with in 
perturbation theory, meaning that, for any given $\slashPT$ observable 
at a given accuracy, only a finite number of neutral pions disappearing 
into the vacuum must be considered \cite{BiraEmanuele}.
For applications, such as the calculation of hadronic \cite{Vri11a} and 
nuclear \cite{Vri11b, Vri12} EDMs, it is however more convenient 
to eliminate the pion tadpoles from the mesonic Lagrangian, 
which can be achieved with field redefinitions of the form discussed 
in Ref. \cite{BiraEmanuele}. 

We define a new pion field $\boldzeta'=\boldpi'/\Fp$ through
\begin{equation}
\zeta_i = 
\frac{1}{d'}\left\{\zeta'_i -\delta_{i3}[2C\zeta'_3 + S(1-\boldzeta'^2)]\right\},
\label{fieldredef}
\end{equation}
where
\begin{equation}
d' = 1-C(1-\boldzeta'^2)+2 S \zeta'_3,
\label{dprime}
\end{equation}
and
\begin{eqnarray}
C=\frac{1}{2}(1-\cos \varphi),\qquad S = \frac{1}{2}\sin \varphi,
\label{CS}
\end{eqnarray}
in terms of an angle $\varphi$. 
This transformation looks complicated, but has the nice property that 
the pion covariant derivative transforms simply as
\begin{equation}
D_\mu \zeta_i = O'_{ij} D'_\mu \zeta_j',
\label{rotation}
\end{equation} 
with an orthogonal matrix
\begin{equation}
O'_{ij}= \delta_{ij}-\frac{2}{d'}
\left\{C\left[(\boldzeta'^2-\zeta_3'^{2})\delta_{ij}
-\ep_{3ik}\ep_{3jl}\zeta'_k\zeta'_l\right]+(C\zeta_3'+S)
\left(\zeta'_i\delta_{3j}-\zeta'_j\delta_{3i}\right)\right\}.
\label{O}
\end{equation}
This simple transformation
ensures that chiral-invariant operators, built from
pion covariant derivatives, are invariant under the field redefinition
in Eq. \eqref{fieldredef}. This is not the case for operators that break 
chiral symmetry and involve the pion field directly. 

Before performing any rotation we summarize here the 
non-derivative, chiral-symmetry-breaking 
Lagrangian in the purely mesonic sector, including 
quark mass and $\slashPT$ operators in the first two orders
(except for the very small qEDM operators, for which
we keep only the least unimportant term).
{}From Eqs. \eqref{eq:pimass}, \eqref{tadchromo}, \eqref{eq:tadpole}, 
\eqref{tadFQLR}, \eqref{tadFQLR2},  \eqref{tadedm},
\eqref{tadgCEDM}, and \eqref{eq:wtadpole},
\begin{eqnarray}
\mathcal L^{}_{\textrm{tadpole}}  &=&  
-\frac{m^2_{\pi}}{2D}\left(1+\frac{\Delta^{}m_\pi^2}{m^2_{\pi}D}\right) \boldpi^2 
+ \frac{\delta^{} m^2_{\pi} }{2D^2} \pi_3^2
+ \left(\bar \Delta^{(-2)}_{\mathrm{LR}1}
+ \bar \Delta_{\tilde q}^{(-2)}+\bar \Delta_q^{(1)}+\bar\Delta^{(-2)}_w\right) 
  \frac{F_{\pi} \pi_3}{2 D}
\nonumber\\
&&
+ \left(\bar\Delta_{\mathrm{LR}}^{(-4)}+\bar \Delta_{\tilde q}^{(0)}
+\bar\Delta^{(0)}_w \right)
\frac{F_\pi \pi_3}{2D}\left(1- \frac{2\boldpi^2}{F^2_{\pi}D}\right) 
+\bar\Delta_{\mathrm{LR}2}^{(-2)} \frac{F_\pi \pi_3}{2D}
\left(1- \frac{2\boldpi^2}{F^2_{\pi}D}\right)^2 .
\label{eq:3.3.1}
\end{eqnarray}
We omit chiral-breaking operators generated by the electromagnetic interaction,
which are not affected by the field redefinitions \cite{BiraEmanuele}.

Our first goal is to remove for each source the dominant tadpole, 
that is, the terms with index $\Delta_6=-4$ for FQLR, $\Delta_6=-2$ 
for qCEDM and $\chi$ISs, and $\Delta_6=1$ for qEDM. 
By performing the field redefinitions in 
Eq. \eqref{fieldredef} with the small angle
\begin{eqnarray}\label{varphiLO}
\varphi\simeq \tan\varphi &=&
- \frac{1}{m^2_{\pi}}
\left(\bar \Delta_{\mathrm{LR}}^{(-4)} + \bar \Delta_{\tilde q}^{(-2)} 
+ \bar\Delta^{(-2)}_w+\bar \Delta_q^{(1)}\right)\nonumber\\
&=&\mathcal O \left(\left( \xi \frac{\MQCD^2}{\mpi^2}+ \tilde\delta_3 
+w\varepsilon+\frac{\alpha_{\mathrm{em}}}{4\pi}(\delta_0 + \delta_3)\right) 
\frac{\MQCD^2}{M^2_{\slashT}}  \right),
\end{eqnarray}
the dominant tadpoles are removed. 
{}From here on we keep only terms linear in $\varphi$.
The effect of this field redefinition 
on the mesonic Lagrangian, apart from canceling the dominant tadpoles, 
is to modify the coefficients of the tensors:
\begin{eqnarray}
\mathcal L^{}_{\textrm{tadpole}}  &=&  
-\frac{m^2_{\pi}}{2D}\left(1+\frac{\Delta^{}m_\pi^2}{m^2_{\pi}D}\right) \boldpi^2 
+ \frac{\delta^{} m^2_{\pi} }{2D^2} \pi_3^2 
- \bar \Delta_{\mathrm{LR}}^{(-4)} \frac{ \pi_3 \boldpi^2}{\Fp D^2} 
+ \bar \Delta_{\mathrm{LR}1}^{(-2)}\frac{\Fp \pi_3}{2 D}
\nonumber\\
&&
+\left(\bar\Delta_{\mathrm{LR}}^{(-2)\prime}+\bar \Delta_{\tilde q}^{(0)\prime}
+\bar\Delta^{(0)\prime}_w\right)
\frac{F_\pi \pi_3}{2D}\left(1- \frac{2\boldpi^2}{F^2_{\pi}D}\right) 
+\bar\Delta_{\mathrm{LR}2}^{(-2)} \frac{F_\pi \pi_3}{2D}
\left(1- \frac{2\boldpi^2}{F^2_{\pi}D}\right)^2\!\!,
\label{tadpoleshift1}
\end{eqnarray}
in terms of the shifted LECs,
\begin{eqnarray}
\bar\Delta_{\mathrm{LR}}^{(-2)\prime}&=&
-\frac{\Delta \mpi^2-\delta \mpi^2}{\mpi^2} \, \bar \Delta_{\mathrm{LR}}^{(-4)} 
= \Or\left(\xi (1+\varepsilon^2)\frac{\mpi^2\MQCD^2}{M_{\slashT}^2}\right),
\\
\bar \Delta_{\tilde q}^{(0)\prime} &=& 
\bar \Delta_{\tilde q}^{(0)}
-\frac{\Delta m^2_{\pi}- \delta m^2_{\pi}}{m^2_{\pi}} \, 
\bar \Delta_{\tilde q}^{(-2)}
= \mathcal O\left(\left(\varepsilon\tilde\delta_0 +(1+\varepsilon^2) 
\tilde \delta_3 \right)  \frac{m_{\pi}^4}{M^2_{\slashT}}\right),
\\ 
\bar\Delta^{(0)\prime}_w  &=& 
\bar\Delta^{(0)}_w
-\frac{\Delta m^2_{\pi}- \delta m^2_{\pi}}{m^2_{\pi}} \, \bar\Delta^{(-2)}_w
= \mathcal O\left(w\varepsilon (1+\varepsilon^2) \frac{m_{\pi}^4}
{M^2_{\slashT}}\right).
\end{eqnarray}
(Here and in the following we neglect terms of higher order than we need 
in the rest of the paper.)
We can thus rotate away the leading tadpoles without introducing 
new interactions in the meson Lagrangian. 
The net effect of the rotation is only to change the dependence 
of the coefficients on the parameters $\tilde\delta_{0}$, $\tilde\delta_3$, 
and $\varepsilon$. 

The remaining tadpoles in Eq. \eqref{tadpoleshift1} can be eliminated by 
a second rotation, now with an even smaller angle
\begin{equation}
\varphi^{\prime} = - \frac{1}{m^2_{\pi}}
\left(\bar\Delta_{\mathrm{LR}1}^{(-2)} + \bar\Delta_{\mathrm{LR}2}^{(-2)} 
+ \bar\Delta_{\mathrm{LR}}^{(-2)\prime} 
+ \bar \Delta_{\tilde q}^{(0)\prime}+ \bar\Delta^{(0)\prime}_w \right),
\label{varphiNLO}
\end{equation}
which leaves us with
\begin{eqnarray}
\mathcal L^{}_{\textrm{tadpole}}  &=&  
-\frac{m^2_{\pi}}{2D}\left(1+\frac{\Delta^{}m_\pi^2}{m^2_{\pi}D}\right) \boldpi^2 
+ \frac{\delta^{} m^2_{\pi} }{2D^2} \pi_3^2
\nonumber\\
&&
-\frac{\pi_3 \boldpi^2}{\Fp D^2} 
\left(\bar \Delta_{\mathrm{LR}}^{(-4)} + \bar \Delta_{\mathrm{LR}}^{(-2)\prime} 
+ \frac{2\bar \Delta_{\mathrm{LR}2}^{(-2)}}{D} + \bar \Delta_{\tilde q}^{(0)\prime}
+ \bar\Delta^{(0)\prime}_w \right)  .
\label{eq:3.3.16}
\end{eqnarray}
Although the tadpoles are removed, residual $\slashPT$ interactions
involving an odd number of pions are left behind. 
In case of the qCEDM and $\chi$ISs these terms carry a high chiral index
and 
such multi-pion vertices contribute to observables at high order. 
In case of the FQLR, due to its complicated $SO(4)$ properties, 
a three-pion interaction remains with a relatively low index. 
The consequences of this interaction are 
discussed in Secs. \ref{FFFQLR}, \ref{FQLREDM}, and \ref{Potential}.  

So far we did not discuss the operators with LECs $\vartheta_i$. 
These operators are not affected by the above rotations up to the order  
we are working. They induce changes in the pion kinetic term,
which nevertheless can be removed by another pion-field redefinition.

\section{Pion-nucleon sector}
\label{piNsector}

For most applications, we can ignore $\slashPT$ effects in the
pion sector, once the tadpoles have been removed.
Tadpole extermination does leave traces, however, in operators
involving nucleons. In this section we construct the $\slashPT$ pion-nucleon
operators, incorporating the new terms arising from the field redefinitions
needed to remove the tadpoles.
$\slashPT$ pion-nucleon interactions are very important for nucleon 
and nuclear EDMs. 

\subsection{Tadpoles strike back}

It is extremely convenient to supplement the pion field redefinition
\eqref{fieldredef} with a nucleon field redefinition \cite{BiraEmanuele}
\begin{equation}
N= U' N',
\label{Nfieldredef}
\end{equation}
in terms of the unitary matrix
\begin{equation}
U^\prime=\frac{1}{\sqrt{d'}}
\left[\sqrt{1-C}+\sqrt{C}(\zeta'_3+i\ep_{3jk}\zeta'_j\tau_k)\right],
\end{equation}
where $d'$, $C$, and $S$ are given by Eqs. \eqref{dprime} and \eqref{CS}.
This redefinition ensures simple transformation laws for covariant
objects such as nucleon covariant derivatives and nucleon bilinears,
\begin{eqnarray}
\mathcal D_\mu N = U' \mathcal D'_\mu N',
\qquad
\bar N N= \bar N' N',
\qquad
\bar N \tau_i N = O'_{ij} \bar N' \tau_j N',
\label{Nrotation}
\end{eqnarray} 
where $O'$ is the orthogonal matrix in Eq. \eqref{O}.
The properties \eqref{Nrotation} extend to the nucleon sector the invariance
under field redefinition of chiral-invariant operators, built from 
nucleon fields and covariant derivatives. 

The two rotations that allow elimination of the pion tadpoles in leading orders 
affect the 
chiral-breaking terms involving nucleons.
Most importantly, the nucleon sigma term and mass splitting
in Eq. \eqref{eq:Nmass}
become, after the rotation of Eq. \eqref{varphiLO},
\begin{equation}
{\mathcal L}_{\slashchi, f=2}= 
\Delta m_N \left(1-\frac{2 \boldpi^2}{\Fp^2 D}
+  \varphi\frac{2\pi_3}{\Fp D}\right) \Nb N 
+ \frac{\delta m_N}{2}
\Nb \left( \tau_3-\frac{ 2 \pi_3 }{\Fp^2 D}\boldtau\cdot\boldpi
+ \varphi\frac{2}{\Fp D} \boldtau\cdot\boldpi  \right)N.
\label{sigmadeltarot}
\end{equation}
The terms linear in $\varphi$ contribute to the $\slashPT$ pion-nucleon 
interactions, in addition to those generated directly from
the chiral structure of the $\slashPT$ dimension-six sources.
Higher-order pion-nucleon operators come from 
rotation with angle $\varphi'$ \eqref{varphiNLO} and/or
higher-order chiral-breaking terms in the $C\!P$-even Lagrangian.

We refrain from giving the boring details here. 
Instead, we turn to the construction
of the $\slashPT$ pion-nucleon interactions from the chiral structure
of the various sources. We indicate below the changes that 
arise from tadpole extermination. 

\subsection{qCEDM}
\label{piNqcedm}

Since the qCEDM breaks chiral symmetry, it gives rise to non-derivative
pion-nucleon interactions. Already at chiral order $\Delta_6 = -1$,
the structures $\tilde d_0\tilde V_4$ and $\tilde d_3\tilde W_3$ give
\begin{equation}
\mathcal L ^{(-1)}_{\tilde q, f=2}= 
-\frac{\bar g_0}{\Fp D}\Nb\boldtau\cdot\boldpi N
-\frac{\bar g_1}{\Fp D}\pi_3 \Nb N,
\label{S4V4W3}
\end{equation}
with 
\begin{equation}
\bar g_0 = \Or\left(\tilde \delta_0 \frac{\mpi^2 \MQCD}{M^2_{\slashT}}\right),
\qquad
\bar g_1 = \Or\left(\tilde \delta_3 \frac{\mpi^2 \MQCD}{M^2_{\slashT}}\right).
\label{LEC0LEC1}
\end{equation}
Comparing Eqs. \eqref{S4V4W3} and \eqref{sigmadeltarot} 
shows that tadpole removal leads to shifts
\begin{eqnarray}
\bar g_0 &\rightarrow& \bar g_0 +\delta m_N 
\frac{\bar \Delta_{\tilde q}^{(-2)}}{\mpi^2} 
= \Or \left( \left(\tilde\delta_0 +\varepsilon \tilde \delta_3\right)
\frac{\mpi^2 \MQCD}{M_{\slashT}^2}\right), \label{g0newa}
\\
\bar g_1 &\rightarrow& \bar g_1 +2\Delta m_N 
\frac{\bar \Delta_{\tilde q}^{(-2)}}{\mpi^2} 
= \Or \left(\tilde \delta_3\frac{\mpi^2 \MQCD}{M_{\slashT}^2}\right).
\label{g0new}
\end{eqnarray}
While $\bar g_1$ gets a change of ${\cal O}(1)$ that 
nevertheless does not affect
its scaling,
$\bar g_0$ now depends on the isovector qCEDM as well. 
Equations \eqref{g0newa} and \eqref{g0new} can be recast in an interesting form 
by making use of chiral symmetry. 
Equation \eqref{tildeVW} shows that the isovector (isoscalar) component of the 
qCEDM
and the isoscalar (isovector) component of the 
quark chromo-magnetic dipole moment (qCMDM) belong to the same chiral vector. 
Chiral symmetry then relates the coefficients of the 
$\slashPT$ operators induced by the 
qCEDM to $PT$ effects from the qCMDM, 
in the same way as operators stemming from the QCD $\tb$ term are related 
to isospin-breaking operators generated by the quark mass difference 
\cite{CDVW79,BiraEmanuele}.   
The qCMDM generates corrections to the pion and nucleon masses and to 
the nucleon mass splitting  of exactly the same form as 
Eqs. \eqref{eq:pimass} and \eqref{eq:Nmass}, with  $m^2_{\pi}$, $\Delta m_N$ 
and $\delta m_N$ replaced, respectively, by the LECs 
$\Delta_{\tilde q} m^2_{\pi}$, $\Delta_{ \tilde q} m_N$, and $\delta_{\tilde q} m_N$.
$\Delta_{\tilde q} m^2_{\pi}$ and $\Delta_{\tilde q} m_N$ are generated by the 
isoscalar qCMDM, and are linear in 
$\tilde c_0 \equiv \textrm{Re} (\tilde \Gamma_0)$, while $\delta_{\tilde q} m_N$
stems from the isovector qCMDM, and is proportional to 
$\tilde c_3 \equiv \textrm{Re} (\tilde\Gamma_3)$.
In terms of these LECs, Eqs. \eqref{g0newa} and \eqref{g0new} become 
\cite{Hockings}
\begin{eqnarray}
\bar g_0 &=&  \left( \delta_{\tilde q} m_N \frac{\tilde d_0}{\tilde c_3}  
+ \delta m_N \frac{\Delta_{\tilde q} m^2_{\pi}}{ m^2_{\pi}} 
\frac{\tilde d_3}{\tilde c_0} \right), 
\label{qCMDM1} \\ 
\bar g_1 &=& - 2 \left(\Delta_{\tilde q} m_N 
- \Delta m_N \frac{\Delta_{\tilde q} m^2_{\pi}}{m^2_{\pi}}\right) 
\frac{\tilde d_3}{\tilde c_0}. 
\label{qCMDM2}
\end{eqnarray}
Equations \eqref{qCMDM1} and \eqref{qCMDM2} are consistent with the results of
Ref. \cite{Pospelov:2001ys}.
Equation \eqref{qCMDM2} is particularly interesting because it signals the 
possibility of a cancellation between the two different contributions 
to $\bar g_1$. While an exact cancellation requires a non-trivial relation 
between the non-perturbative, vacuum and nucleon matrix elements of the 
isoscalar qCMDM operator, and it is not to be expected on general grounds, 
accidental, partial cancellations could suppress $\bar g_1$ with respect 
to its NDA value.
A first-principle calculation of the 
corrections to the pion and nucleon masses induced by the qCMDM, 
$\Delta_{\tilde q} m_N$ and $\Delta_{\tilde q} m^2_{\pi}$, is, then, 
of foremost importance for a robust estimate of $\bar g_1$.
With this in mind,
for notational simplicity in the following we absorb the terms 
$\propto \bar\Delta_{\tilde q}^{(-2)}$ in $\bar g_i$.

Equation \eqref{S4V4W3} shows the first important difference between 
$\slashPT$ from dimension-six operators and the QCD $\tb$ term,
namely, the presence of the $\slashPT$ isospin-breaking 
interaction with LEC $\bar g_1$ at leading order in the $f=2$ Lagrangian. 
This fact is particularly relevant for the $\slashPT$ moments of the 
deuteron \cite{Lebedev, Vri11b, Vri12}.
The $\tb$ term also generates this interaction but it is suppressed by 
$(\mpi/\MQCD)^2$ compared to leading-order interactions \cite{BiraEmanuele}.

Increasing $\Delta_6$ by one, we find operators with the same
chiral structures but one covariant derivative,
\begin{eqnarray}
\mathcal L ^{(0)}_{\tilde q, f=2}&=& 
\frac{2\bar \bt_1}{\Fp^2 D}(\boldpi\cdot\mathcal D_\mu \boldpi) \Nb S^\mu N 
+ \frac{2\bar \bt_2}{\Fp^2 D} (\pi_3 \mathcal D_\mu \boldpi)\cdot 
\Nb S^\mu \boldtau N 
\nonumber\\
&&+\frac{\bar \bt_3}{\Fp} \left(\delta_{i3}-\frac{2\pi_i \pi_3}{\Fp^2 D}\right) 
\Nb (\boldtau\times v\cdot D \boldpi)_i N,
\label{chromo1der}
\end{eqnarray}
where the LECs are
\begin{equation}\label{LEC2}
\bar\bt_1 = \Or\left(\tilde \delta_0 \frac{\mpi^2}{M^2_{\slashT}}\right),
\qquad
\bar \bt_{2,3} = \Or\left(\tilde \delta_3 \frac{\mpi^2}{M^2_{\slashT}}\right). 
\end{equation}

Increasing $\Delta_6$ by another unit, the chiral structures
proliferate significantly, and so does the number of different
interactions.
With still the same chiral structure as before,
but two covariant derivatives, we find 
\begin{eqnarray}
\mathcal L ^{(1)}_{\tilde q, f=2}&=& 
\bigg\{ \frac{\bar \zeta_1}{ \Fp} (\mathcal D_{\mu,\perp} D^\mu_\perp \boldpi)\cdot 
\Nb \boldtau  N 
+ \frac{\bar \zeta_2}{\Fp^2} (D_\mu \boldpi\times v\cdot D\boldpi)\cdot 
\Nb S^\mu \boldtau N
\nonumber\\
&&+  \frac{\bar \zeta_3}{\Fp}  (v\cdot\mathcal D v\cdot D \boldpi)\cdot 
\Nb \boldtau  N + \frac{\bar \zeta_4}{2\Fp}  (D_\nu \boldpi)\cdot 
\Nb [S^\mu,S^\nu] \boldtau \mathcal D_{\mu,-} N 
\bigg\} \frac{1}{D}\left(1-\frac{\boldpi^2}{\Fp^2}\right)
\nonumber\\
&& +\bigg\{\frac{\bar \zeta_5}{4} \Nb \boldtau (v\cdot \mathcal D_-)^2 N
+ \frac{\bar \zeta_6}{4} \Nb \boldtau \mathcal D_{\perp,-}^2 N 
- \frac{\bar \zeta_7}{\Fp} 
\Nb S^\mu [\boldtau\times(v\cdot \mathcal D D_\mu \boldpi)]N
\nonumber\\
&&- \frac{i\bar \zeta_8}{\Fp} (v\cdot D \boldpi)\Nb S\cdot \mathcal D_- N
- \frac{2i\bar \zeta_9}{\Fp^2} (D_\mu \boldpi \times D_\nu \boldpi)
\Nb [S^\mu,S^\nu] N
\nonumber\\
&&-\frac{\bar \zeta_{10}}{\Fp^2}(D_{\mu,\perp} \boldpi)(D^{\mu}_\perp\boldpi)\cdot 
\Nb \boldtau N
-\frac{\bar \zeta_{11}}{\Fp^2}(v\cdot D \boldpi) (v\cdot D \boldpi)\cdot 
\Nb \boldtau N
\nonumber\\
&&- \frac{\bar \zeta_{12}}{\Fp^2}(D_{\mu,\perp} \boldpi)^2\Nb \boldtau N 
-\frac{\bar \zeta_{13}}{\Fp^2}(v\cdot D \boldpi)^2\Nb \boldtau N\bigg\}\cdot 
\frac{\boldpi}{\Fp D}
\nonumber\\
&&+\bigg\{ \frac{\bar \xi_{1}}{\Fp} (\mathcal D_{\mu,\perp} D^\mu_\perp \pi_i)\Nb N 
+ \frac{\bar \xi_2}{\Fp^2} (D_\mu \boldpi\times v\cdot D\boldpi)_i 
\Nb S^\mu  N
+ \frac{\bar \xi_{3}}{\Fp}(v\cdot \mathcal Dv\cdot D\pi_i)\Nb N
\nonumber\\
&&+ \frac{\bar \xi_{4}}{\Fp}(D_\nu \pi_i)\Nb[S^\mu,S^\nu]\mathcal D_{\mu,-}N
+\frac{i\bar \xi_{5}}{\Fp} 
\Nb ((D_\mu \boldpi)\times \boldtau)_i \mathcal D^\mu_{\perp,-} N \bigg\}
\left(\delta_{i3}-\frac{2\pi_3\pi_i}{\Fp^2 D}\right)
\nonumber\\
&&+ \bigg\{\frac{\bar \xi_6}{4} \Nb (v\cdot \mathcal D_-)^2 N 
+\frac{\bar \xi_7}{4} \Nb \mathcal D_{\perp,-}^2 N 
+ \frac{i\bar \xi_8}{2 \Fp} (v\cdot D \boldpi)\cdot
\Nb \boldtau S\cdot \mathcal D_- N
\nonumber\\
 &&+ \frac{i\bar \xi_9}{2\Fp^2} (D_\mu \boldpi \times D_\nu \boldpi)\cdot 
\Nb [S^\mu,S^\nu]\boldtau N 
+\frac{\bar \xi_{10}}{\Fp^2}(D_{\mu,\perp} \boldpi)^2\Nb  N 
+ \frac{\bar \xi_{11}}{\Fp^2}(v\cdot D \boldpi)^2\Nb  N\bigg\}
\frac{2\pi_3}{\Fp D}.
\nonumber \\
\label{eq:2der.1eq:2der.2}
\end{eqnarray}
The scaling of the LECs $\bar \zeta_i$ and $\bar \xi_i$ is given by
\begin{eqnarray}\label{LEC3}
\bar \zeta_i = \Or\left(\tilde \delta_0 \frac{\mpi^2}{M^2_{\slashT}\MQCD}\right),
\qquad
\bar \xi_i = \Or\left(\tilde \delta_3 \frac{\mpi^2}{M^2_{\slashT}\MQCD}\right).
\end{eqnarray}
Reparametrization invariance relates some of these operators
to operators of lower order,
\begin{eqnarray}
&&\bar \zeta_4 = \bar \zeta_6 = \frac{\bar  g_0}{2 m_N^2},
\qquad 
\bar \zeta_8 =  \frac{g_A \bar  g_0}{ m_N^2}-\frac{\bar  \bt_1}{ m_N},
\nonumber\\
&& \bar  \xi_4 =  \bar  \xi_7 = \frac{\bar  g_1}{4 m_N^2},
\qquad 
\bar  \xi_5 = -\frac{\bar  \bt_3}{2 m_N},
\qquad 
\bar  \xi_8 = -\frac{g_A\bar  g_1}{m_N^2}-\frac{\bar  \bt_2}{m_N}.
\label{eq:rpirelations.1}
\end{eqnarray}

In these subleading pion-nucleon Lagrangians, the effects of the 
field redefinition
on operators that transform as $\tilde W_3$, 
like those with LECs
$\bar \beta_{2}$ and $\bar\beta_{3}$ in Eq. \eqref{chromo1der} 
and $\bar \xi_i$ in Eq. \eqref{eq:2der.1eq:2der.2},   
can be absorbed in a redefinition of the coefficients, 
whose scaling is still given by Eqs. \eqref{LEC2} and \eqref{LEC3}. 
For operators that transform as $\tilde V_4$, 
the scaling of the coefficients is modified, and
they get a contribution from the 
isospin-symmetry-breaking Lagrangian at order $\Delta = 1,2$ 
(listed, for example, in Ref. \cite{BiraEmanuele}). Schematically,
\begin{equation}
\bar\beta_{1} \rightarrow \bar\beta^{\prime}_{1}  =   
\mathcal O\left(\left(\tilde\delta_0 + \varepsilon \tilde\delta_3 \right) 
\frac{m^2_{\pi}}{M^2_{\slashT} }\right), \qquad
\bar\zeta_{i} \rightarrow  \bar\zeta^{\prime}_{i}  =  
\mathcal O\left(\left(\tilde\delta_0 + \varepsilon \tilde\delta_3\right) 
\frac{m^2_{\pi}}{M^2_{\slashT} \MQCD}\right).
\label{eq:3.3.27}
\end{equation}
 
Still at order $\Delta_6=1$,
interactions arise from the combined effect of the qCEDM 
and the QCD mass terms. The resulting operators transform as 
$(\bar m  S_4-\varepsilon \bar m  P_3)\otimes
(- \tilde d_0 \tilde V_4 +\tilde d_3 \tilde W_3)$.
They generate
\begin{eqnarray}
\mathcal L ^{(1)}_{\tilde q, f=2}&=&
- \frac{\delta \bar g_0 }{\Fp D}\left(1-\frac{2\boldpi^2}{\Fp^2}\right)
\Nb \boldtau\cdot \boldpi N
-\frac{\delta \bar g_1}{\Fp D}\left(1-\frac{2\boldpi^2}{\Fp^2}\right)\pi_3\Nb N 
\nonumber\\
&&
-\frac{\bar g_2}{\Fp D} \pi_3 \Nb \left(\tau_3 -\frac{2 \pi_3}{\Fp D} 
\boldtau\cdot\boldpi\right)N.
\label{cedmmass1cedmmass2}
\end{eqnarray}
The first two terms are corrections to the isoscalar and isovector
$\slashPT$ couplings in 
Eq. \eqref{S4V4W3}, from which they differ only by terms with three or more 
pions.
The corrections scale as 
\begin{equation}
\delta \bar g_0=\Or\left(\tilde \delta_0\frac{\mpi^4}{M^2_{\slashT}\MQCD}\right),
\qquad 
\delta \bar g_1= 
\Or\left(\left(\varepsilon \tilde \delta_0+ \tilde \delta_3\right) 
\frac{\mpi^4}{M^2_{\slashT}\MQCD}\right).
\end{equation}
For most practical purposes $\delta \bar g_0$ and $\delta \bar g_1$ can be 
absorbed into $\bar g_0$ and $\bar g_1$. 
The third operator 
is the most interesting, as it is the first contribution of the qCEDM to the 
isospin-breaking pion-nucleon interaction $\pi_3 \Nb \tau_3  N$. 
It 
has the scaling
\begin{equation}
\bar g_2 = 
\Or\left(\varepsilon \tilde \delta_3 \frac{\mpi^4}{M^2_{\slashT}\MQCD}\right).
\end{equation}
Thus, two orders above lowest all three possible non-derivative pion-nucleon 
$\slashPT$ interactions receive a contribution from the isovector qCEDM. 
Just like the $\tb$ term \cite{BiraEmanuele},
the isoscalar qCEDM generates $\bar g_2$ as well, 
but this requires a photon exchange so that $\bar g_2$ is suppressed 
by $\alpha_{\mathrm{em}}/4\pi$ and enters three orders above lowest.
This provides the second difference between isovector qCEDM
and $\tb$ term (and isoscalar qCEDM).
However, since for all sources $\bar g_2$ enters in the subleading Lagrangian, 
this difference is of little phenomenological interest.

Tadpole elimination affects subleading non-derivative couplings in two ways. 
First, the elimination of subleading tadpoles 
shifts the coefficients 
$\bar g_0$ and $\bar g_1$ in Eqs. \eqref{g0newa} and \eqref{g0new}, 
$\bar g_{0,1} \rightarrow \bar g_{0,1}+ \delta^{\prime} \bar g_{0,1}$, with 
\begin{equation}
\delta^{\prime} \bar g_0 = 
\mathcal O\left(\left(\tilde \delta_0 + \varepsilon \tilde{\delta}_3\right)
(1+\varepsilon^2)\frac{m_{\pi}^4}{M^2_{\slashT } \MQCD}\right), \quad
\delta^{\prime} \bar g_1 = \mathcal O\left(\left(\varepsilon \tilde \delta_0 
+ (1+\varepsilon^2) \tilde{\delta}_3\right)\frac{m_{\pi}^4}{M^2_{\slashT } \MQCD}
\right)
\end{equation}
contributing to the $\Delta_6 = 1$ Lagrangian.
Second, the operators in Eq. \eqref{cedmmass1cedmmass2} receive corrections 
from the transformation of $C\!P$-even, chiral-breaking operators that 
contribute to the chiral Lagrangian at order $\Delta = 3$. 
These operators are proportional to two powers of the quark masses, 
and transform as the tensor products 
$(\bar m S_4 - \bar m \varepsilon P_3) \otimes 
(\bar m S_4 - \bar m \varepsilon P_3)$. 
The rotation needed to eliminate the leading tapdoles causes them to generate 
corrections to the coefficients $\delta \bar g_{0,1}$ and $\bar g_2$. 
Now $\delta \bar g_0$ also depends on $\varepsilon \tilde \delta_3$
and $\delta \bar g_1$ receives a $\varepsilon^2 \tilde\delta_3$ correction, 
\begin{equation}
\delta \bar g_0 \to \delta \bar g_0^{\prime} =
\mathcal O\left(\left(\tilde \delta_0 + \varepsilon \tilde{\delta}_3\right)
\frac{m_{\pi}^4}{M^2_{\slashT } \MQCD}\right), \quad
\delta \bar g_1 \to \delta \bar g_1^{\prime} =
\mathcal O\left(\left(\varepsilon \tilde \delta_0 
+ (1+\varepsilon^2) \tilde{\delta}_3\right)\frac{m_{\pi}^4}{M^2_{\slashT } \MQCD}
\right),
\end{equation}
while the scaling of $\bar g_2$ is unchanged.

\subsection{FQLR}
\label{FQLRpiN}

The FQLR operator also can generate non-derivative pion-nucleon couplings.
Pion-nucleon operators start at 
$\Delta_6 = - 3$, and at this order there is only one operator
with the structure of $X_{34}$, the isovector pion-nucleon interaction.
It appears as if the FQLR generates only $\bar g_1$ at LO, 
in stark contrast with the $\tb$ term and qCEDM which generate, 
respectively, $\bar g_0$ and $\bar g_{0,1}$ at LO. However, 
the removal of the tadpole in Eq. \eqref{tadFQLR} effectively causes
the appearance of the $\bar g_0$ interaction at $\Delta_6 =-3$.  
At this order, 
\begin{equation}
\mathcal L ^{(-3)}_{\mathrm{LR}, f=2}= 
-\frac{\bar g_0}{\Fp D}\Nb\boldtau\cdot\boldpi N
-\frac{\bar g_1}{\Fp D}\left(1-\frac{2 \boldpi^2}{\Fp^2 D} \right) \pi_3 \Nb N
-\frac{\bar g_1'}{\Fp D}\pi_3 \Nb N,
\label{FQLR0}
\end{equation}
where
\begin{equation}
\bar g_1 = \Or\left(\xi \frac{\MQCD^3}{M^2_{\slashT}}\right)
\label{LECFQLR0}
\end{equation}
arises directly, and
\begin{equation}
\bar g_0 = \delta m_N\frac{\bar\Delta_{\mathrm{LR}}^{(-4)}}{\mpi^2}  
= \Or \left( \varepsilon \xi \frac{\MQCD^3}{M_{\slashT}^2}\right),
\qquad
\bar g_1' = 2 \Delta m_N\frac{\bar\Delta_{\mathrm{LR}}^{(-4)} }{\mpi^2} 
= \Or \left(\xi \frac{\MQCD^3}{M_{\slashT}^2}\right)
\label{g0FQLR}
\end{equation}
stem from tadpole removal via Eq. \eqref{sigmadeltarot}.
As in the case of the qCEDM, there can be an accidental cancellation
between $\bar g_1$ and $\bar g_1^{\prime}$. 
One can eliminate $\bar\Delta_{\mathrm{LR}}^{(-4)}$ and write 
\begin{equation}
\bar g_0 = \frac{\delta m_N}{2 \Delta m_N} \bar g_1^{\prime}.
\label{ratiog0g1}
\end{equation}
A lattice evaluation found $\delta m_N = 2.26$ MeV \cite{latticedeltamN}, 
while values for 
$\Delta m_N$ range between 45 and 60 MeV  
\cite{Gasser:1990ce}.
Therefore, for FQLR $\bar g_0$ is actually only 
a few percent of $\bar g_1^{\prime}$.

At higher orders FQLR generates some interactions that resemble
those of the isovector qCEDM, although with a more complicated
structure, plus some new interactions.
At $\Delta_6 = - 2$ and $\Delta_6 = - 1$
we find interactions similar to those with LECs
$\bar\beta_{2,3}$ in Eq. \eqref{chromo1der} 
and $\bar\xi_i$ in Eq. \eqref{eq:2der.1eq:2der.2},
but with the replacements
\begin{eqnarray}
\frac{2\pi_3}{\Fp D} &\rightarrow& \frac{2\pi_3}{\Fp D}
\left(1-\frac{2\boldpi^2}{\Fp^2 D}\right)
\nonumber\\
\left(\delta_{i3}-\frac{2\pi_3\pi_i}{\Fp^2 D}\right)&\rightarrow& 
\left[\delta_{i3}\left(1-\frac{2 \boldpi^2}{\Fp^2 D}\right)
-\frac{2\pi_i \pi_3}{\Fp^2 D}\left(3-\frac{4 \boldpi^2}{\Fp^2 D}\right)\right].
\end{eqnarray}
The additional interactions are
\begin{equation}
\mathcal L ^{(-2)}_{\mathrm{LR}, f=2}= 
\frac{2\bar \bt_4}{\Fp^2 D}
\left[(D_\mu \pi_3)\pi_i\ +(\boldpi\cdot D_\mu \boldpi)
\left(\delta_{i3}-\frac{4 \pi_i\pi_3 }{\Fp^2 D}\right)\right]\Nb S^\mu \tau_i N,
\label{FQLR1}
\end{equation}
and
\begin{eqnarray}
\mathcal L ^{(-1)}_{\mathrm{LR}, f=2}&=&  
\frac{2\bar \xi_{12}}{\Fp^3 D} \left[ D^\mu \pi_3 
-\frac{2\pi_3}{\Fp^2 D} (\boldpi\cdot D_\mu \boldpi)\right]
(\boldpi\cdot D_\mu \boldpi)\Nb\!N
\nonumber\\
&& + \frac{2\bar \xi_{13}}{\Fp^3 D} 
\left[ \dt_{i3}\pi_j + \dt_{j3}\pi_i -\frac{4\pi_3 \pi_i\pi_j }{\Fp^2 D} \right] 
(D_\mu \boldpi\times D_\nu \boldpi)_i \, \Nb  i [S^\mu, S^\nu]\tau_j  N 
\nonumber\\
&& + \frac{2i\bar \xi_{14}}{\Fp^2 D} 
\left[ \dt_{i3}\pi_j + \dt_{j3}\pi_i -\frac{4\pi_3 \pi_i\pi_j }{\Fp^2 D} \right] 
(v\cdot D\pi_i) \Nb \tau_j S\cdot \mathcal D_-  N,
\end{eqnarray}
where
\begin{equation}
\bar \beta_{2,3,4} = \Or\left(\xi \frac{\MQCD^2}{M^2_{\slashT}}\right),
\qquad
\bar \xi_{i} = \Or\left(\xi \frac{\MQCD}{M^2_{\slashT}}\right).
\label{LECFQLR1LECFQLR2}
\end{equation}
The RPI relations are the same as in Eq. \eqref{eq:rpirelations.1}, with
the additional relation
\begin{eqnarray}\label{xi14}
\bar\xi_{14} = -\frac{\bar  \bt_4}{m_N}.
\end{eqnarray}

Just like for qCEDM,
tadpole removal induces the subleading isospin-conserving 
$\slashPT$ operators with LECs $\bar \beta_1$ in Eq. \eqref{chromo1der} 
and $\bar\zeta_i$ in Eq. \eqref{eq:2der.1eq:2der.2},
but with scalings
\begin{equation}
\bar\beta_{1}= 
\mathcal O\left(\varepsilon \xi\frac{\MQCD^2}{M^2_{\slashT} }\right), 
\qquad
\bar\zeta_{i}= 
\mathcal O\left(\varepsilon \xi \frac{\MQCD}{M^2_{\slashT}}\right).
\label{eq:3.3.28}
\end{equation}

Again, two orders above the lowest order, 
operators appear due to an insertion of the quark mass (difference),
\begin{eqnarray}
\mathcal L ^{(-1)}_{\mathrm{LR}, f=2}&=& 
- \frac{\delta \bar g_0}{F_{\pi} D} \, \Nb \boldtau\cdot \boldpi N
-\frac{1}{\Fp D}\left[\delta_1\bar g_1  
+ \delta_2\bar g_1 \left(1-\frac{2\boldpi^2}{\Fp^2 D}\right)^2\right]\pi_3\Nb N 
\nonumber\\
&&-\frac{\bar g_2}{\Fp D} \left(1-\frac{2\boldpi^2}{\Fp^2 D}\right)  
\pi_3\Nb \left(\tau_3 -\frac{2 \pi_3}{\Fp D} \boldtau\cdot\boldpi\right)N
-\frac{\bar g_3}{\Fp^3 D^3}\pi_3^2 \Nb \boldtau\cdot\boldpi N,
\label{FQLRmass}
\end{eqnarray}
where the LECs scale as 
\begin{eqnarray}
\delta_{1,2}\bar g_1 = \Or\left(\xi \frac{\mpi^2\MQCD}{M_{\slashT}^2}\right),
\qquad
\delta \bar g_0 = \Or\left(\varepsilon\xi\frac{\mpi^2\MQCD}{M_{\slashT}^2}\right),
\qquad
\bar g_{2,3} = \Or\left(\varepsilon\xi\frac{\mpi^2\MQCD}{M_{\slashT}^2}\right). 
\label{FQLRmass2}
\end{eqnarray}
Although the operators in Eq. \eqref{FQLRmass} appear to be very complicated, 
if we ignore terms with three or more pions 
the $\delta_{1,2} \bar g_1$ and $\delta \bar g_0$ couplings are contributions
to the standard non-derivative pion-nucleon interactions. 
Just as for the isovector qCEDM, the third pion-nucleon coupling 
$\bar g_2$ comes in two orders higher than $\bar g_{0,1}$. 

As in the case of the qCEDM, the elimination of the tadpoles modifies 
Eq. \eqref{FQLRmass}. 
The elimination of the subleading tadpoles $\bar \Delta^{(-2)}_{\textrm{LR}1}$ 
and $\bar \Delta^{(-2)}_{\textrm{LR}2}$
shifts $\delta \bar g_0$ and $\delta_1 \bar g_1$, 
but does not modify their dependence on $\xi$ and $\varepsilon$. 
On the other hand, the effect of the elimination of the leading tadpole 
on chiral-breaking operators proportional to two powers of the quark masses 
$\bar m$ and $\bar m\, \varepsilon$ is to generate a contribution identical 
to Eq. \eqref{cedmmass1cedmmass2}, with the coefficients replaced by  
\begin{equation}\label{tadsubFQLR}
\delta^{\prime} \bar g_1 = \mathcal O\left(\left(1+\varepsilon^2\right) \xi 
\frac{m^2_{\pi} M_{\textrm{QCD}}}{M^2_{\slashT}}\right), 
\qquad 
\delta^{\prime} \bar g_0 = \mathcal O\left(\varepsilon \xi 
\frac{m^2_{\pi} M_{\textrm{QCD}}}{M^2_{\slashT}}\right), 
\qquad
\bar g_2^{\prime} = \mathcal O\left(\varepsilon \xi 
\frac{m^2_{\pi} M_{\textrm{QCD}}}{M^2_{\slashT}}\right).
\end{equation}

\subsection{qEDM} 

The pion-nucleon interactions originating from the qEDM
arise from the tensor product 
$(- d_0  V_4 +d_3  W_3)\otimes e (I^{\mu}/6+T^{\mu}_{34}/2)$,
and thus have a rich chiral structure:
\begin{equation}
\mathcal L ^{(2)}_{q, f=2}= 
-\frac{\bar g_0}{\Fp D}\Nb\boldtau\cdot\boldpi N 
- \frac{\bar g_1}{\Fp D}\pi_3 \Nb\!N
-\frac{\bar g_2}{\Fp D}\pi_3 \Nb \left[\tau_3 + \frac{2}{\Fp^2 D}
\left(\pi_3 \boldtau\cdot\boldpi-\boldpi^2 \tau_3\right)\right]N,
\label{eq:edm1}
\end{equation}
the scaling of the LECs being
\begin{eqnarray}
 &&\bar g_0 = \Or\left( \left(\delta_0 + \delta_3\right) 
\frac{\alpha_{\textrm{em}}}{4\pi} \frac{\mpi^2 \MQCD }{M^2_{\slashT}}\right),
\qquad
\bar g_1 = \Or\left( \left(\delta_0 + \delta_3\right) 
\frac{\alpha_{\textrm{em}}}{4\pi} \frac{\mpi^2 \MQCD }{M^2_{\slashT}}\right),
\nonumber\\
&&\bar g_2 = \Or\left(\delta_3 \frac{\alpha_{\textrm{em}}}{4\pi}  
\frac{\mpi^2 \MQCD }{M^2_{\slashT}}\right).
\label{LEC5}
\end{eqnarray}

The qEDM also generates tadpoles, as given in Eq. \eqref{tadedm}, 
which can be removed in the same way as for the other sources. 
This removal alters the  $\varepsilon$ dependence of $\bar g_0$, 
in a way that can be summarized by replacing $(\delta_0 + \delta_3)$ 
with $(\delta_0+\delta_3)(1+\varepsilon)$ in Eq. \eqref{LEC5}.
Since, as  Eq. \eqref{LEC5} shows, $\bar g_{0,1}$ 
already receive contributions from both the isoscalar and isovector qEDMs,
the added contributions are
not particularly relevant. 

Note that all possible non-derivative pion-nucleon interactions appear 
at the same order. (This only holds for the isovector qEDM, since
for the isoscalar qEDM $\bar g_2$ is suppressed.) 
Only the qEDM has this property: for all other $\slashPT$ dimension-six 
sources $\bar g_2$ is suppressed with respect to $\bar g_0$ and/or $\bar g_1$. 
However, the $\alpha_{\textrm{em}}/4\pi$ suppression
ensures that the qEDM pion-nucleon couplings are not significant 
for most purposes. 

\subsection{$\chi$ISs}
\label{piNchi}

In the pion-nucleon sector, no chiral-invariant $\slashPT$ operator
can be constructed with zero or one covariant derivative. 
The first operators therefore start at $\Delta_6 =-1$ and 
have two covariant derivatives or one insertion of the quark mass (difference): 
\begin{eqnarray}
\mathcal L ^{(-1)}_{w, f=2}&=& 
\frac{\bar \zeta_1}{\Fp} (\mathcal D_{\mu,\perp} D^\mu_\perp \boldpi)\cdot 
\Nb \boldtau  N 
+ \frac{\bar \zeta_2}{\Fp^2} (D_\mu \boldpi\times v\cdot D\boldpi)\cdot 
\Nb S^\mu \boldtau N
\nonumber\\
&&
-\frac{\bar g_{0 }}{\Fp D}\Nb\boldtau\cdot\boldpi N 
-\frac{\bar g_{1}}{\Fp D}\pi_3 \Nb\!N ,
\label{eq:Wein1}
\end{eqnarray} 
with scalings
\begin{equation}
\bar \zeta_{i }=\Or\left(w\frac{\MQCD}{M^2_{\slashT}}\right),
\qquad 
\bar g_{0 } = \Or\left(w\frac{\mpi^2 \MQCD}{M^2_{\slashT}}\right), 
\qquad 
\bar g_{1 } = \Or\left(\varepsilon w \frac{\mpi^2 \MQCD}{M^2_{\slashTsub}}\right).
\label{scaleWein1}
\end{equation}
It is interesting to notice that rotation invariance alone would allow 
a two-derivative operator of the form
$\bar \zeta_4 (D_\nu \boldpi) \cdot\Nb\boldtau [S^\mu, S^\nu] \mathcal D_{\mu -}N$
in Eq. \eqref{eq:Wein1}. However, this operator is not RPI by itself, and 
its variation cannot be absorbed by any other operator in the
leading-order Lagrangian. RPI, therefore, 
forces $\bar\zeta_4$  to vanish in leading order.
The operator $\bar\zeta_4$ appears at NNLO, and 
it is linked to $\bar g_0$ by RPI, as in Eq. \eqref{eq:rpirelations.1}.      

Here the tadpole rotation induces the shifts
\begin{eqnarray}\label{tadweinLO}
\bar g_0 &\rightarrow& \bar g_0 +\delta m_N \frac{\bar \Delta^{(-2)}_w}{\mpi^2} 
= \Or \left(\left(1 +\varepsilon^2 \right) w
\frac{\mpi^2 \MQCD}{M_{\slashT}^2}\right),
\\
\bar g_1 &\rightarrow& \bar g_1 +\Delta m_N \frac{\bar \Delta^{(-2)}_w}{\mpi^2} 
= \Or \left( \varepsilon w \frac{\mpi^2 \MQCD}{M_{\slashT}^2}\right),
\end{eqnarray}
which are less interesting,
because the dependence on $w$ is not fundamentally changed. 
As before, we absorb contributions from tadpole elimination in $\bar g_i$

{}From Eq. \eqref{eq:Wein1} we see that the $\chi$ISs, just as the 
qCEDM and in contrast to the $\tb$ term, induce $\bar g_1$ 
at the same order as $\bar g_0$. 
Differently from all the other sources, the $\chi$ISs also generate 
two-derivative operators of the same importance as $\bar g_0$ and $\bar g_1$. 
 
One order higher we find various types of interactions:
chiral-invariant operators with three covariant derivatives;
chiral-symmetry-breaking operators with one power of quark masses and 
one covariant derivative; 
and electromagnetic operators coming from the tensor product 
$ d_W \bar I \otimes e (I^{\mu}/6+T^{\mu}_{34}/2) 
\otimes e (I^{\mu}/6+T^{\mu}_{34}/2)$. 
The chiral-invariant operators are
\begin{eqnarray}
\mathcal L ^{(0)}_{w, f=2}&=& 
 - \frac{i\bar\varkappa_{1}}{\Fp} \mathcal D_\mu (v\cdot D \boldpi)\cdot 
\Nb \boldtau \mathcal D^\mu_{\perp -} N
+ \frac{i\varkappa_2}{2\Fp^2}( D_\mu \boldpi\times D_\nu \boldpi)\cdot 
\Nb \boldtau S^\mu \mathcal D^\nu_{\perp -} N 
\nonumber\\
&&+\frac{\bar\varkappa_{3}}{\Fp^2}
\left(\mathcal D_\mu (D\boldpi)^2\right) \Nb S^\mu N 
+ \frac{\bar \varkappa_{4}}{\Fp^2} 
\left(\mathcal D_\perp^\mu(D_\mu \boldpi \cdot D_\nu\boldpi)\right)
  \Nb S^\nu N,
\label{wein3der}
\end{eqnarray} 
where 
\begin{equation}
\bar \varkappa_{3,4}= \Or\left(\frac{w}{M^2_{\slashTsub}}\right),
\label{scaleWein1b}
\end{equation}
while RPI gives
\begin{equation}
\bar\varkappa_{1,2}=\frac{\bar\zeta_{1,2}}{m_N}.
\end{equation}
The quark-mass insertions give rise to 
\begin{eqnarray}
\mathcal L ^{(0)}_{ w, f=2}&=&\frac{2 \bar \bt_{1}}{\Fp^2 D} 
(\boldpi\cdot\mathcal D_\mu \boldpi) \Nb S^\mu N 
+ \frac{2 \bar \bt_{2 }}{\Fp^2 D}  (\pi_3 \mathcal D_\mu \boldpi)\cdot 
\Nb S^\mu \boldtau N 
\nonumber\\
&& + \frac{\bar \bt_{3 }}{\Fp} 
\left(\delta_{i3}-\frac{2\pi_i \pi_3}{\Fp^2 D}\right) 
\Nb (\boldtau\times v\cdot D \boldpi)_i N,
\label{weinmass1der}
\end{eqnarray}
with
\begin{equation}
\bar \bt_{1 } = \Or\left(w \frac{\mpi^2}{M^2_{\slashTsub}}\right),
\qquad
\bar \bt_{2,3} = \Or\left(\varepsilon w  \frac{\mpi^2}{M^2_{\slashTsub}}\right). 
\end{equation}
The removal of tadpoles does not affect the chiral-invariant operators 
in Eq. \eqref{wein3der}, while it slightly modifies the $\varepsilon$ 
dependence of $\beta_1$ in Eq. \eqref{weinmass1der}, 
in a way analogous to Eq. \eqref{tadweinLO}.

Finally, integrating out a hard photon gives the single operator
\begin{equation}
\mathcal L ^{(0)}_{w, f=2}=
-\frac{\bar g_{2}}{\Fp D} \Nb \left\{ \pi_3 \tau_3 
- \boldtau\cdot \boldpi\left[1 - \frac{2}{F^2_{\pi} D} 
\left(\boldpi^2 - \pi_3^2\right)\right]\right\}N.
\label{weinintphoton}
\end{equation}
We see that at this order the first contribution to the non-derivative 
pion-nucleon interaction $\bar g_{2}$ appears. It scales as 
\begin{equation}
\bar g_{2} = \Or\left(w \frac{\alpha_{\textrm{em}}}{4\pi} 
\frac{\MQCD^3}{M^2_{\slashTsub}}\right).
\end{equation}
With the usual counting 
$\alpha_{\textrm{em}}/4\pi \sim m^3_{\pi}/\MQCD^3$, 
this interaction is
suppressed by one order
compared to the $\bar g_{0,1}$ 
interactions.

\section{The pion-nucleon form factor}
\label{PNFF}

An important element in the evaluation of hadronic and nuclear EDMs is 
the $\slashPT$ pion-nucleon coupling.
In this section, we summarize the  pion-nucleon interactions for the 
different $\slashPT$ sources we have considered,
by calculating the $\slashPT$ PNFF
with the Lagrangian derived above. For convenience, we consider the 
Lagrangian after tadpole extermination.

We consider the three-point Green's function for an incoming (outgoing) nucleon
of momentum $p^\mu$ ($p^{\prime\, \mu }$) and an outgoing pion of momentum
$q^{\, \mu} = p^{\mu} - p^{\prime\, \mu }$ and isospin $a$. 
We take the incoming and outgoing nucleon to be
nonrelativistic and on-shell, so 
\begin{equation}
p^0 = \frac{\vec p^{\,2 }}{2 m_{N}} - \Delta m_N \mp \frac{\delta m_N}{2}
+\ldots, 
\quad p^{\prime 0}
 = \frac{\vec p^{\, \prime 2}}{2m_N} - \Delta m_N \mp \frac{\delta m_N}{2}+\ldots, 
\label{onshell}
\end{equation}
where the $- (+)$ sign holds for protons (neutrons) and the arrow denotes 
vectors in three-dimensional Euclidean space,  
$p^{\mu} = (p^0,\, \vec p\,)$, $p_{\mu} = (p^0,\, - \vec p\,)$.
It is clear that $p^0$ enters at one order higher than $\vec{p}$.
The Green function for on-shell nucleons can be parameterized by three 
form factors, corresponding to three different isospin structures,
\begin{equation}
V_a(q,  K) = -\frac{i}{\Fp}
\left[F_1(q,K)\tau_a +F_2(q,K) \delta_{a3} + F_3(q,K)\delta_{a3}\tau_3\right],
\end{equation}
in terms of the functions $F_{1,2,3}$ of $q^\mu$ and 
$ K^\mu = ( p^\mu + p^{\prime\, \mu} )/2$.
In what follows, we give $F_1$, $F_2$, and $F_3$ for each of the dimension-six 
$\slashPT$ sources. We calculate the PNFFs up to the order where each of them 
gets a non-vanishing contribution. As derived in Sec. \ref{piNsector}, 
this implies going to NNLO for qCEDM and FQLR,
and to NLO for $\chi$ISs. 
For qEDM we need the LO contributions only.

At LO and NLO, for all sources but FQLR contributions arise exclusively 
from tree diagrams.
At NLO for the FQLR and at NNLO for the other sources loops appear. 
We use dimensional regularization in $d$ spacetime dimensions, 
which introduces the renormalized scale $\mu$ and 
\begin{equation}
L=\frac{2}{4-d} - \gamma_E + \ln 4\pi,
\end{equation}
where $\gamma_E \simeq 0.557$ is the Euler-Mascheroni constant. 
At these orders
we use $q^0=0$. 

\subsection{qEDM}

The PNFFs from the qEDM are very simple to the order we are interested in, 
since all three PNFFs appear at the same, leading order:
$\Delta_6=2$. We read off from the Lagrangian directly,
\begin{eqnarray}
F_1=\bar g_{0},\qquad F_2=\bar g_{1},\qquad F_3=\bar g_{2}.
\end{eqnarray}
Since these interactions contain suppression factors of
$\alpha_{\mathrm{em}}/4\pi \sim \mpi^3/\MQCD^3$, 
they are most likely not important in calculations of nuclear EDMs, 
which are
dominated by short-range contributions to the nucleon EDM
\cite{Vri12}.

\subsection{$\chi$ISs}

For the gCEDM and chiral-invariant four-quark operators, 
the PNFFs receive contributions at LO and NLO
from 
interactions 
with LECs, respectively, 
$\bar g_0$, $\bar g_1$, and $\bar\zeta_{1}$ in Eq. \eqref{eq:Wein1},
and $\bar g_2$ in Eq. \eqref{weinintphoton}.
The contributions to the PNFFs can be read off easily,
\begin{eqnarray}
F_1(\vec q)&=&\bar g_{0 } - \bar g_{2 } - \bar\zeta_1 \vec q^{\,2},
\\
F_2&=&\bar g_{1 }, 
\\
F_3&=&\bar g_{2 }.
\end{eqnarray}
We conclude that, for the chiral-invariant $\slashPT$ sources, 
$F_1$ and $F_2$ appear at the same order
and $F_3$ appears one order down in the $Q/\MQCD$ expansion. 
Apart from contributing to $F_3$,
$\bar g_{2}$ also contributes to $F_1$. 
This is of no phenomenological interest, since $F_1$ receives
a larger contribution from $\bar g_{0}$. 
At LO $F_1$ depends on the pion momentum because of the
presence of the chiral-invariant operator  $ \bar\zeta_1$.  
The consequences for the $\slashPT$ $N\!N$
potential have been worked out in Ref. \cite{Mae11}, 
where it was shown that $ \bar\zeta_1$  
generates a long-range $\slashPT$ potential, 
which can be accounted for by a redefinition of $\bar g_0$,
and a short-range $\slashPT$ potential, 
which can be absorbed in a nucleon-nucleon contact interaction.

\subsection{qCEDM}

In LO ($\Delta_6=-1$) and NLO ($\Delta_6=0$) the PNFF arises from tree
diagrams where the pion-nucleon vertices are,
respectively, 
the two non-derivative interactions with LECs $\bar g_0$ and $\bar g_1$
in Eq. \eqref{S4V4W3}, 
and the single-derivative interaction with LEC $\bar \beta_3$
in Eq. \eqref{chromo1der}.
At NNLO ($\Delta_6=1$) there are further tree-level contributions
from the two-derivative interactions with LECs $\bar \zeta_1$, 
$\bar \zeta_4$, $\bar \zeta_6$, $\bar \xi_1$, $\bar \xi_4$,
$\bar \xi_5$ and $\bar \xi_7$ 
in Eq. \eqref{eq:2der.1eq:2der.2}, 
and the non-derivative interactions with LECs 
$\delta\bar g_0$, $\delta \bar g_1$ and $\bar g_2$
in Eq. \eqref{cedmmass1cedmmass2}.
Use of the RPI relations \eqref{eq:rpirelations.1} 
shows that the PNFFs can be expressed in terms of
$\bar g_0$, $\bar g_1$, $\bar \beta_3$,
$\bar \zeta_1$, $\bar \xi_1$,
$\delta\bar g_0$, $\delta \bar g_1$ and $\bar g_2$.
The operators with LECs $\bar g_0$, $\bar \zeta_1$ and $\delta\bar g_0$
contribute to $F_1$,
those with LECs $\bar g_1$, $\bar \xi_1$, $\delta \bar g_1$ to $F_2$,
and that with LEC $\bar g_2$ to $F_3$.
The operator with LEC $\bar \beta_3$ and its recoil correction 
contribute to the PNFF
\begin{equation}
V_a(\vec q, \vec K) = \frac{\bar \bt_3}{\Fp}\ep_{3ab}\tau_b 
\left(q^0 -\frac{\vec K\cdot \vec q}{m_N}\right).
\end{equation}
When we use the nucleon on-shell conditions \eqref{onshell},
this contributes to $F_1$ and $F_3$.

\begin{figure}
\centering
\includegraphics[scale = 0.7]{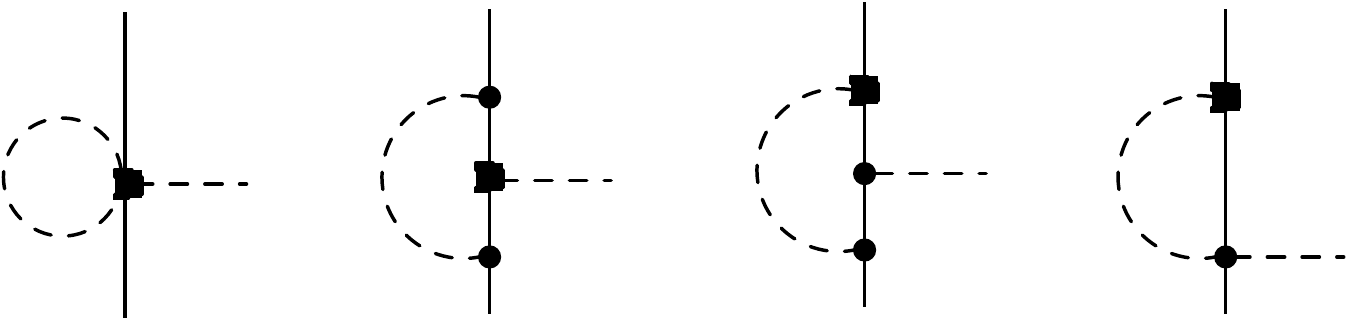}
\caption{One-loop contributions of relative order  
$\mathcal O(m_\pi^2/(2\pi F_\pi)^2)$
to the pion-nucleon form factors 
$F_1(\vec q, \vec K)$ and $F_{2}(\vec q, \vec K)$.
A nucleon (pion) is represented by a solid (dashed) line;
the $\slashPT$ vertices from Eq. \eqref{S4V4W3}
are indicated by a square. The other vertices represent LO 
interactions from Eq. \eqref{LagrCons}.
For simplicity only one possible ordering is shown.} 
\label{Fig:FF.1}
\end{figure}

At NNLO there are also one-loop diagrams
where the $\slashPT$ vertex is one 
of the two non-derivative interactions with LECs $\bar g_0$ and $\bar g_1$
in Eq. \eqref{S4V4W3}, other vertices coming from the LO chiral Lagrangian,
Eq. \eqref{LagrCons}.  These loops are shown in Fig. \ref{Fig:FF.1}.
The structure of the diagrams 
is such that the momentum
of the external pion never flows into the loop and the only scale 
in the integral is the pion mass. As a
consequence, the diagrams do not yield any non-trivial momentum dependence 
and simply renormalize the LECs $\delta \bar g^\prime_{0,1}$ and 
$\delta^\prime \bar g_{0,1}$. 
We define the renormalized LECs
\begin{eqnarray}\label{ct}
\bar {\delta} \bar g_0&=& \delta\bar g^\prime_0 + \delta^\prime \bar g_0
+ \frac{\bar g_0}{4}\frac{ \mpi^2}{ (2\pi \Fp)^2}\left[(1+3g_A^2)
\left( L + 1 - \log \frac{\mpi^2}{\mu^2}\right) - 2g_A^2\right], 
\nonumber\\
\bar {\delta }\bar g_1&=& \delta \bar g^\prime_1 
+ \delta^\prime \bar g_1 +\frac{\bar g_1}{4}
\frac{ \mpi^2}{ (2\pi \Fp)^2}\left[\left(5-9 g_A^2\right)
\left( L + 1 - \log \frac{\mpi^2}{\mu^2}\right) + 6 g_A^2\right].
\end{eqnarray}

With this 
definition the PNFFs for on-shell nucleons read, up to NNLO,
\begin{eqnarray}
F_1 (\vec q, \vec K) &=& \bar g_0 
\left\{1 -\frac{1}{2 m_N^2} 
\left[\vec K^{\,2} +\vec S\cdot  \left(\vec K\times \vec q\right)\right]
\right\} 
+ \bar {\delta} \bar g_0 
+  \delta m_N \bar \bt_3
- \bar\zeta_1 \vec q^{\,2}  ,
\label{F1}\\
F_2 (\vec q, \vec K) &=& \bar g_1 
\left\{1 -\frac{1}{2 m_N^2}
\left[\vec K^{\,2} +\vec S\cdot \left(\vec K\times \vec q\right)\right]
\right\} 
+ \bar {\delta} \bar g_1
- \bar \xi_1 \vec q^{\,2}  
,\label{F2}\\
F_3  &=& \bar g_2 -  \delta m_N \bar \bt_3. 
\label{FFqCEDM}
\end{eqnarray}
We see that $F_1$ and $F_2$ receive contributions at the same order. 
Two orders down we find momentum dependence of these PNFFs, 
as well as the first static contribution to $F_3$.
The implication of the momentum dependence to the 
two-nucleon potential was discussed in Ref. \cite{Mae11}.

\subsection{FQLR}
\label{FFFQLR}

The main part of the PNFFs from FQLR is very similar to 
those from qCEDM.
After replacing the scalings of the LECs, 
one sees that contributions to $F_1$ and $F_2$ start
at LO ($\Delta_6=-3$),
and contributions to $F_3$ at NNLO ($\Delta_6=-1$);
at this order $F_1$ and $F_2$ obtain analytic momentum dependence.

Again, the diagrams in Fig. \ref{Fig:FF.1} only renormalize LECs.
The different chiral structure of the isovector vertices $\bar g_1$ and 
$\bar g_1^{\prime}$ in Eq. \eqref{FQLR0} only affects the first diagram in 
Fig. \ref{Fig:FF.1}, with the minor consequence of modifying the 
counterterm $\bar\delta \bar g_1$ with respect to Eq. \eqref{ct}.
The results in Eqs. \eqref{F1}, \eqref{F2}, and \eqref{FFqCEDM} 
give the dominant
contributions to the three PNFFs, 
with the replacement $\bar g_1 \rightarrow \bar g_1 + \bar g_1^{\prime}$. 
and 
\begin{eqnarray}
\bar {\delta }\bar g_1&=& \delta_1 \bar g_1 + \delta_2 \bar g_1 
+ \delta^{\prime} \bar g_1 
+ \frac{5 }{4}\left(3 \bar g_1 + \bar g_1^{\prime} \right)
\frac{ \mpi^2}{ (2\pi \Fp)^2}
\left(L + 1 - \log \frac{\mpi^2}{\mu^2}\right)
\nonumber \\
& & + \frac{3 g_A^2}{4}\left( \bar g_1 + \bar g_1^{\prime}\right)
\frac{ \mpi^2}{ (2\pi \Fp)^2}\left[-3
\left(L + 1 - \log \frac{\mpi^2}{\mu^2}\right) + 2 \right].
\end{eqnarray}

\begin{figure}
\centering
\includegraphics[scale = 0.7]{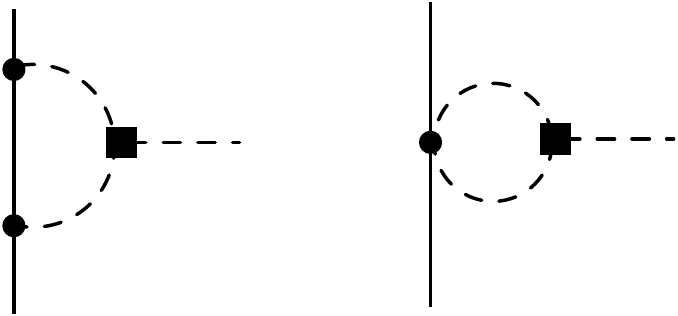}
\caption{One-loop contributions from purely mesonic $\slashPT$ 
interactions to the pion-nucleon form factors. 
The square denotes the $\slashPT$ vertex from Eq. \eqref{eq:3.3.16}.
The other notation is as in Fig. \ref{Fig:FF.1}.} 
\label{FQLRff}
\end{figure}

However, this is not the whole story. 
As shown in Sec. \ref{tad}, the elimination of the tadpoles in
case of the FQLR leaves 
the three-pion vertex with LEC $\bar \Delta_{\mathrm{LR}}^{(-4)}$ 
in Eq. \eqref{eq:3.3.16}
with a lower chiral index than the dominant pion-nucleon interactions. 
The one-loop diagrams in Fig. \ref{FQLRff}
contribute to the PNFFs at NLO ($\Delta_6=-2$) 
and add to $F_2$ a non-analytic momentum dependence not present for the qCEDM.
This PNFF becomes, instead of Eq. \eqref{F2}, 
\begin{eqnarray}
F_2 (\vec q, \vec K) &=& \left(\bar g_1 + \bar g_1^{\prime}\right) 
\left\{1 -\frac{1}{2 m_N^2}
\left[\vec K^{\,2} +\vec S\cdot \left(\vec K\times \vec q\right)\right]
\right\} 
+ \bar {\delta} \bar g_1 
- \bar \xi_1 \vec q^{\,2} 
\nonumber\\
&&
-5 \pi g_A^2  \frac{\mpi\bar \Delta_{\mathrm{LR}}^{(-4)}}{(2\pi\Fp)^2}
\,\, f_2\left(\frac{|\vec{q}|}{2m_\pi}\right),
\label{F2FQLR}
\end{eqnarray}
where
\begin{equation}
f_2 (x) =  1+ \frac{1 +2x^2}{2x}\arctan x .
\label{F2LR}
\end{equation}
Using Eq. \eqref{g0FQLR} we can eliminate $\Delta_{\mathrm{LR}}^{(-4)}$ in favor of
$\bar g_1^{\prime}$ or $\bar g_0$.
Precise measurements on the deuteron $\slashPT$ electromagnetic form factors, 
which depend strongly on 
$F_2$ \cite{Lebedev, Khriplovich:1999qr,
Vri11b}, 
could, in principle, measure this momentum dependence and separate the 
FQLR from the qCEDM. 

\section{Electromagnetic interactions}
\label{Ngammasector}

The experimental interest in EDMs brings $\slashPT$
electromagnetic interactions to the forefront.
Some interactions of hadrons with soft photons can be obtained using 
the $U(1)$-gauge covariant derivatives \eqref{eq:minimal.1} 
in existing operators.
More interesting are the interactions that arise through
the field strength $F_{\mu\nu}$, which we describe here. 

Since the pion has spin $0$, we cannot construct an EDM operator
in the $f = 0$ sector.
In contrast, there are plenty of 
$\slashPT$ interactions
in the $f=2$ sector.

\subsection{qCEDM}

In the case of the qCEDM, operators containing the electromagnetic 
field strength have the chiral properties of the tensor product of 
the $\slashPT$ source and the $C\! P$-even electromagnetic
interactions.
In lowest orders we need
only 
$(-\tilde d_0 \tilde V_4 +\tilde d_3 \tilde W_3)\otimes 
e (I^{\mu}/6+T^{\mu}_{34}/2)$. 

Interactions start at $\Delta_6=1$ and transform as the fourth component 
of a vector, or as the product of a vector and an anti-symmetric tensor,
\begin{eqnarray}
\mathcal L ^{(1)}_{\tilde q, f=2\;\mathrm{em}} &=& 
-2 \Nb \left[\bar d_0 \left(1-\frac{2\boldpi^2}{\Fp^2 D}\right) 
+ \bar d_1\left(\tau_3-\frac{2\pi_3}{\Fp^2 D}\boldtau\cdot\boldpi\right)\right]
S^\mu  N\, v^\nu \Fmu
\nonumber\\
&&- 2 \bar d'_1 \left(1-\frac{2\boldpi^2}{\Fp^2 D}\right)\Nb 
\left[\tau_3-\frac{2}{\Fp^2 D}
\left(\boldpi^2\tau_3-\pi_3 \boldpi\cdot\boldtau\right)\right]
S^\mu  N\, v^\nu \Fmu
\nonumber\\
&&+\frac{1}{\Fp D} \ep^{\mu \nu \alpha \beta} v_\al 
\Nb \left(\bar c_0 \boldtau\cdot \boldpi +\bar c_1 \pi_3\right) S_\bt N \,\Fmu
\nonumber\\
&&+ \bar c_2 \frac{\pi_3}{\Fp D}  \ep^{\mu \nu \alpha \beta} v_\al 
\Nb \left[\tau_3 -\frac{2}{\Fp^2 D}
\left(\boldpi^2\tau_3-\pi_3 \boldpi\cdot\boldtau\right)\right] S_\bt N  \,\Fmu,
\label{EMthetad0EMd1}
\end{eqnarray}
where the electromagnetic LECs scale as
\begin{eqnarray}
&&\qquad\quad
\bar d_{0,1},\bar c_{0,1} = 
\Or\left( e \left(\tilde \delta_0+  \tilde \delta_3\right) 
\frac{\mpi^2}{M^2_{\slashT}\MQCD}\right),
\nonumber\\
&&\bar d'_{1} = 
\Or\left( e \tilde \delta_0 \frac{\mpi^2}{M^2_{\slashT}\MQCD}\right),
\qquad
\bar c_{2} = 
\Or\left( e \tilde \delta_3 \frac{\mpi^2}{M^2_{\slashT}\MQCD}\right).
\label{EMqcedm0qCEDMLECem}
\end{eqnarray}
The first term in Eq. \eqref{EMthetad0EMd1} is a short-range contribution 
to the isoscalar nucleon EDM. The second and third terms both contribute 
to the isovector nucleon EDM, and they differ only by interactions 
involving two or more pions, such that separating them is practically 
impossible. The last three terms in Eq. \eqref{EMthetad0EMd1} are 
pion-nucleon-photon interactions which play a role in the calculation 
of the deuteron 
MQM \cite{Vri11b, Liu12}.

As always, the isoscalar qCEDM generates the same interactions as the 
$\tb$ term \cite{BiraEmanuele}. 
But, while in the pion and pion-nucleon  
sectors the isoscalar qCEDM generated only isoscalar interactions in LO, 
in the electromagnetic 
sector both isoscalar and isovector interactions appear, 
due to breaking of chiral symmetry by the quark electric charge. 
The main extra feature of the isovector qCEDM
is the appearance of the $\bar c_2$ interaction, which has a 
$\tau_3 \pi_3$ structure. However, it is unlikely that this operator 
has any important physical consequences.

The elimination of the leading tadpole modifies to coefficients in 
Eq. \eqref{EMthetad0EMd1}  in a way that can be schematically summarized 
by the replacement 
$\tilde\delta_0\rightarrow \tilde\delta_0 + \varepsilon \tilde\delta_3$ 
in Eq. \eqref{EMqcedm0qCEDMLECem}. Since most operators already receive 
contributions from both the isoscalar and isovector qCEDM these shifts 
are not particularly interesting. 

The Schiff moment of the nucleon \cite{Vri11a, Mer11} and EDMs 
of light nuclei \cite{Vri12} are dominated by 
$\slashPT$ pion-nucleon interactions. 
The LO $\slashPT$ nucleon-photon operators 
have little impact on these observables, so we do not construct operators 
with higher chiral index, 
which would be even more suppressed.

\subsection{FQLR}

The interactions stemming from FQLR are similar to those
from the qCEDM. 
Already at LO ($\Delta_6=-1$) FQLR generates operators with 
a more complicated pion structure 
due to the $X_{34}\otimes T^\mu_{34}/2$ tensor product
in $\mathrm{Im}\Xi\, X_{34}\otimes e (I^{\mu}/6+T^{\mu}_{34}/2)$. 
However, apart from terms with two or more pions,
the Lagrangian is identical to Eq. \eqref{EMthetad0EMd1}:
\begin{equation}
\mathcal L ^{(-1)}_{\textrm{LR}, f=2\;\mathrm{em}} = 
\mathcal L ^{(1)}_{\tilde q, f=2\;\mathrm{em}}+\ldots,
\label{EMFQLR}
\end{equation}
but with the scaling
\begin{equation}
\bar d_{0,1},\,\bar d'_{1},\, \bar c_{0,1},\,\bar c_{2} = 
\Or\left( e \xi \frac{\MQCD}{M^2_{\slashT}}\right).
\label{FQLRLECem}
\end{equation}
Tadpole extermination brings in no important new features.
Like for the qCEDM, operators with additional pions and/or higher chiral index
have no obvious phenomenological interest.

\subsection{qEDM}
\label{NgammaqEDM}

Since the qEDM contains a photon field, it yields nucleon-photon operators that 
transform like itself, namely
as the third and fourth components of, respectively, the vectors $V$ and $W$ 
in Eq. \eqref{VW}. At LO this generates
\begin{eqnarray}
\mathcal L ^{(1)}_{q, f=2\;\mathrm{em}} &=& 
-2 \Nb \left[\bar d_0 \left(1-\frac{2\boldpi^2}{\Fp^2 D}\right) 
+\bar d_1\left(\tau_3-\frac{2\pi_3}{\Fp^2 D} \boldtau\cdot\boldpi\right)\right]
S^\mu 
N\, v^\nu  \Fmu
\nonumber\\
&&+ \frac{1}{\Fp D} \ep^{\mu \nu \alpha \beta} v_{\alpha}  
\Nb (\bar c_0 \boldtau\cdot \boldpi +\bar c_1 \pi_3) S_{\beta}
N  \, \Fmu,
\label{quarkedm}
\end{eqnarray}
with
\begin{equation}
\bar d_0, \bar c_0 = 
\Or\left( e \delta_0  \frac{\mpi^2}{M^2_{\slashT}\MQCD}\right),
\qquad 
\bar d_1, \bar c_1 = 
\Or\left( e \delta_3  \frac{\mpi^2}{M^2_{\slashT}\MQCD}\right).
\end{equation}
In contrast to the qCEDM, the isoscalar (isovector) 
qEDM generates isoscalar (isovector) nucleon-photon interactions,
since the symmetry properties of the qEDM need
not be mixed with 
chiral-symmetry breaking due to the quark charge.

In the case of the qEDM, long-range physics propagated by pions is suppressed 
by powers of $\alpha_{\mathrm{em}}$, and $\slashPT$ observables are dominated 
by short-range nucleon-photon interactions. 
Since the operators in Eq. \eqref{quarkedm} contribute to the nucleon EDM only
and not to the momentum-dependent part of the corresponding form factor, 
for the latter we need to construct electromagnetic operators with 
higher chiral index. 
It turns out that momentum dependence arises only at NNLO, so that we need 
to construct the Lagrangians with $\Delta_6=2, 3$.

At $\Delta_6=2$, the $\slashPT$ electromagnetic operators that are not 
constrained by Lorentz invariance contain at least one pion,
\begin{eqnarray}
\mathcal L ^{(2)}_{q, f=2\;\mathrm{em}} &=& 
-\frac{i}{m_N}\Nb \left[\bar d_0 \left(1-\frac{2\boldpi^2}{\Fp^2 D}\right) 
+
 \bar d_1\left(\tau_3-\frac{2\pi_3}{\Fp^2 D} \boldtau\cdot\boldpi\right)\right]
S^\mu \mathcal D^{\nu}_{\perp\, -} N\,  \Fmu
\nonumber\\
&&+ \frac{i}{2\Fp m_N D} \ep^{\mu \nu \alpha \beta} 
\Nb (\bar c_0 \boldtau\cdot \boldpi +\bar c_1 \pi_3) S_{\beta}
\mathcal D_{\alpha\, -} N  \,  \Fmu
\nonumber\\
&&
+\frac{1}{F_{\pi}}\Nb \left[ \bar \wp_1
\left(1-\frac{2\boldpi^2}{\Fp^2 D}\right)\left(D^\mu \boldpi\right)\cdot\boldtau
+\bar \rho_1 \left(D^\mu \pi_3 - \frac{2\pi_3}{F^2_{\pi} D} \boldpi \cdot 
D^{\mu} \boldpi \right)  \right] N\, v^\nu \Fmu 
\nonumber\\ 
&& 
+ \frac{1}{\Fp D} \Nb \left( \bar \wp_2 \boldpi\cdot \boldtau 
+ \bar \rho_2 \pi_3 \right)  N\, v^\nu \partial^{\mu} \Fmu 
\nonumber\\
&& 
+ \frac{1}{\Fp^2 D} \ep^{\al\bt\mu\nu} v_\al   
\Nb  \left[\bar \wp_3(\boldpi\cdot D_\bt \boldpi) 
+ \bar \rho_3 \pi_3 \left(D_\bt \boldpi\right) \cdot \boldtau\right] N\, \Fmu 
\nonumber \\ 
&&
+ \frac{1}{\Fp^2 D} \Nb S^\mu \left [ \bar \wp_4(\boldpi\times D^\nu \boldpi)
+ \bar \wp_5 (\boldpi\times v\cdot D \boldpi)v^\nu\right]\cdot\boldtau N \,\Fmu
\nonumber\\
&&
+\frac{1}{\Fp} \ep^{\al\bt\mu\nu} \,\Nb\left[ ( \bar \rho_4 D_\al \boldpi 
+\bar \rho_5 v_\al v\cdot D \boldpi) \times \boldtau\right]_i S_\bt N 
\left(\delta_{i3}-\frac{2\pi_3\pi_i}{\Fp^2 D}\right) \, \Fmu,
\label{delta2}
\end{eqnarray}
where $\bar \wp_i$ ($\bar \rho_i$) originate from the isoscalar (isovector) 
qEDM, with
\begin{equation}
\bar \wp_i = \Or\left( e \delta_0  \frac{\mpi^2}{M^2_{\slashT}\MQCD^2}\right),
\qquad 
\bar \rho_i = \Or\left( e \delta_3  \frac{\mpi^2}{M^2_{\slashT}\MQCD^2}\right).
\end{equation}
In Eq. \eqref{delta2} we already incorporated RPI,
and the first two sets of interactions are recoil
corrections to Eq. \eqref{quarkedm}. 

At $\Delta_6=3$, we encounter operators with two covariant derivatives 
or one insertion of the quark mass. Since it is unlikely that, 
at this order in the chiral series, operators containing pions are 
of any phenomenological use, we focus on terms without pions:
\begin{eqnarray}
{\cal L}_{q, f =2\,\mathrm{em}}^{(3)}&=&
-2\Nb \left(\delta \bar{d}_{0}+\delta \bar{d}_{1}\tau_3\right) S^\mu N
\, v^\nu \Fmu 
\nonumber\\
&& 
+\frac{1}{4 m_N^2} \Nb  \left(\bar{d}_{ 0}+\bar{d}_{1}\tau_3\right)
S \cdot \mathcal D_{\perp -}\mathcal D_{\perp -}^{\mu} N \, v^{\nu} \Fmu 
\nonumber\\
&& 
+ \bar N \left({\bar S}'_{ 0}+{\bar S}'_{ 1}\tau_3\right)
\left(S \cdot \mathcal D_{\perp +} \mathcal D_{\perp +}^{\mu}
      +S^\mu \mathcal D_{\perp +}^2 \right)N \, v^\nu \Fmu 
+\ldots
\label{delta3}
\end{eqnarray}
The first two terms are corrections to the isoscalar and isovector nucleon EDMs 
due the quark mass, the third and fourth terms are relativistic corrections,
and the operators with LECs ${\bar S}'_{0}$ and ${\bar S}'_{ 1}$ 
are the first qEDM contributions to, respectively, the isoscalar and isovector 
Schiff moments. The dots denote multi-pion components of the listed operators, 
and other operators that start at one pion, which we neglect. 
The LECs scale as
\begin{eqnarray}
\delta \bar{d}_{0} =
\Or\left(e\left(\delta_0 + \varepsilon\delta_3 \right)
\frac{\mpi^4}{M_{\slashT}^2\MQCD^{3}}\right),
&& 
\qquad
\delta \bar{d}_{1} =
\Or\left(e\left(\varepsilon\delta_0  + \delta_3 \right)
\frac{\mpi^4}{M_{\slashT}^2\MQCD^{3}}\right),
\nonumber \\
{\bar S}'_{ 0}=
\Or\left(e\delta_0\frac{\mpi^2}{M_{\slashT}^2\MQCD^{3}}\right),
&&
\qquad {\bar S}^{\prime\, }_{ 1}= 
\Or\left(e\delta_3\frac{\mpi^2}{M_{\slashT}^2 \MQCD^{3}}\right).
\label{eq:3.2.18}
\end{eqnarray}

Here tadpoles are much smaller than the nucleon-photon interactions,
and their removal does not affect the operators above.

\subsection{$\chi$ISs}

The $\slashPT$ chiral-invariant sources generate electromagnetic interactions 
that transform as $w\bar I \otimes e (I^\mu/6 + T^\mu_{34}/2)$ 
with chiral index $\Delta_6 = -1$,
\begin{eqnarray}
\mathcal L ^{(-1)}_{w, f=2\; \mathrm{em}}&=& 
-2\Nb \left\{ \bar{d}_{0 }+ \bar{d}_{1 } \left[\tau_3+\frac{2}{\Fp^2 D}
\left(\pi_3\boldpi\cdot\boldtau-\boldpi^{\,2}\tau_3\right)
\right]\right\} S^\mu 
N\,v^\nu \Fmu,
\label{weinem1}
\end{eqnarray} 
where the LECs scale as
\begin{equation}
\bar d^{(-1)}_{0,1} = \mathcal O\left( e w \frac{\MQCD}{M^2_{\slashT}}\right).
\end{equation}
Differently from qCEDM and FQLR, 
for chiral-invariant $\slashPT$ sources the short-distance
EDM operators have the same chiral index as the leading 
pion-nucleon coupling, Eq. \eqref{eq:Wein1}.
Since the latter contributes to the nucleon EDM only via loops, it follows 
that, for $\chi$ISs,
the nucleon EDM is mainly determined by short-distance physics.

A second consequence of the enhancement of short-distance vs. 
long-distance physics
for $\chi$ISs is that the nucleon EDFF does not depend on the momentum transfer
at leading order in $\chi$PT. 
Momentum dependence only arises at NNLO, for which accuracy 
we need to consider the power-suppressed $\Delta_6 = 0, 1$  Lagrangian.
We construct the complete chiral $\Delta_6 = 0$ Lagrangian,
\begin{eqnarray}
\mathcal L ^{(0)}_{w, f=2 \; \mathrm{em}}&=& 
-\frac{i}{m_N} \Nb \left\{ \bar{d}_{0 }+ \bar{d}_{1 } 
\left[\tau_3+\frac{2}{\Fp^2 D}
\left(\pi_3\boldpi\cdot\boldtau-\boldpi^{\,2}\tau_3\right)
\right]\right\} S^\mu \mathcal D_{\perp -}^\nu N\,\Fmu
\nonumber\\
&&+\frac{{\bar\chi_0}}{\Fp} (D^\mu\boldpi) \cdot \Nb \boldtau N\, v^\nu\Fmu 
\nonumber\\
&&+ \frac{1}{\Fp^2 D}\left[\bar\chi_1(\boldpi\times D^\mu \boldpi)_3 
+ \bar\chi_2(\boldpi\times v\cdot D \boldpi)_3 v^\mu\right]  
\Nb S^\nu N \Fmu
\nonumber\\
&&+\frac{\bar{\chi}_3}{\Fp^2 D}\ep^{\al\bt\mu\nu}v_\al 
\Nb\left[ (D_\bt \pi_3)\boldtau\cdot\boldpi
-(\boldpi\cdot D_\bt \boldpi)\tau_3\right] N \,\Fmu 
\nonumber\\
&&+ \frac{1}{\Fp}\left\{\bar{\chi}_4  (D^\mu \pi_i)\,\Nb N v^\nu 
+ \ep^{\al\bt\mu\nu} \Nb\left[\bar{\chi}_5(D_\al \boldpi \times \boldtau)_i 
+ \bar{\chi}_6 (v\cdot D \boldpi \times \boldtau)_i v_\al \right] S_\bt N\right\}
\nonumber\\
&&
\times 
\left[\dt_{i3} 
+ \frac{2}{\Fp^2 D}\left(\pi_3 \pi_i -\boldpi^2 \dt_{i3}\right)\right]\Fmu 
+ \frac{\bar{\chi}_7}{\Fp}\ep^{\al\bt\lambda\mu}v_\al 
\Nb(\boldtau\times \boldpi)^3\mathcal D_{\bt+} S_\lambda N\,v^\nu \Fmu,
\nonumber\\
\label{chi0}
\end{eqnarray}
where the LECs scale as
\begin{equation}
\bar \chi_{i} = \mathcal O\left( \frac{e w}{M^2_{\slashT}}\right).
\end{equation}

At $\Delta_6 = 1$ we only construct the operators that start without pions, 
since operators with
pions are of little phenomenological use. We have
\begin{eqnarray}
\mathcal L ^{(1)}_{w, f=2 \; \mathrm{em}}&=& 
-2\Nb\left[\delta \bar{d}_{0 }\left(1-\frac{2\boldpi^2}{F_\pi^2 D} \right)
+\delta \bar{d}_{1 }
 \left(\tau_3-\frac{2\pi_3}{F_\pi^2 D}\boldpi\cdot\boldtau\right)
\right] S^\mu N \, v^\nu \Fmu
\nonumber\\
&&- 2\, \bar\delta {d}_{1 }^{\prime} \left(1-\frac{2\boldpi^2}{F_\pi^2D}\right)
\Nb\left[\tau_3
         +\frac{2}{F_\pi^2D}\left(\pi_3\boldpi\cdot\boldtau-\boldpi^2\tau_3
         \right)\right]
S^\mu N \, v^\nu \Fmu
\nonumber\\
&&+\frac{1}{4m_N^2}  
\Nb \left\{ \bar{d}_{0 } + \bar{d}_{ 1 } \left[\tau_3+\frac{2}{\Fp^2 D}
      \left(\pi_3\boldpi\cdot\boldtau-\boldpi^{\,2}\tau_3\right)
           \right] \right\}  
 S \cdot \mathcal D_{\perp -}
\mathcal D_{\perp -}^{\mu}N \,  v^{\nu} \Fmu 
\nonumber\\
&&+
\bar N \left({\bar S}^{\prime}_{0 }+{\bar S}^{\prime}_{1 }\tau_3\right)
\left(S \cdot \mathcal D_{\perp +} \mathcal D_{\perp +}^{\mu}
      +S^\mu \mathcal D_{\perp +}^2\right) N 
\, v^\nu \Fmu +\ldots
\label{weindelta1}
\end{eqnarray}
As for qEDM,
the first 
operators 
(with LECs $\delta \bar{d}_{0,1}$ and $\bar\delta {d}_{1 }^{\prime} $) 
are corrections   
to the isoscalar and isovector nucleon EDMs
proportional to the quark masses.
The next interactions
are relativistic corrections to Eq. \eqref{weinem1}. 
Finally,
the 
last operators (with LECs ${\bar S}^{\prime}_{0,1}$) 
are  short-distance contributions to the first derivative of the nucleon EDFF,
the Schiff moment. The scaling of the LECs is
\begin{equation}
\delta \bar d_{0,1} = 
\mathcal O\left(e (1+\varepsilon) w\frac{m^2_{\pi}}{M^2_{\slashT}\MQCD} \right),
\qquad  
\delta \bar d_{1 }^{\prime}= 
\mathcal O\left(e w\frac{m^2_{\pi}}{M^2_{\slashT}\MQCD}\right),
\qquad
\bar S_{0,1}^{\prime} = \mathcal O\left(\frac{e w}{M^2_{\slashT}\MQCD}\right).
\end{equation}

The leading nucleon-photon interactions 
in Eq. \eqref{weinem1} are chiral invariant or transform as 
the 34 component of an antisymmetric tensor. 
In both cases, they are not affected by the field redefinition 
in Eq. \eqref{rotation}.
Only the NNLO operators $\delta \bar d_{0,1}$ and $\delta \bar d'_{1}$, 
which are proportional to the quark masses, are affected by the elimination 
of the tadpoles, in a way that leads to the shift
$w\rightarrow w (1+\varepsilon^2)$.

\section{Nucleon EDM from the FQLR}
\label{FQLREDM}

The electromagnetic interactions of the previous section
allow the calculation of EDFFs.
In fact,
the chiral Lagrangians from the qEDM, qCEDM, and $\chi$ISs 
derived in this article
have already been used to calculate the $\slashPT$ form factors of the nucleon 
and several light nuclei in Refs. \cite{Vri11a, Mer11, Vri11b, Vri12, Liu12}. 
However, in these references the contributions
from the FQLR operator were omitted. In this section we update our previous
work by giving the FQLR contributions to the nucleon EDM to NLO.
Nuclear issues are discussed in the next section.

The nucleon EDFF, following Refs. \cite{BiraHockings, Vri11a}, 
is decomposed as
\begin{equation}
F_i(Q^2) = d_i -S'_i Q^2+ H_i(Q^2),
\label{eq:Gdef}
\end{equation}
where $d_i$ is the isospin-$i$ component of the EDM,
$S'_i$ is the corresponding Schiff moment \cite{Thomas:1994wi}, 
and $H_i(Q^2)$ accounts for the remaining dependence on $Q^2=-q^2>0$,
$q$ being the outgoing momentum of the photon.
The EDFF of the proton (neutron) is $F_0 + F_1$ ($F_0 - F_1$).

The LO $\slashPT$ pion-nucleon and nucleon-photon interactions 
from the FQLR are 
very similar to those from an isovector qCEDM, apart from interactions 
involving 
multiple pions, which contribute at higher orders.
Therefore, the LO nucleon EDFF from the FQLR is of identical form as that 
from the isovector qCEDM calculated in Ref. \cite{Vri11a}, but, of course, with 
different scalings for the LECs. 
The LO nucleon EDFF 
gets contributions from the short-range interactions in Eq. \eqref{EMFQLR}
and from one-loop diagrams involving the LO isoscalar $\slashPT$ 
pion-nucleon vertex $\bar g_0$ in Eq. \eqref{FQLR0} 
({\it cf.} Ref. \cite{CDVW79, Thomas:1994wi, BiraHockings}). 
At NLO, there are additional one-loop contributions, involving both $\bar g_0$ 
and the isovector $\slashPT$ vertex $\bar g_1$, and subleading couplings 
in the $PT$-even Lagrangian 
({\it cf.} Refs. \cite{Narison,ottnad,Mer11}).
One has in principle to consider also the effects of the three-pion coupling 
$\bar \Delta_{\mathrm{LR}}^{(-4)}$ 
in Eq. \eqref{eq:3.3.16}, which  contributes to the nucleon EDM 
via two-loop diagrams. 
However, it is easy to see that all the two-loop diagrams vanish because of 
their isospin structure, so that the nucleon EDM at NLO stemming from the FQLR 
is identical to that generated by the qCEDM.
In the evaluation of loop diagrams we use dimensional regularization,
as in Sec. \ref{PNFF}.
 
The leading loop diagrams do not generate an isoscalar EDFF, which is 
therefore 
purely tree-level and static at LO. The first non-analytic contribution to 
the isoscalar EDM arises at NLO, and it has both $\bar g_0$ and 
$\bar g_1$ pieces. At the same order one also finds the first momentum 
dependence of the isoscalar EDFF, which is proportional to the hadronic 
part of the nucleon mass splitting, $\delta m_N$ in Eq. \eqref{eq:Nmass}, 
and the isoscalar coupling $\bar g_0$. The isoscalar EDFF to NLO is given by
\begin{eqnarray}
d_{0} &=& {\bar d}_{0}  
+ \pi \frac{e g_A \bar g_0}{(2\pi F_{\pi})^2}  
\left[  \frac{3 m_{\pi}}{ 4 m_N} \left(1 + \frac{\bar g_1}{3\bar g_0}\right) 
-\frac{\delta m_N}{m_{\pi}}\right],
\label{EDM0}
\\
S_{0}'&=&  -\frac{\pi }{12 m^2_{\pi}} \frac{e g_A \bar g_0}{(2\pi F_{\pi})^2} 
\frac{\delta m_N}{m_{\pi}},
\label{Schiffchromo0}
\\
H_{0}(Q^2)&=& - \frac{\pi  }{5} \frac{e g_A \bar g_0}{(2\pi F_{\pi})^2}
\frac{\delta m_N}{m_{\pi}} \; h_0^{(1)} \left(\frac{Q^2}{4 m_{\pi}^2}\right),
\label{scalarqCEDM}
\end{eqnarray}
with the function 
\begin{equation}
h_0^{(1)}(x) \equiv 5 \left( \frac{1}{\sqrt{x}}\arctan \sqrt{x} - 1 
+ \frac{x}{3}\right),
\end{equation}
such that $h_0^{(1)}(x) = x^2 + \mathcal O(x^3)$ for $x\ll 1$.

In contrast, for the isovector EDM loop diagrams 
do 
renormalize 
short-distance operators,
in addition to
generating a non-trivial momentum dependence.
Instead of the nucleon mass splitting from the quark masses,
there is a contribution from the electromagnetic
pion mass splitting  $\breve\delta m^2_{\pi}$
in Eq. \eqref{hardEMLag}.
To NLO, the isovector EDM is found to be
\begin{equation}
d_{1} = \bar{d}_1+\bar{d}_1^{\prime}
+\frac{eg_A\bar{g}_0}{(2\pi\Fp)^2}
\left[ L-\ln\frac{\mpi^2}{\mu^2}  
+\frac{5\pi}{4} \frac{m_{\pi}}{m_N} \left(1 + \frac{\bar g_1}{5\bar g_0}\right) 
-\frac{\breve{\delta}m^2_{\pi}}{m^2_{\pi}} \right],
\label{nEDMLOFQLR}
\end{equation}
while the momentum dependence is given by  
\begin{eqnarray}
S_{1}'&=& \frac{1}{6\mpi^2}
\frac{eg_A\bar{g}_0}{(2\pi\Fp)^2} \left[ 1 - \frac{5\pi}{4} \frac{m_{\pi}}{m_N} 
-\frac{\breve{\delta} m^2_{\pi}}{m^2_{\pi}}\right],
\label{Schiffchromo1} 
\\
H_{1}(Q^2)&=& \frac{4}{15}\frac{eg_A\bar{g}_0}{(2\pi\Fp)^2}  
\left[ h_1^{(0)}\left(\frac{Q^2}{4\mpi^2}\right) 
-\frac{7\pi}{8} \frac{m_{\pi}}{m_N} \; h_1^{(1)}\left(\frac{Q^2}{4 m^2_{\pi}}\right)
-\frac{2\breve\delta m^2_{\pi}}{m^2_{\pi}} \; 
\breve h^{(1)}_1\left(\frac{Q^2}{4 m^2_{\pi}}\right)\right].
\label{momdepvectorqCEDM}
\end{eqnarray}
Here the functions 
\begin{eqnarray}
h_1^{(0)}(x) & \equiv &  
-\frac{15}{4}\left[\sqrt{1+\frac{1}{x}}
\ln \left(\frac{\sqrt{1+\frac{1}{x}}+1}{\sqrt{1+\frac{1}{x}}-1}\right) 
- 2\left(1 +\frac{x}{3}\right)\right],  
\\
h_1^{(1)}(x) & \equiv & -\frac{1}{7} \left[3(1+2x)\; h_1^{(0)}(x) - 10 x^2\right], 
\\
\breve{h}_1^{(1)}(x) & \equiv & -\frac{1}{4(1+x)} \left(  h_1^{(0)}(x)-5x^2\right)
\label{eq:function}
\end{eqnarray}
are defined 
so that they satisfy $h_1^{(i)}(x\ll 1)=x^2+{\cal O}(x^3)$.

The dependence on the arbitrary scale $\mu$ from the loops
is compensated by the counterterm $\bar{d}_1+\bar{d}_1^{\prime}$.
In fact, the loop contributions cannot be separated from the short-range
pieces in a model-independent way.
However, one does not expect cancellations between the non-analytic
dependence in the pion mass and the analytic dependence of the short-range part.
Taking the long-range part in Eq. \eqref{nEDMLOFQLR} at $\mu=m_N$
as an estimate for the neutron EDM, and using relation \eqref{g0FQLR}
with $\delta m_N\simeq 2.3$ MeV \cite{latticedeltamN},
\begin{equation}
\abs{d_n}
\simge 
\frac{eg_A\delta m_N \abs{\bar\Delta_{\mathrm{LR}}^{(-4)}}}{2\pi^2 m_{\pi}^2 \Fp^2 }
\ln\frac{m_N}{\mpi}
\simeq 0.003  \frac{\abs{\bar\Delta_{\mathrm{LR}}^{(-4)}}}{F^2_{\pi}} 
\; e\,\mathrm{fm}.
\label{nEDMFQLR}
\end{equation}
The expected scaling of $\bar \Delta_{\mathrm{LR}}^{(-4)}$ 
in Eq. \eqref{tadscalFQLR} 
with $\MQCD \sim 2\pi\Fp$ 
gives
$\abs{d_n}
\sim 0.8\,\mathrm{GeV}\, e\abs{\xi}/M^2_{\slashT}$,
which is about two thirds of the direct NDA estimate 
for $\bar d_1$ and $\bar d_1'$,
Eq. \eqref{FQLRLECem}. We therefore expect the latter to
provide a better estimate for the neutron EDM, and 
\begin{equation}
\abs{d_n}
\sim 1.2\,\mathrm{GeV}\,  \frac{e\abs{\xi}}{M^2_{\slashT}}.
\end{equation}

Note that the neutron EDM was computed from general $\slashPT$ 
four-quark operators in Ref. \cite{ji}, where it was found to be
$\abs{d_n} \simeq 0.15\, \mathrm{GeV}\, e\, \abs{C_4}$
in terms of $C_4$, the coupling constant of the four-quark operators. 
We compare to our result by writing $C_4 = (4\pi)^2 \xi/M_{\slashT}^2$, 
and find that 
it is about an order of magnitude larger than our estimate.
This discrepancy can be traced \cite{youngpeople}
to the use of a relativistic formalism
in Ref. \cite{ji}.
In this reference, a large contribution to the nucleon EDM 
comes from pion loops in which the photon couples to the nucleon 
via the nucleon anomalous magnetic moment, while $\slashPT$ 
is provided by the coupling $\bar g_1$. 
While a chiral logarithm $\log m_{\pi}^2/m_N^2$
appears with a suppression of $m_{\pi}^2/m_N^2$,
the contributions from
loop momenta $|\vec k| \sim m_N$ give rise to a large EDM, 
which does not have such a suppression.
In the heavy-baryon formalism, the nucleon magnetic moment does 
contribute to the nucleon EDM, but only at NNLO, and should 
reproduce 
the $\log m_{\pi}^2$ dependence of the relativistic calculation 
\cite{youngpeople}. 
Unfortunately, $\chi$PT cannot be trusted for momenta where
the nucleon is relativistic.
Thus, we 
interpret the result of Ref. \cite{ji} as a model-dependent 
estimate of the size of the counterterms $\bar d_{1}$, $\bar d_{1}^{\prime}$ 
and $\bar d_0$, which, in this calculation, appears to be somewhat 
larger than NDA.

\section{Nuclear effects}
\label{NNsector}

So far we have focused on interactions involving at most one nucleon.
Power counting is more complicated for processes involving more than one 
nucleon:
infrared enhancements due to purely nucleonic states require a resummation
of leading interactions \cite{original}, 
and this resummation in turn leads to the enhancement
of certain $C\! P$-even operators that one would naively think are 
of high order \cite{nogga}.
Thus, the chiral index $\Delta$ in Eq. \eqref{Delta} 
is not particularly 
useful due to non-perturbative renormalization.

On the other hand, $\slashPT$ operators should be treated perturbatively 
and might well be unaffected by this subtlety, although this remains to be 
thoroughly investigated.
It is convenient then
to continue to organize them according to the index $\Delta_6$ 
in Eq. \eqref{Deltadim6},
where $f$ can now take even values larger than 2.
Since the index increases with increasing $f$, the operators most
likely to be relevant have $f=4$.
The most important four-nucleon $\slashPT$ operators 
have been constructed in Ref. \cite{Mae11}.
We give here some of the details for dimension-six sources.
The consequences for the $\slashPT$ nuclear potential induced 
by the qCEDM, $\chi$ISs and qEDM  can be found in Ref. \cite{Mae11}.
Ref. \cite{Mae11} did  not discuss the $\slashPT$ potential stemming 
from the FQLR operator. 
In this section, after constructing the short-range $N\!N$ and $N\!N\pi$ 
operators induced by each $\slashPT$ source, we remedy this omission.

\subsection{$\chi$ISs}
\label{NNchi}

While in the pion-nucleon sector one needs at least either two derivatives 
or one insertion of the quark mass to generate operators from the $\chi$ISs, 
in the nucleon-nucleon
sector one derivative is enough, and it need not include a pion field:
\begin{equation}
\mathcal L ^{(-1)}_{w, f=4}= 
\bar C_{1} \bar N N \, \mathcal \partial_{\mu} (\bar N S^{\mu} N)  
+ \bar   C_{2} \bar N \boldtau N \cdot 
  \mathcal D_{\mu} (\bar N S^{\mu} \boldtau N),
\end{equation}
with scalings
\begin{equation}
\bar C_{1,2}=\mathcal O\left(w \frac{\MQCD}{\Fp^2 M_{\slashT}^2}\right).
\end{equation}
We see that for the $\chi$ISs, the chiral index of the LO $f=4$ LECs 
is the same as that of the LO $f=2$ LECs, in contrast with 
sources that are chiral breaking.
Thus, whereas for $\tb$, qCEDM, and FQLR the potential is dominated
by pion exchange (and photon exchange for qEDM), 
the potential for the $\chi$ISs  gets
contributions of the same order from pion exchange 
and short-range 
$N\!N$ 
interactions \cite{Mae11}.

The first terms linear in the pion field appear one order higher,
\begin{equation}
\mathcal L ^{(0)}_{w, f=4}= 
\frac{\mathcal D_\mu \boldpi}{\Fp}\cdot 
\left(\bar G_{1}\Nb S^\mu \boldtau N \, \mathcal D_\nu (\Nb S^\nu N) 
+ \bar G_{2} \Nb S^\mu N \, \mathcal D_\nu (\Nb \boldtau  S^\nu N)\right),
\end{equation}
with scalings
\begin{equation}
\bar G_{1,2}=\mathcal O\left(\frac{w}{\Fp^2 M_{\slashT}^2}\right).
\end{equation}
When the pion is attached to another nucleon, a three-nucleon force results.

The interactions above are not affected by tadpole removal because 
the operators are chiral invariant.
At NNLO we find additional $\slashPT$ 
nuclear interactions, which we do not construct. 
At this order the first contributions to the isovector 
$\slashPT$ 
nucleon-nucleon interactions appear as well, 
originating from an insertion of the quark mass difference.

\subsection{qCEDM}
\label{NNqCEDM}

Because it is chiral breaking, the qCEDM generates pion interactions
at lower order than short-range, purely nucleonic interactions.
At $\Delta_6=0$,
\begin{eqnarray}
\mathcal L^{(0)}_{\tilde q, f=4} &=& 
-\frac{1}{F_{\pi} D} \boldpi \cdot 
\left(\bar\gamma_{1} \bar N \boldtau N \,\bar N N
+\bar\gamma_{2} \bar N \boldtau S_{\mu} N \,\bar N S^{\mu}N\right)
\nonumber\\
&& 
-\frac{\pi_3}{F_{\pi} D} 
\left(\bar\gamma_{3} \bar N N \,\bar N N 
+ \bar\gamma_{4} \bar N S_\mu N \,\bar N S^{\mu} N\right) ,
\label{LTVpiNNNN0LTVpiNNNN1}
\end{eqnarray}
where, after tadpole extermination, the LECs scale as 
\begin{equation}
\bar \gamma_{1,2}=\mathcal O\left( \left(\tilde\delta_0
+ \varepsilon \tilde \delta_3\right)\frac{m^2_{\pi}}
{F^2_{\pi}M^2_{\slashT}}\right),
\qquad
\bar\gamma_{3,4}
=\mathcal O\left(\tilde\delta_3 
\frac{m^2_{\pi}}{F^2_{\pi}M^2_{\slashT}}\right).
\end{equation}

One order down, we find the first short-range interactions contributing 
directly to nucleon-nucleon
scattering, 
\begin{eqnarray}
\mathcal L^{(1)}_{\tilde q, f=4} &=& 
\left(1 - \frac{2\boldpi^2}{F^2_{\pi}D}\right) 
\left[\bar  C_{1} \bar N N \, \partial_{\mu} (\bar N S^{\mu} N)  
+ \bar C_{2} \bar N \boldtau N \cdot 
  \mathcal D_{\mu} (\bar N S^{\mu} \boldtau N) \right]
\nonumber\\
&&+\left(\delta_{3i} - \frac{2\pi_3\pi_i}{F^2_{\pi}D}\right)
\left[\bar C_{3} \bar N \tau_i N \,\partial_{\mu} 
\left(\bar N S^{\mu} N\right)
+ \bar  C_{4} 
\bar N  N \, \mathcal D_{\mu} \left(\bar N \tau_i S^{\mu} N \right) \right],
\end{eqnarray}
with the scaling
\begin{equation}
\bar C_{1,2}=\mathcal O\left(\left(\tilde\delta_0 
+\varepsilon \tilde \delta_3\right)
\frac{m^2_{\pi}}{F^2_{\pi}M^2_{\slashT} \MQCD}\right),
\qquad 
\bar  C_{3,4}=
\mathcal O\left(\tilde\delta_3 \frac{m^2_{\pi}}{F^2_{\pi}M^2_{\slashT}\MQCD}
\right).
\label{CqCEDM}
\end{equation}
Here, as before,
the LECs of the operators 
induced by the isoscalar qCEDM 
get a contribution from the isovector qCEDM as well, proportional
to $\varepsilon$. 
At this order, for all sources, there appear operators that start 
with one or more pions. We do not list them.

A comparison between the 
pion-nucleon and nuclear sectors shows that for qCEDM
the most important $\slashPT$ pion-nucleon
interactions are larger by a factor $\MQCD^2/Q^2$ than the
short-range  
$\slashPT$ 
nucleon-nucleon interactions, implying that for these sources the 
$\slashPT$ 
nuclear potential is dominated by pion exchange \cite{Mae11}. 
This observation justifies, \textit{a posteriori}, 
the assumption often made in the literature, that $\slashPT$ 
nuclear observables can be calculated solely from 
$\slashPT$ pion exchange.
However, as we have seen, 
for 
$\chi$ISs this assumption is not valid.

\subsection{FQLR}
\label{NNFQLR}

As we saw already in the pion and pion-nucleon sectors, 
the isovector qCEDM and FQLR generate very similar interactions, 
if one neglects the more complicated pion structure of the FQLR.

Here again,
\begin{equation}
\mathcal L^{(-2)}_{\textrm{LR}, f=4} = \mathcal L^{(0)}_{\tilde q, f=4} 
+ \dots,
\label{LTVpiFQLR}
\end{equation}
where ``$\ldots$'' represent interactions with more pion fields,
which are of little importance. The LECs scale as 
\begin{equation}
\bar \gamma_{1,2}=\mathcal O\left( 
\varepsilon \xi 
\frac{\MQCD^2}{F^2_{\pi}M^2_{\slashT}}\right),
\qquad
\bar\gamma_{3,4}
=\mathcal O\left(\xi \frac{\MQCD^2}{F^2_{\pi}M^2_{\slashT}}\right)
\end{equation}
after tadpole removal.
Similarly,
\begin{equation}
\mathcal L^{(-1)}_{\textrm{LR}, f=4} = \mathcal L^{(1)}_{\tilde q, f=4} 
+\ldots,
\end{equation}
with
\begin{equation}
\bar C_{1,2}=\mathcal O\left(\varepsilon\xi 
\frac{\MQCD}{F^2_{\pi}M^2_{\slashT} }\right),
\qquad 
\bar  C_{3,4}=
\mathcal O\left(\xi \frac{\MQCD}{F^2_{\pi}M^2_{\slashT}}
\right).
\label{CFQLR}
\end{equation}

As for qCEDM, one-pion exchange is more important than 
these short-range interactions \cite{Mae11},
which we do not pursue further.

\subsection{qEDM}

The qEDM, while sharing with qCEDM and FQLR the property
that interactions involving at least one pion dominate \cite{Mae11},
produces a richer isospin structure for the short-range
nucleon-nucleon interactions, due to the integration
of hard photon.

At $\Delta_6=3$ we find interactions that transform as the tensor
product $(- d_0  V_4 +d_3  W_3)\otimes e (I^{\mu}/6+T^{\mu}_{34}/2)$,
\begin{eqnarray}
\mathcal L^{(3)}_{q, f=4} &=& 
-\frac{1}{F_{\pi} D} \boldpi \cdot 
\left(\bar\gamma_{1} \bar N \boldtau N \,\bar N\!N
+\bar\gamma_{2} \bar N \boldtau S_{\mu} N \,\bar N S^{\mu}N\right) 
\nonumber\\
&&-\frac{\pi_3}{F_{\pi} D} \left(\bar\gamma_{3} \bar N\!N \,\bar N\!N 
+ \bar\gamma_{4} \bar N S_\mu N \,\bar N S^{\mu} N\right)
\nonumber\\
&& - \frac{\pi_3}{\Fp D}
\left[\delta_{3i} +  \frac{2}{\Fp^2 D}
\left(\pi_3 \pi_i-\boldpi^2 \delta_{3i}\right)\right] 
\left[\bar \gamma_5 \Nb \tau_i N \Nb N 
+\bar\gamma_6 \Nb \tau_i S_\mu N \Nb S^\mu N\right] ,
\label{LTVpiNNNNqEDM}
\end{eqnarray}
with
\begin{equation}
\bar\gamma_{1,2,3,4}=\mathcal O\left(\left(\delta_0 +  \delta_3\right) 
\frac{\alpha_{\mathrm{em}}}{4\pi}
\frac{m^2_{\pi}}{F^2_{\pi}M^2_{\slashT}}\right),
\qquad  
\bar\gamma_{5,6}=\mathcal O\left(\delta_3 \frac{\alpha_{\mathrm{em}}}{4\pi}
\frac{m^2_{\pi}}{F^2_{\pi}M^2_{\slashT} }\right).
\end{equation}

One order higher we find
\begin{eqnarray}
\mathcal L^{(4)}_{ q, f=4} &=& 
\left(1 - \frac{2\boldpi^2}{F^2_{\pi}D}\right) 
\left[\bar C_{1} \bar N N \, \partial_{\mu} (\bar N S^{\mu} N)  
+ \bar   C_{2} \bar N \boldtau N \cdot 
\mathcal D_{\mu} (\bar N S^{\mu} \boldtau N) \right] 
\nonumber\\
& & +
\left(\delta_{3i} - \frac{2\pi_3  \pi_i}{F^2_{\pi}D}\right)
\left[\bar C_{3} \bar N \tau_i N \,\partial_{\mu} \left(\bar N S^{\mu} N\right)
+ \bar C_{4} \bar N  N \, \mathcal D_{\mu} \left(\bar N \tau_i S^{\mu} N \right) 
\right]\nonumber\\
&& + 
\left\{\left(1 - \frac{2\boldpi^{2}}{F^2_{\pi} D}\right) 
\left[
\bar C_5 \bar N \tau_l N \, \partial_{\mu} \left(\bar N S^{\mu} N\right) 
+
\bar C_6  \bar N  N\, \mathcal D_{\mu} \left(\bar N \tau_l S^{\mu} N \right) 
\right]
\right.
\nonumber \\
& & \left.
+ \bar C_7 
\left(\delta_{3k} - \frac{2\pi_3 \pi_k }{F^2_{\pi}D}\right)
\bar N \tau_k N \,
\mathcal D_{\mu} \left( \bar N S^{\mu}\tau_l N\right) \right\}
\left[\delta_{l 3} + \frac{2}{F^2_{\pi} D} 
\left(\pi_3 \pi_l - \boldpi^{2} \delta_{3 l}\right)\right]
\nonumber\\
&& +\ldots,
\label{LTVNNNNqEDM}
\end{eqnarray} 
where we have ignored operators that start with one or more pions, and
\begin{eqnarray} \label{CqEDM}
&&\bar C_{1,2,3,4}=\mathcal O\left((\delta_0 +  \delta_3) 
\frac{\alpha_{\mathrm{em}}}{4\pi}
\frac{m^2_{\pi}}{F^2_{\pi}M^2_{\slashT}\MQCD}\right),
\qquad  
\bar C_{5,6}=\mathcal O\left(\delta_0 
\frac{\alpha_{\mathrm{em}}}{4\pi}\frac{m^2_{\pi}}{F^2_{\pi}M^2_{\slashT} \MQCD}
\right),\nonumber\\
&&  
\bar C_{7}=\mathcal O\left(\delta_3 \frac{\alpha_{\mathrm{em}}}{4\pi}
\frac{m^2_{\pi}}{F^2_{\pi}M^2_{\slashT} \MQCD }\right).
\end{eqnarray}
Since these operators are all
suppressed by $\alpha_{\mathrm{em}}/4\pi$, 
their phenomenological impact is minimal, and we do
not construct operators with higher chiral index.

\subsection{The $\slashPT$ potential from the FQLR}
\label{Potential}

In Ref. \cite{Mae11} 
it was found that for the qCEDM the two-body $\slashPT$ potential 
is dominated by one-pion exchange (OPE),
where one of the pion couplings is a
$\slashPT$ coupling from Sec. \ref{piNsector}. 
Short-range nucleon-nucleon interactions play a role at NNLO.
On the other hand, for $\chi$ISs
short-range interactions appear at LO, 
while for the qEDM both pion and photon exchange contribute 
to the LO potential, which, however, 
does not usually play an important role in the calculation of EDMs.
For qCEDM, qEDM and $\chi$ISs, few-body $\slashPT$ forces are suppressed. 
For FQLR, we will see that the forces are somewhat different.
We will show that, in LO, the two-nucleon $\slashPT$ potential is 
dominated by OPE and it is very similar to the potential induced by the qCEDM.
However, 
the first loop correction appears already at NLO. 
The most striking feature of the FQLR is, however,
that the three-pion vertex $\bar \Delta_{\mathrm{LR}}^{(-4)}$ induces 
a three-nucleon $\slashPT$ force at LO, which impacts the calculation of the 
EDMs of $^3$He and $^3$H. 

We write the potential in terms of the spin, isospin, and 
incoming (outgoing) momentum of nucleon $i$,
$\vec \sigma^{(i)}$, $\boldtau^{(i)}$, and $\vec p_i$ ($\vec p^{\,\prime}_i$)
respectively.
We denote the transferred momenta by
$\vec q_i = \vec p_i - \vec p^{\,\prime}_i$.
In the two-nucleon case, $\vec q_2 =-\vec q_1$,
while for three nucleons, $\vec q_3 =-(\vec q_1+\vec q_2)$.
The coordinate-space version of the potential can be
obtained straightforwardly following the procedure of,
for example, Ref. \cite{Mae11}.

At LO in $\chi$PT, the $\slashPT$ two-nucleon potential is given by OPE, 
with $T$ violation provided by the pion-nucleon 
couplings of Sec. \ref{FQLRpiN}. 
The LO two-nucleon potential is thus identical to that generated 
by the qCEDM, and it is given by
\begin{eqnarray}\label{pot1}
V^{(-3)}_{2,\textrm{FQLR}}(\vec q_1)  &=& 
i \frac{g_A}{F^2_{\pi}} 
\frac{\vec q_1}{\vec q_1^{\, 2} + m_{\pi}^2} \cdot 
\left\{\bar g_0 \; \boldtau^{(1)} \cdot \boldtau^{(2)} 
\left(\vec{\sigma}^{(1)}-\vec{\sigma}^{(2)}\right)\right.
\nonumber \\ 
& & \left.
+ \frac{(\bar g_1 + \bar g_1^{\prime})}{2} 
\left[\left(\tau_3^{(1)}+ \tau_3^{(2)}\right) 
\left( \vec{\sigma}^{(1)} -\vec{\sigma}^{(2)} \right) 
+ \left(\tau_3^{(1)} - \tau_3^{(2)}\right) 
\left(\vec\sigma^{(1)}+ \vec\sigma^{(2)}\right) \right]
\right\} .
\end{eqnarray}
Since the couplings $\bar g_0$ and $\bar g_1^{\prime}$ are not independent, 
this potential 
depends on two low-energy constants, 
which we can choose to be $\bar g_1$ and $\bar g_1^{\prime}$
using Eq. \eqref{ratiog0g1}.
The isoscalar piece of the potential is suppressed by 
the smallness of the ratio of the nucleon mass difference and 
the sigma term.
Unless $\bar g_1$ is unnaturally tuned to cancel the contribution of 
$\bar g_1^{\prime}$, the two-nucleon potential is mainly isovector.

In Sec. \ref{FFFQLR} we showed that the isovector PNFF $F_2$ receives 
loop corrections, and in particular non-analytic momentum dependence, at NLO. 
These corrections are proportional to the three-pion vertex 
$\bar \Delta_{\mathrm{LR}}^{(-4)}$, which is related to $\bar g_1^{\prime}$ 
by Eq. \eqref{g0FQLR}, and are enhanced by a factor of $\pi$ 
with respect to NDA. Their contribution to the isovector potential 
can thus be sizable, and, furthermore, it comes with a different 
momentum dependence with respect to the LO OPE.

The deuteron EDM and MQM induced by the FQLR at LO can therefore be read off  
the calculations for the qCEDM of Refs. \cite{Vri12, Liu12}, 
after one accounts for the suppression of $\bar g_0$. 
At NLO, the deuteron EDM and MQM receive contributions from the 
three-pion vertex $\bar\Delta_{\mathrm{LR}}^{(-4)}$, and the NLO corrections 
could be important.
While we postpone a detailed analysis of the $\bar\Delta_{\mathrm{LR}}^{(-4)}$ 
correction 
to the deuteron EDM, we expect the main qualitative conclusion of 
Ref. \cite{Vri12} to continue to hold
---that is, that the deuteron EDM induced by an isospin-breaking 
$\slashPT$ source is expected to be significantly larger than the 
isoscalar nucleon EDM.

\begin{figure}
\centering
\includegraphics[scale = 0.7]{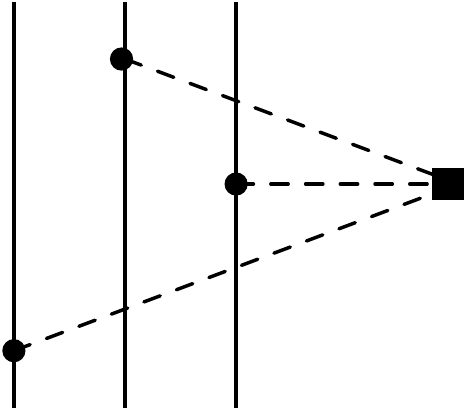}
\caption{Contribution from the purely mesonic $\slashPT$ 
interaction to the three-nucleon potential. 
Notation as in Fig. \ref{FQLRff}.} 
\label{3body}
\end{figure}

The three-pion vertex $\bar \Delta_{\mathrm{LR}}^{(-4)}$ 
has even more striking consequences 
in the three-nucleon system.
This interaction induces a three-nucleon potential, depicted in 
Fig. \ref{3body}, which in the power counting of Ref. \cite{Vri12} 
enters at the same order as the LO two-nucleon potential. 
The three-nucleon potential is of the form
\begin{eqnarray}
V^{(-3)}_{3,\textrm{FQLR}}(\vec q_1, \vec q_2)  &=& 
\frac{2g_A^3 \bar \Delta_{\mathrm{LR}}^{(-4)}}{\Fp^4}
\left(\tau_3^{(1)}\,\boldtau^{(2)}\cdot \boldtau^{(3)}
+ \tau_3^{(2)}\,\boldtau^{(1)}\cdot \boldtau^{(3)}
+ \tau_3^{(3)}\,\boldtau^{(1)}\cdot \boldtau^{(2)}\right)
\nonumber\\
&&\times 
\frac{\vec \sigma^{(1)}\cdot \vec q_1 \; \vec \sigma^{(2)}\cdot \vec q_2\; 
\vec \sigma^{(3)}\cdot (\vec q_1+\vec q_2)}
{(\vec q_1^{\, 2} +\mpi^2)(\vec q_2^{\, 2} +\mpi^2)
[(\vec q_1 + \vec q_2)^2+\mpi^2]}.
\label{3bodypotential}
\end{eqnarray}
The size of the contributions from Eq. \eqref{3bodypotential} can be compared 
to those from the LO two-nucleon potential \eqref{pot1} by using the 
power-counting rules outlined in Ref. \cite{Vri12}. 
These rules indicate that the three-body contributions scale exactly 
as the LO two-body contributions.
Equation \eqref{3bodypotential} provides a correction to the 
helion and triton EDMs, 
which, by power counting, is of the same size as the two-body terms 
calculated in Ref. \cite{Vri12}. 
While we expect that the main qualitative conclusions of Ref. \cite{Vri12}
---that for an isospin-breaking source the EDMs of helion and triton
are significantly different from the EDMs of their constituents--- to hold, 
we leave a quantitative analysis to future work.

Notice that, as always, power counting provides only a guide for what 
should be included in a calculation. In particular, the power counting 
for few-nucleon forces has not been significantly probed. 
The one adopted in Ref. \cite{Vri12}, which is in line with the current 
understanding of the two-body system, agrees with the one in 
Ref. \cite{Friar:1996zw}. If, instead, the power counting of 
Refs. \cite{original, vanKolck:1994yi} is employed, 
then Eq. \eqref{3bodypotential} becomes instead an NLO correction.

The discussion of the 
two- and three-nucleon potentials shows that 
for $\slashPT$ observables in light nuclei 
the three-pion vertex $\bar\Delta_{\mathrm{LR}}^{(-4)}$,
which survives tadpole extermination because of the tensor nature of the FQLR,
plays an important role.
Another handle to disentangle the qCEDM and the FQLR could be provided by 
the measurement of the $T$-odd correlation coefficient in $\beta$ decay, 
$D$ \cite{Ng:2011ui}. 
Indeed, the $\Xi_1$ operator in Eq. \eqref{qqHH}, which generates 
at tree level the FQLR operator, 
also contributes to an operator that couples right-handed quarks to 
left-handed electron and neutrino, with coefficient of $\mathcal O(\Xi_1)$.
The qCEDM contributions to the same operator are suppressed by $G_F$ 
due to the need for the exchange of a $W$ boson and, 
because of 
an extra chirality flip, 
by an insertion of the light-quark mass.
Therefore, the contribution of the qCEDM to the operator scales as 
$\tilde d_{0,3} G_F \bar m/ 4\pi \sim G_F \bar m^2 \tilde\delta_{0,3}/M^2_{\slashT}$,
which is suppressed by a factor $\bar m^2/M_W^2$. 
The size of the $D$ coefficient with respect to nuclear EDMs is 
therefore very different for isovector qCEDM and FQLR operators. 
In the case that the observation of nucleon and deuteron EDM points 
to an isospin-breaking $\slashPT$ source,
the measurement of $D$ could help to discriminate between these two 
operators.

\section{Discussion and conclusion}
\label{discussion}

As we have seen, the chiral structure of the various $\slashPT$ sources
is influential in the form and expected magnitude of the low-energy
$\slashPT$ interactions.
Even without going into detailed 
results for $\slashPT$ hadronic observables we can draw some 
qualitative conclusions by looking at the Lagrangian that we constructed. 
In Ref. \cite{BiraEmanuele} it was found that for the Standard Model $\tb$ 
term, 
all LECs are proportional to negative powers of the scale $\MQCD$. 
The reasons for this are twofold: the $\tb$ term 
(\emph{i}) can be seen as a complex quark mass term,
which brings in at least one power of $\mpi^2$, and
(\emph{ii}) transforms, once
vacuum stability is imposed, as the fourth component of an $SO(4)$ vector, 
which means that 
a pion tadpole can be eliminated. 
Just like isospin violation \cite{vanKolck},
time reversal is therefore an accidental symmetry in the SM,
in the sense 
that it would be somewhat suppressed (by at least one power of $\mpi/\MQCD$) 
even if $\tb$ were not small. 

Extending the analysis to the dimension-six operators, we see
that positive powers of $\MQCD$ do appear, but they are of course
overcompensated by two negative powers of the much larger $M_{\slashT}$.
The main result of tadpole extermination
is 
that the vacuum misalignment signaled by the pion tadpoles causes the 
isoscalar operators, like $\bar g_0$ or $\bar C_{1,2}$, to receive 
additional contributions from isospin-breaking sources, 
like the isovector qCEDM and the FQLR operator, of the same importance 
as the contributions of isoscalar sources. 
These contributions can be schematically obtained by replacing 
$\tilde\delta_0$ before tadpole removal
with $\tilde\delta_0 + \varepsilon \tilde \delta_3 
+ \varepsilon \xi M^2_{QCD}/m^2_{\pi}$ 
in the power counting estimates of isoscalar operators. 
Shifts in isovector quantities 
do not change the dependence of the LECs on 
$\xi$, $\tilde \delta_{0,3}$, and $w$, even if possible accidental 
cancellations could affect the numerical values of the isovector LECs.
It is in the case of FQLR that tadpole rotation is 
qualitatively
most important
because, first, FQLR does not have an isoscalar piece and, second,
a relatively important three-pion vertex survives.

We notice that the different
chiral properties and field content of the $\tb$ term, qCEDM, and FQLR 
on the one side, and qEDM and $\chi$ISs on the other, imply very 
different relations between long-distance
and short-distance $\slashPT$ effects. 
The $\tb$ term, qCEDM, and FQLR all violate chiral symmetry
and thus generate $\slashPT$ pion-nucleon interactions 
in which the pion couples to the nucleon non-derivatively. 
As a consequence, the first
$\slashPT$ pion-nucleon couplings appear in the Lagrangian 
two orders before short-range
contributions to the nucleon EDM. For the nucleon EDM and EDFF,
this fact implies that even though pion-nucleon $\slashPT$ couplings 
can only contribute to
the nucleon EDM via loops, which bring in a $\mpi^2/(2\pi\Fp)^2$ 
suppression, they are still
as important as the short-distance operators \cite{BiraHockings, Vri11a}. 
For light nuclei, such as the deuteron \cite{Khriplovich:1999qr,
Vri11b} 
and 
helion \cite{Stetcu:2008vt, Vri12},
the most important contribution to $\slashPT$ electromagnetic moments 
comes from the
$\slashPT$ OPE potential ---which causes the nucleus wavefunction to mix 
with states of
different parity--- unless the admixed component has quantum numbers 
that cause the
dipole matrix element to vanish. 
The application of chiral EFT to study effects
of $\tb$ term and qCEDM in 
systems with $A\geq 2$ nucleons is thus particularly
promising, since $\slashPT$ observables are likely 
to depend on few LECs from
the $f = 2$ $\slashPT$ Lagrangian. 
Once these constants are fixed in experiments, one is
in the position to make testable, model-independent predictions. 
For the FQLR, there appear additional complications for EDMs of nuclei 
due to a $\slashPT$ three-pion interaction, which induces an NLO correction 
to the two-nucleon potential and, more importantly, a LO 
three-nucleon potential.
The contributions of this three-nucleon force to light-nuclear EDMs have, 
so far, not been calculated.

For the chiral-invariant $\slashPT$ sources, instead, 
the pion-nucleon $\slashPT$ couplings
appear in the Lagrangian at the same order 
as short-distance nucleon EDM
operators. This happens because it is not possible
to write a $\slashPT$ chiral-invariant pion-nucleon coupling 
with only one derivative. The
first chiral-invariant $\slashPT$ pion-nucleon coupling must have 
two derivatives, while nonderivative
couplings can be generated by considering the combined effects 
of chiral-invariant $\slashPT$ sources and the chiral-breaking quark mass; 
in any case, pion-nucleon
couplings receive a further suppression of $Q^2/\MQCD^2$. 
It should be noted that this difficulty does not affect
the nucleon-photon and nucleon-nucleon sectors, where operators with
a minimal number of derivatives can be constructed.
The consequence for the nucleon EDM is that it is dominated in this case 
by short-distance
contributions. For light nuclei, $\slashPT$ corrections to the wavefunction now
are not only due to $\slashPT$ pion exchange, but also to short-distance 
nucleon-nucleon interactions. In
general, the increased role of short-distance interactions in the 
case of $\chi$ISs reduces the predictive power of our analysis, 
because of the appearance of more LECs.

Finally, pion physics is suppressed also in the case of $\slashPT$ 
from the qEDM. In this
case the suppression comes from the need to integrate out the photon 
to produce purely hadronic operators,
which leads to a factor of $\alpha_{\mathrm{em}}/4\pi$. 
In this case, EDMs of light nuclei are dominated by the nucleon EDM.

Note that the formalism developed here and in
Ref. \cite{BiraEmanuele} can be extended to the construction of
interactions involving the Delta (1232) isobar explicitly. 
Because the Delta-nucleon mass splitting
is only 290 MeV, some LECs can take unnaturally large
values when the Delta is not included.
To avoid the concomitant limitation in convergence 
\cite{Pandharipande:2005sx,paulo}, 
the Delta 
should be taken into account as an explicit degree of freedom in the
effective Lagrangian, in which case it appears in $\slashPT$
observables at high orders.

Traditionally, when discussing parity and time-reversal violation 
between nucleons and pions, three
non-derivative pion-nucleon interactions are considered on the same footing:
\begin{equation}
\mathcal L_{\slashT, \pi N}= 
-\frac{\bar g_0}{\Fp} \Nb \boldtau\cdot\boldpi N 
-\frac{\bar g_1}{\Fp} \pi_3 \Nb N 
- \frac{\bar g_2}{\Fp} \pi_3 \Nb \tau_3 N.
\end{equation}
When one takes into account the chiral properties of the fundamental sources 
of $\slashPT$, the $\tb$ term and the dimension-six sources in 
Eq. \eqref{dim6}, this picture changes. 
First, these interactions are accompanied by others with further
pion fields.
Second, and more important, the three interactions
are not of the same size, as noted in Ref. \cite{Pospelov:2001ys}. 
For example, in the case of the chiral-symmetry breaking 
but isospin-conserving $\tb$ term, at LO only $\bar g_0$ appears \cite{CDVW79}. 
The couplings $\bar g_1$ and $\bar g_2$ are respectively suppressed by 
two and three powers of $\mpi/\MQCD$ \cite{BiraEmanuele},
although, once values from the connection to isospin violation
in the quark masses \cite{BiraEmanuele} are taken into account,
$\bar g_0$ is numerically smaller than expected \cite{bsaisou}.
If one wants to study observables sensitive 
to $\bar g_1$ one needs to take into account the full NNLO Lagrangian 
constructed above, which includes, apart from $\bar g_1$, 
also derivative pion-nucleon 
and multi-pion-nucleon interactions. An example of this is the $\slashPT$ 
two-nucleon potential which, at the order where $\bar g_1$ appears, has
a rich and non-trivial momentum dependence in the isoscalar channel 
\cite{Mae11}. 

For qCEDM, arguably the most natural case is the one where the isoscalar 
and isovector 
components are of similar size,
$\abs{\tilde \delta_0}\simeq \abs{\tilde \delta_3}$. 
In this scenario the $\bar g_0$ and $\bar g_1$ interactions appear 
at the same order. 
The third pion-nucleon coupling $\bar g_2$ comes in two orders down 
in the chiral expansion. 
The case of a dominant isoscalar qCEDM,
$\abs{\tilde \delta_0} \gg \abs{\tilde \delta_3}$, generates 
an identical low-energy $\slashPT$ Lagrangian as the $\tb$ term, 
making it impossible 
to separate these two scenarios from low-energy $\slashPT$ observables alone. 
To this goal, more input from techniques like lattice QCD is needed.
The appearance of a dominant isovector qCEDM,
$\abs{\tilde \delta_3} \gg \abs{\tilde \delta_0}$, implies that 
$\bar g_0$ and $\bar g_1$ are approximately of the same order, 
although the former is expected to be somewhat smaller due to 
the extra factor $\varepsilon$.

The pattern of non-derivative pion-nucleon interactions 
in the case of the chiral- and 
isospin-symmetry-breaking four-quark operator FQLR is very similar to that of 
a dominant isovector qCEDM. The first difference is the appearance of 
interactions 
involving multiple pions, which are hard to isolate. 
A second difference is the appearance of $\slashPT$ mesonic operators 
which, through loops, 
add to the $\slashPT$ PNFF momentum dependence, however at subleading order. 
The same mesonic operator induces a LO three-nucleon potential,
which could be relevant for nuclei with $A\geq3$.
Regardless, separating the FQLR from an isovector qCEDM with 
hadronic observables requires 
very precise measurements and is, at this point, not very likely. 
The search for $T$ violation in $\beta$ decay could provide additional 
clues \cite{Ng:2011ui}. 

The $\chi$ISs ---gCEDM and two four-quark interactions--- give rise to 
a similar hierarchy 
between the non-derivative pion-nucleon couplings as an isovector qCEDM,
although in this case 
it is $\bar g_1$ that is expected to be smaller than $\bar g_0$ 
by a factor $\varepsilon$.
The LEC $\bar g_2$ is suppressed by only one power of $\mpi/\MQCD$. 
Also, at the same order as $\bar g_0$ and $\bar g_1$,
derivative pion-nucleon interactions appear. 
Within our framework a separation of the different chiral-invariant operators 
themselves is not 
possible. For that more advanced techniques than NDA are required to 
estimate the size of the LECs. 

Only for qEDM are the three non-derivative pion-nucleon couplings
expected to be of the same order. But in this case all these couplings
are 
irrelevant for most applications.
The scalings of these couplings for all sources are summarized
in the first three rows of Table \ref{table}.

\begin{table}[t]
\caption{The LO scaling of important $\slashPT$
LECs for the different $\slashPT$ sources.
The $\bar g_i$ 
are the non-derivative 
$\pi N$ couplings,  
$\bar d_0$ and $\bar d_1$ isoscalar and isovector short-range nucleon EDMs,  
$\bar c_0$ and $\bar c_1$ isoscalar and isovector 
magnetic 
$\g\pi N$ interactions, 
and $\bar C_{1,2}$ and $\bar C_{3,4}$ isoscalar 
and isovector 
$N\!N$ interactions. 
(The $+$ in some of the entries should not be taken literally but only
as an indication that the LECs get contributions from two sources.)}
\begin{center}
\begin{tabular}{||c|ccccc||} \hline
\rule{0pt}{3ex}
Source & $\tb$ & qCEDM &FQLR&  qEDM  & $\chi$ISs  
\tabularnewline
\hline 
\rule{0pt}{4ex}
$\bar g_0$  
& $\tb \frac{\mpi^2}{\MQCD}$    
& $(\tilde\delta_0+\varepsilon \tilde\delta_3)\frac{\mpi^2\MQCD}{M_{\slashTsub}^2}$ 
& $\varepsilon\xi \frac{\MQCD^3}{M_{\slashTsub}^2} $  
& $(\delta_0+\delta_3)\frac{\alpha_{\mathrm{em}}}{4\pi}
\frac{\mpi^2\MQCD}{M_{\slashTsub}^2}$ 
& $w \frac{\mpi^2 \MQCD}{M_{\slashTsub}^2}$ 
\tabularnewline
\rule{0pt}{4ex}
$\bar g_1/\bar g_0$  
& $\varepsilon \frac{\mpi^2}{\MQCD^2}$ 
& $\frac{\tilde \delta_3}{\tilde \delta_0 +\varepsilon \tilde \delta_3}$
& $\frac{1}{\varepsilon}$ 
& $ 1$ 
& $\varepsilon $ 
\tabularnewline 
\rule{0pt}{4ex}
$\bar g_2/\bar g_0$  
& $\frac{\alpha_{\mathrm{em}}}{4\pi} $ 
& $\frac{\varepsilon \tilde \delta_3}{\tilde\delta_0 +\varepsilon\tilde\delta_3}
\frac{\mpi^2}{\MQCD^2}$
& $\varepsilon\frac{\mpi^2}{\MQCD^2}$ 
& $\frac{\delta_3}{\delta_0+\delta_3}$ 
& $ \frac{\alpha_{\mathrm{em}}}{4\pi}\frac{\MQCD^2}{\mpi^2}$
\tabularnewline
\hline
\rule{0pt}{4ex}
$\bar d_0$   
& $e \tb \frac{\mpi^2}{\MQCD^3}$ 
& $e (\tilde \delta_0+ \tilde \delta_3)\frac{\mpi^2}{M_{\slashTsub}^2 \MQCD}$
& $e \xi \frac{\MQCD}{M_{\slashTsub}^2}$ 
& $e \delta_0\frac{\mpi^2}{M_{\slashTsub}^2 \MQCD}$ 
& $e w \frac{\MQCD}{M_{\slashTsub}^2}$
\tabularnewline
\rule{0pt}{4ex}
$\bar d_1/\bar d_0$   
& $1$
&$1$ 
& $ 1$ 
& $\frac{\delta_3}{\delta_0}$ 
& $1$
\tabularnewline
\rule{0pt}{4ex}
$\bar c_0/\bar d_0$   
& $1$
&$1$ 
& $1$ 
& $1$ 
& $\frac{\mpi^2}{\MQCD^2}$
\tabularnewline
\rule{0pt}{4ex}
$\bar c_1/\bar d_0$   
& $1$
&$1$ 
& $1$ 
& $\frac{\delta_3}{\delta_0}$ 
& $\frac{\mpi^2}{\MQCD^2}$
\tabularnewline
\hline
\rule{0pt}{4ex}
$\bar C_{1,2}$  
& $ \tb \frac{\mpi^2}{\Fp^2 \MQCD^3}$ 
& $(\tilde \delta_0+\varepsilon\tilde \delta_3)
\frac{\mpi^2}{\Fp^2 M_{\slashTsub}^2 \MQCD}$
&$\xi\frac{\MQCD }{\Fp^2 M_{\slashTsub}^2 }$ 
& $(\delta_0+\tilde \delta_3)\frac{\alpha_{\mathrm{em}}}{4\pi}
\frac{\mpi^2}{\Fp^2 M_{\slashTsub}^2 \MQCD}$ 
& $w\frac{\MQCD}{\Fp^2 M_{\slashTsub}^2}$
\tabularnewline
\rule{0pt}{4ex}
$\bar C_{3,4}/\bar C_{1,2}$  
& $\varepsilon\frac{\mpi^2}{ \MQCD^2}$ 
& $\frac{\tilde \delta_3}{\tilde \delta_0 +\varepsilon \tilde \delta_3}$
&$\frac{1}{\varepsilon}$ 
& $1$ 
& $\varepsilon\frac{\mpi^2}{\MQCD^2}$
\tabularnewline
\hline 
\end{tabular}
\end{center}
\label{table}
\end{table}

The fact that different sources of $\slashPT$ are responsible for different 
hierarchies between
the non-derivative $\slashPT$ couplings has important implications for
the $\slashPT$ electromagnetic moments of nuclei, of which the best example 
is the deuteron EDM. 
In calculations where the non-derivative pion-nucleon interactions 
are assumed to be equally sized,
it is found that the deuteron EDM is dominated by $\bar g_1$ 
and is larger than the nucleon EDM \cite{Khriplovich:1999qr}. 
The MQM gets contributions of similar size from $\bar g_0$ and $\bar g_1$. 
Similarly, it is found that for the helion EDM both $\bar g_0$ and $\bar g_1$
are important \cite{Stetcu:2008vt}.
Taking into account the chiral properties of the fundamental $\slashPT$ 
sources, these conclusions only hold for the isovector qCEDM and FQLR, 
since for these sources $\bar g_0$ and $\bar g_1$ interactions appear 
at leading order. 
More generally, we see no particular reason to insist
on including $\bar g_2$ for any source, at least in calculations
of light systems.

Instead, what
our analysis shows is that
one should include other types of $\slashPT$ operators, 
such as nucleon-photon, pion-nucleon-photon, and nucleon-nucleon interactions. 
In Ref. \cite{Vri12}, we argued that the EDMs of light nuclei depend on 
six different LECs. 
In addition to
the two $\slashPT$ pion-nucleon interactions $\bar g_0$ and $\bar g_1$, 
they consist of
the short-range isoscalar and isovector nucleon EDMs $\bar d_0$ and $\bar d_1$, 
\begin{equation}
\mathcal L_{\slashT, \gamma N} = 
\Nb \left(\bar d_0 + \bar d_1 \tau_3\right) S^\mu  N\, v^\nu \Fmu+\ldots
\end{equation}
and the short-range isoscalar $\slashPT$ nucleon-nucleon interactions 
$\bar C_1$ and $\bar C_2$,
\begin{equation}
\mathcal L_{\slashT, NN}  = 
\bar C_{1} \bar N N \, \mathcal \partial_{\mu} (\bar N S^{\mu} N)  
+ \bar   C_{2} \bar N \boldtau N \cdot 
  \mathcal D_{\mu} (\bar N S^{\mu} \boldtau N)+\ldots
\end{equation}
The dots in these equations denote terms with additional pions whose forms
depend on the $\slashPT$ source. 
In Ref. \cite{Vri12} the FQLR source was not considered.
As we saw above, the surviving three-pion vertex introduces a dependence
of observables on $\bar \Delta_{\mathrm{LR}}^{(-4)}$. 
However, $\bar \Delta_{\mathrm{LR}}^{(-4)}$
can be eliminated in favor of $\bar g_0$ and $C\!P$-even LECs 
via Eq. \eqref{g0FQLR}, so we 
have at LO for this source also
\begin{eqnarray}
\mathcal L_{\slashT, \pi^3}  =  
-\frac{\mpi^2 \, \bar g_0 }{\delta m_N\Fp }  \pi_3 \boldpi^2 +\ldots
\end{eqnarray}
We can still consider LO nuclear EDMs
to be expressed in terms of those six LECs.
Higher electromagnetic moments
can depend on additional interactions, such as the deuteron MQM
\cite{Vri11a,Liu12} on the
$\slashPT$ pion-nucleon-photon vertices,
\begin{equation}
\mathcal L_{\slashT, \gamma \pi N} = \frac{1}{\Fp} \ep^{\mu \nu \alpha \beta} v_{\alpha}  
\Nb \left(\bar c_0 \boldtau\cdot \boldpi +\bar c_1 \pi_3\right) S_{\beta}
N  \, \Fmu +\ldots
\end{equation}
We summarize the size of these important LECs 
in Table \ref{table}.

The Lagrangians derived in this work for each of the dimension-six 
$\slashPT$ sources  serve as the basis for the calculations of 
hadronic and nuclear
observables.  
In fact, the $\slashPT$ moments of the nucleon 
\cite{BiraHockings,Vri11a,Mer11}
and of light nuclei \cite{Vri11b,Vri12,Liu12,bsaisou}
have already
been calculated with some of the interactions constructed here
and, for the theta term, in Ref.  \cite{BiraEmanuele}.
Other observables such as 
the EDMs and Schiff moments of heavier nuclei
can now be tackled with the same method.

\section*{Acknowledgements}
We thank D. Boer, J. Bsaisou, C. Hanhart, C.-P. Liu, 
I. Stetcu, and A. Wirzba for helpful discussions. 
We thank M. Ramsey-Musolf for helpful discussions and 
comments on the manuscript, and for pointing out a misprint in 
Eq. \eqref{Fierz}.
J. de Vries and E. Mereghetti
acknowledge discussions with W. den Dunnen and W. Dekens, and 
with H. Murayama, respectively.
U. van Kolck acknowledges discussions with W. Hockings
at very early stages of this work,
and the hospitality of the KVI Groningen on many occasions. 
This research was supported  by the Dutch Stichting FOM
under programs 104 and 114 (JdV, RGET), and by the US DOE under contract DE-AC02-05CH11231 with
the Director, Office of Science, Office of High Energy Physics (EM),
and under grant DE-FG02-04ER41338 (UvK).

\end{document}